\documentclass[english]{article}
\usepackage[T1]{fontenc}
\usepackage[latin9]{inputenc}
\usepackage{verbatim}
\usepackage{slashed}
\usepackage{amsmath}
\usepackage{amssymb}
\usepackage{graphicx}
\usepackage{subfig}
\usepackage{float}

\makeatletter
\@ifundefined{date}{}{\date{}}
\usepackage{multicol}
\floatstyle{ruled}
\newfloat{algorithm}{tbp}{loa}

\makeatother

\usepackage{babel}
\begin{document}

\title{Efficient Adaptive Detection of Complex Event Patterns}

\maketitle
\begin{multicols}{2}

\begin{center}
Ilya Kolchinsky\\
Technion, Israel Institute of Technology
\par\end{center}

\begin{center}
Assaf Schuster\\
Technion, Israel Institute of Technology
\par\end{center}

\end{multicols}
\begin{abstract}
Complex event processing (CEP) is widely employed to detect occurrences
of predefined combinations (patterns) of events in massive data streams.
As new events are accepted, they are matched using some type of evaluation
structure, commonly optimized according to the statistical properties
of the data items in the input stream. However, in many real-life
scenarios the data characteristics are never known in advance or are
subject to frequent on-the-fly changes. To modify the evaluation structure
as a reaction to such changes, adaptation mechanisms are employed.
These mechanisms typically function by monitoring a set of properties
and applying a new evaluation plan when significant deviation from
the initial values is observed. This strategy often leads to missing
important input changes or it may incur substantial computational
overhead by over-adapting.

In this paper, we present an efficient and precise method for dynamically
deciding whether and how the evaluation structure should be reoptimized.
This method is based on a small set of constraints to be satisfied
by the monitored values, defined such that a better evaluation plan
is guaranteed if any of the constraints is violated. To the best of
our knowledge, our proposed mechanism is the first to provably avoid
false positives on reoptimization decisions. We formally prove this
claim and demonstrate how our method can be applied on known algorithms
for evaluation plan generation. Our extensive experimental evaluation
on real-world datasets confirms the superiority of our strategy over
existing methods in terms of performance and accuracy.
\end{abstract}

\section{Introduction}

\label{sec:Introduction}

Real-time detection of complex data patterns is one of the fundamental
tasks in stream processing. Many modern applications present a requirement
for tracking data items arriving from multiple input streams and extracting
occurrences of their predefined combinations. Complex event processing
(CEP) is a prominent technology for providing this functionality,
broadly employed in a wide range of domains, including sensor networks,
security monitoring and financial services. CEP engines represent
data items as \textit{events} arriving from event sources. As new
events are accepted, they are combined into higher-level \textit{complex
events} matching the specified patterns, which are then reported to
end users.

One of the core elements of a CEP system is the \textit{evaluation
mechanism}. Popular evaluation mechanisms include non-deterministic
finite automata (NFAs) \cite{WuDR06}, evaluation trees \cite{MeiM09},
graphs \cite{AkdereMCT08} and event processing networks (EPNs) \cite{EtzionN10}.
A CEP engine uses an evaluation mechanism to create an internal representation
for each pattern $P$ to be monitored. This representation is constructed
according to the \textit{evaluation plan}, which reflects the structure
of $P$. The evaluation plan defines how primitive events are combined
into partial matches. Typically, a separate instance of the internal
representation is created at runtime for every potential pattern match
(i.e., a combination of events forming a valid subset of a full match).

As an example, consider the following scenario.

\textbf{Example 1.} A system for managing an array of smart security
cameras is installed in a building. All cameras are equipped with
face recognition software, and periodical readings from each camera
are sent in real time to the main server. We are interested in identifying
a scenario in which an intruder accesses the restricted area via the
main gate of the building rather than from the dedicated entrance.
This pattern can be represented as a sequence of three primitive events:
1) camera A (installed near the main gate) detects a person; 2) later,
camera B (located inside the building's lobby) detects the same person;
3) finally, camera C detects the same person in the restricted area.

\begin{figure}
	\centering
	\subfloat[]{\includegraphics[width=.75\linewidth]{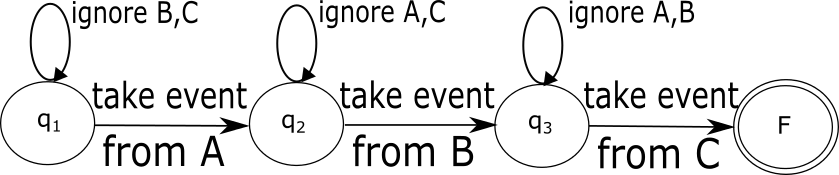}\label{fig:nfa-no-reordering}}\\[2ex]
	\subfloat[]{\includegraphics[width=.75\linewidth]{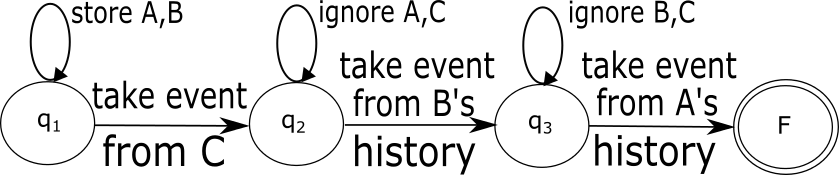}\label{fig:nfa-with-reordering}}\\[2ex]
    \caption{Evaluation structures for a sequence of events from streams \textit{A,B,C}: \protect\subref{fig:nfa-no-reordering} NFA without reordering; \protect\subref{fig:nfa-with-reordering} Lazy NFA with reordering.}
	\label{fig:nfa-sequence}
\end{figure}

Figure \ref{fig:nfa-sequence}\subref{fig:nfa-no-reordering} demonstrates
an example of an evaluation mechanism (a non-deterministic finite
automaton) for detecting this simple pattern by a CEP engine. This
NFA is created according to the following simple evaluation plan.
First, a stream of events arriving from camera A is inspected. For
each accepted event, the stream of B is probed for subsequently received
events specifying the same person. If found, we wait for a corresponding
event to arrive from camera C.

Pattern detection performance can often be dramatically improved if
the statistical characteristics of the monitored data are taken into
account. In the example above, it can be assumed that fewer people
access the restricted area than pass through the main building entrance.
Consequently, the expected number of face recognition notifications
arriving from camera C is significantly smaller than the expected
number of similar events from cameras A and B. Thus, instead of detecting
the pattern in the order of the requested occurrence of the primitive
events (i.e., $A\rightarrow B\rightarrow C$), it would be beneficial
to employ the ``lazy evaluation'' principle \cite{KolchinskySS15}
and process the events in a different order, first monitoring the
stream of events from C, and then examining the local history for
previous readings of B and A. This way, fewer partial matches would
be created. Figure \ref{fig:nfa-sequence}\subref{fig:nfa-with-reordering}
depicts the NFA constructed according to the improved plan.

Numerous authors proposed methods for defining evaluation plans based
on the statistical properties of the data, such as event arrival rates
\cite{AkdereMCT08,KolchinskySS15,MeiM09,Schultz-MollerMP09}. It was
shown that systems tuned according to the a priori knowledge of these
statistics can boost performance by up to several orders of magnitude,
especially for highly skewed data.

Unfortunately, in real-life scenarios this a priori knowledge is rarely
obtained in advance. Moreover, the data characteristics can change
rapidly over time, which may render an initial evaluation plan extremely
inefficient. In Example 1, the number of people near the main entrance
might drop dramatically in late evening hours, making the event stream
from camera A the first in the plan, as opposed to the event stream
from C.

To overcome this problem, a CEP engine must continuously estimate
the current values of the target parameters and, if and whenever necessary,
adapt itself to the changed data characteristics. We will denote systems
possessing such capabilities as \textit{Adaptive CEP (ACEP) }systems.

\begin{figure}
	\centering
	\includegraphics[width=.9\linewidth]{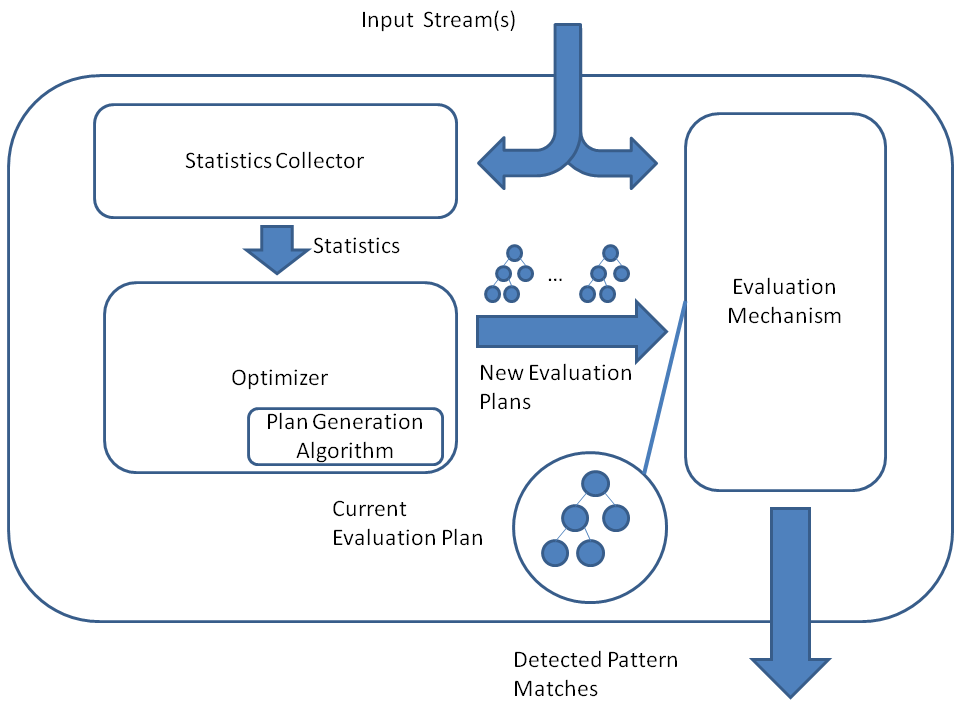}
    \caption{General structure of an adaptive CEP system.}
	\label{fig:cep-evaluation}
\end{figure}

A common structure of an ACEP system is depicted in Figure \ref{fig:cep-evaluation}.
The evaluation mechanism starts processing incoming events using some
initial plan. A dedicated component calculates up-to-date estimates
of the statistics (e.g., event arrival rates in Example 1) and transfers
them to the optimizer. The optimizer then uses these values to decide
whether the evaluation plan should be updated. If the answer is positive,
a plan generation algorithm is invoked to produce a new plan (e.g.,
a new NFA), which is then delivered to the evaluation mechanism to
replace the previously employed structure. In Example 1, this algorithm
simply sorts the event types in the ascending order of their arrival
rates and returns a chain-structured NFA conforming to that order.

Correct decisions by the optimizer are crucial for the successful
operation of an adaptation mechanism. As the process of creating and
deploying a new evaluation plan is very expensive, we would like to
avoid \textit{``false positives,''} that is, launching reoptimizations
that do not improve the currently employed plan. \textit{``False
negatives,''} occurring when an important shift in estimated data
properties is missed, are equally undesirable. A flawed decision policy
may severely diminish or even completely eliminate the gain achieved
by an adaptation mechanism.

The problem of designing efficient and reliable algorithms for reoptimization
decision making has been well studied in areas such as traditional
query optimization \cite{DeshpandeIR07}. However, it has received
only limited attention in the CEP domain (\cite{KolchinskySS15,MeiM09}).
In \cite{KolchinskySS15}, the authors present a structure which reorganizes
itself according to the currently observed arrival rates of the primitive
events. Similarly to Eddies \cite{AvnurH00}, this system does not
adopt a single plan to maintain, but rather generates a new plan for
each newly observed set of events regardless of the performance of
the current one. The main strength of this method is that it is guaranteed
to produce the optimal evaluation plan for any given set of events.
However, it can create substantial bottlenecks due to the computational
overhead of the plan generation algorithm. This is especially evident
for stable event streams with little to no data variance, for which
this technique would be outperformed by a non-adaptive solution using
a static plan.

The second approach, introduced in \cite{MeiM09}, defines a constant
threshold $t$ for all monitored statistics. When any statistic deviates
from its initially observed value by more than $t$, plan reconstruction
is activated. This solution is much cheaper computationally than the
previous one. However, some reoptimization opportunities may be missed.

Consider Example 1 again. Recall that we are interested in detecting
the events by the ascending order of their arrival rates, and let
the rates for events generated by cameras A, B and C be $rate_{A}=100,\, rate_{B}=15,\, rate_{C}=10$
respectively. Obviously, events originating at A are significantly
less sensitive to changes than those originating at B and C. Thus,
if we monitor the statistics with a threshold $t>6$, a growth in
C to the point where it exceeds B will not be discovered, even though
the reoptimization is vital in this case. Alternatively, setting a
value $t<6$ will result in detection of the above change, but will
also cause the system to react to fluctuations in the arrival rate
of A, leading to redundant plan recomputations.

No single threshold in the presented scenario can ensure optimal operation.
However, by removing the conditions involving $t$ and monitoring
instead a pair of constraints $\left\{ rate_{A}>rate_{B},rate_{B}>rate_{C}\right\} $,
plan recomputation would be guaranteed if and only if a better plan
becomes available.

This paper presents a novel method for making efficient and precise
on-the-fly adaptation decisions. Our method is based on defining a
tightly bounded set of conditions on the monitored statistics to be
periodically verified at runtime. These conditions, which we call
\textit{invariants}, are generated during the initial plan creation,
and are constantly recomputed as the system adapts to changes in the
input. The invariants are constructed to ensure that a violation of
at least one of them guarantees that a better evaluation plan is available.

To the best of our knowledge, our proposed mechanism is the first
to provably avoid false positives on reoptimization decisions. It
also achieves notably low numbers of false negatives as compared to
existing alternatives, as shown by our empirical study. This method
can be applied to any deterministic algorithm for evaluation plan
generation and used in any stream processing scenario.

The contributions and the structure of this paper can thus be summarized
as follows:

\textbullet{} We formally define the reoptimizing decision problem
for the complex event processing domain (Section \ref{sec:Notations-and-Problem}).

\textbullet{} We present a novel method for detecting reoptimization
opportunities in ACEP systems by verifying a set of invariants on
the monitored data characteristics and formally prove that no false
positives are possible when this method is used. We also extend the
basic method to achieve a balance between computational efficiency
and precision (Section \ref{sec:Invariant-Based-Method-for}).

\textbullet{} We demonstrate how to apply the invariant-based method
on two known algorithms for evaluation structure creation, the greedy
order-based algorithm (an extended version of \cite{KolchinskySS15})
and ZStream algorithm \cite{MeiM09}, and discuss the generalization
of these approaches to broader categories of algorithms (Section \ref{sec:Applications-of-the}).

\textbullet{} We conduct an extensive experimental evaluation, comparing
the invariant-based method to existing state-of-the-art solutions.
The results of the experiments, performed on two real-world datasets,
show that our proposed method achieves the highest accuracy and the
lowest computational overhead (Section \ref{sec:Experimental-Evaluation}).

\section{Preliminaries}

\label{sec:Notations-and-Problem}

This section presents the notations used throughout this paper, outlines
the event detection process in an ACEP system, and provides a formal
definition of the reoptimizing decision problem, which will be further
discussed in the subsequent sections.

\subsection{Notations and Terminology}

\label{sub:Events-and-Patterns}

A pattern recognized by a CEP system is defined by a combination of
primitive events, operators, a set of predicates, and a time window. 
The patterns are formed using declarative specification languages (\cite{WuDR06,CugolaM10,DemersGHRW06}).

Each event is represented by a type and a set of attributes, including
the occurrence timestamp. Throughout this paper we assume that each
primitive event has a well-defined type, i.e., the event either contains
the type as an attribute or it can be easily inferred from the event
attributes using negligible system resources. We will denote the pattern 
size (i.e., the number of distinct primitive events in a pattern) by $n$.

The predicates to be satisfied by the participating events are usually 
organized in a Boolean formula. Any condition can be specified on any 
attribute of an event, including the timestamp (e.g., for supporting multiple 
time windows).

The operators describe the relations between the events comprising
a pattern match. Among the most commonly used operators are sequence
(SEQ), conjunction (AND), disjunction (OR), negation (typically marked
by '\textasciitilde{}', requires the absence of an event from the
stream) and Kleene closure (marked by '{*}', accepts multiple appearances
of an event in a specified position). A pattern may include an arbitrary
number of operators.

To illustrate the above, consider Example 1 again. We will define
three event types according to the identifiers of the cameras generating
them: A, B and C. For each primitive event, we will set the attribute
\textit{person\_id} to contain a unique number identifying a recognized
face. Then, to detect a sequence of occurrences of the same person
in three areas in a 10-minute time period, we will use the following
pattern specification syntax, taken from SASE \cite{WuDR06}:

\[
\begin{array}{l}
PATTERN\: SEQ\left(A\: a,B\: b,C\: c\right)\\
WHERE\:(\left(a.person\_id=b.person\_id\right)\wedge\\
\quad\quad\quad\quad\quad\left(b.person\_id=c.person\_id\right))\\
WITHIN\:10\: minutes.
\end{array}
\]%

On system initialization, the pattern declaration is passed to the
\textit{plan generation algorithm} $\mathcal{A}$ to create the \textit{evaluation
plan}. The evaluation plan provides a scheme for the CEP engine, according
to which its internal pattern representation is created. The plan
generation algorithm accepts a pattern specification $P$ and a set
of statistical data characteristic values $Stat$. It then returns
the evaluation plan to be used for detection. If these values are
not known in advance, a default, empty $Stat$, is passed. Multiple
plan generation algorithms have been devised, efficiently supporting
patterns with arbitrarily complex combinations of the aforementioned
operators \cite{KolchinskySK17Long,MeiM09}.

In Example 1, $Stat$ contains the arrival rates of event types A,
B and C, the evaluation plan is an ordering on the above types, and
$\mathcal{A}$ is a simple sorting algorithm, returning a plan following
the ascending order of the arrival rates. The CEP engine then adheres
to this order during pattern detection. Another popular choice for
a statistic to be monitored is the set of selectivities (i.e., the
probabilities of success) of the inter-event conditions defined by
the pattern. Examples of plan generation algorithms requiring the
knowledge of condition selectivities are presented in Section \ref{sec:Applications-of-the}.

The plan generation algorithm attempts to utilize the information
in $Stat$ to find the best possible evaluation plan subject to some
predefined set of performance metrics, which we denote as $Perf$.
These metrics may include throughput, detection latency, network communication
cost, power consumption, and more. For instance, one possible value
for $Perf$ in Example 1 is $\left\{ throughput,memory\right\} $,
as processing the events according to the ascending order of their
arrival rates was shown to vastly improve memory consumption and throughput
of a CEP system \cite{KolchinskySS15}.

In the general case, we consider $\mathcal{A}$ to be a computationally
expensive operation. We also assume that this algorithm is optimal;
that is, it always produces the best possible solution for the given
parameters. While this assumption rarely holds in practice, the employed
techniques usually tend to produce empirically good solutions.

\begin{figure}
	\centering
	\subfloat[]{\includegraphics[width=.45\linewidth]{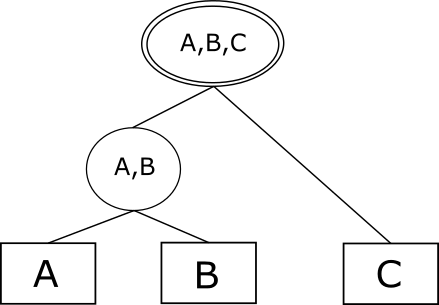}\label{fig:zstream-abc-left}}\quad 	\subfloat[]{\includegraphics[width=.45\linewidth]{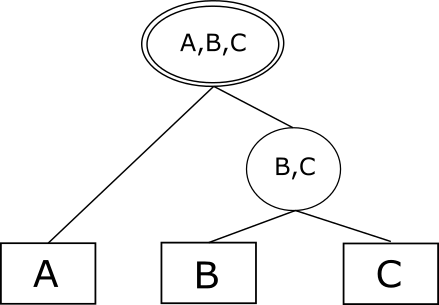}\label{fig:zstream-abc-right}}
    \caption{Evaluation trees produced by ZStream for the sequence of events from streams \textit{A,B,C}: \protect\subref{fig:zstream-abc-left} a left-deep tree; \protect\subref{fig:zstream-abc-right} a right-deep tree.}
	\label{fig:zstream-example}
\end{figure}

An evaluation plan is not constrained to be merely an order. Figure
\ref{fig:zstream-example} demonstrates two possible tree-structured
plans as defined by ZStream \cite{MeiM09}. An evaluation structure
following such a plan accumulates the arriving events at their corresponding
leaves, and the topology of the internal nodes defines the order in
which they are matched and their mutual predicates are evaluated.
Matches reaching the tree root are reported to the end users. From
this point on, we will denote such plans as \textit{tree-based plans},
whereas plans similar to the one used for Example 1 will be called
\textit{order-based plans}. While the methods discussed in this paper
are independent of the specific plan structure, we will use order-based
and tree-based plans in our examples.

\subsection{Detection-Adaptation Loop}

\label{sub:Detection-Adaptation-Loop}

During evaluation, an ACEP system constantly attempts to spot a change
in the statistical properties of the data and to react accordingly.
This process, referred to as the \textit{detection-adaptation loop},
is depicted in Algorithm \ref{alg:Detection-adaptation-loop-in}. 

The system accepts events from the input stream and processes them
using the current evaluation plan. At the same time, the values of
the data statistics in $Stat$ are constantly reestimated by the dedicated
component (Figure \ref{fig:cep-evaluation}), often as a background
task. While monitoring simple values such as the event arrival rates
is trivial, more complex expressions (e.g., predicate selectivities)
require advanced solutions. In this paper, we utilize existing state-of-the-art
techniques from the field of data stream processing \cite{BabcockDMO03,DatarGIM02}.
These histogram-based methods allow to efficiently maintain a variety
of stream statistics over sliding windows with high precision and
require negligible system resources.

Opportunities for adaptation are recognized by the \textit{reoptimizing
decision function} $\mathcal{D}$, defined as follows:
\[
\mathcal{D}:\: STAT\rightarrow\left\{ true,false\right\} ,
\]
where $STAT$ is a set of all possible collections of the measured
statistic values. $\mathcal{D}$ accepts the current estimates for
the monitored statistic values and decides whether reoptimization
is to be attempted%
\footnote{In theory, nothing prevents $\mathcal{D}$ from using additional information,
such as the past and the current system performance. We will focus
on the restricted definition where only data-related statistics are
considered.%
}. Whenever $\mathcal{D}$ returns $true$, the detection-adaptation
loop invokes $\mathcal{A}$. The output of $\mathcal{A}$ is a new
evaluation plan, which, if found more efficient than the current plan
subject to the metrics in $Perf$, is subsequently deployed.

Methods for replacing an evaluation plan on-the-fly without significantly affecting
system performance or losing intermediate results are a major focus of current research
 \cite{DeshpandeIR07}. Numerous advanced techniques were proposed
in the field of continuous query processing in data streams \cite{AlyAOM14,KramerYCSP06,ZhuRH04}.
In our work, we use the CEP-based strategy introduced in \cite{KolchinskySS15}.
Let $t_{0}$ be the time of creation of the new plan. Then, partial
matches containing at least a single event accepted before $t_{0}$
are processed according to the old plan $p_{old}$, whereas the newly
created partial matches consisting entirely of ``new'' events are
treated according to the new plan $p_{new}$. Note that since $p_{old}$
and $p_{new}$ operate on disjoint sets of matches, there is no duplicate processing 
during execution. At time $t_{0}+W$ (where $W$ is the time window of
the monitored pattern), the last ``old'' event expires and the system
switches fully to $p_{new}$.

In general, we consider the deployment procedure to be a costly operation
and will attempt to minimize the number of unnecessary plan replacements.

\begin{algorithm}
\caption{\label{alg:Detection-adaptation-loop-in}Detection-adaptation loop
in an ACEP system}
Input: pattern specification $P$, plan generation algorithm $\mathcal{A}$,
reoptimizing decision function $\mathcal{D}$, initial statistic values
$in\_stat\in STAT$

\hfill

$curr\_plan\Leftarrow\mathcal{A}\left(P,in\_stat\right)$

while more events are available:

\quad process incoming events using $curr\_plan$

\quad$curr\_stat\Leftarrow$ estimate current statistic values

\quad if $\mathcal{D}\left(curr\_stat\right)$: 

\quad\quad$new\_plan\Leftarrow\mathcal{A}\left(P,curr\_stat\right)$

\quad\quad if $new\_plan$ is better than $curr\_plan$: 

\quad\quad\quad$curr\_plan\Leftarrow new\_plan$

\quad\quad\quad apply $curr\_plan$
\end{algorithm}

\subsection{Reoptimizing Decision Problem}

\label{sub:Re-Optimizing-Decision-Problem}

The reoptimizing decision problem is the problem of finding a function
$\mathcal{D}$ that maximizes the performance of a CEP system subject
to $Perf$. It can be formally defined as follows: \textit{given the
pattern specification $P$, the plan generation algorithm $\mathcal{A}$,
the set of monitored statistics $Stat$, and the set of performance
metrics $Perf$, find a reoptimizing decision function $\mathcal{D}$
that achieves the best performance of the ACEP detection-adaptation
loop (Algorithm 1) subject to $Perf$.}

In practice, the quality of $\mathcal{D}$ is determined by two factors.
The first factor is the correctness of the answers returned by $\mathcal{D}$.
Wrong decisions can either fall into the category of \textit{false
positives} (returning $true$ when the currently used plan is still
the best possible) or \textit{false negatives} (returning $false$
when a more efficient plan is available). Both cases cause the system
to use a sub-optimal evaluation plan. The second factor is the time
and space complexity of $\mathcal{D}$. As we will see in Section
\ref{sec:Experimental-Evaluation}, an accurate yet resource-consuming
implementation of $\mathcal{D}$ may severely degrade system performance
regardless of its output.

We can now analyze the solutions to the reoptimizing decision problem
implemented by the adaptive frameworks which we discussed in Section
\ref{sec:Introduction}. The tree-based NFA \cite{KolchinskySS15}
defines a trivial decision function $\mathcal{D}$, unconditionally
returning $true$. In ZStream \cite{MeiM09} this functions loops
over all values in the input parameter $curr\_stat$ and returns $true$
if and only if a deviation of at least $t$ is detected.

\section{Invariant-Based Method for the Reoptimizing Decision Problem}

\label{sec:Invariant-Based-Method-for}

As illustrated above, the main drawback of the previously proposed
decision functions is their coarse granularity, as the same condition
is verified for every monitored data property. We propose a different
approach, based on constructing a set of fine-grained \textit{invariants}
that reflect the existing connections between individual data characteristics.
The reoptimizing decision function $\mathcal{D}$ will then be defined
as a conjunction of these invariants.

In this section, we present the invariant-based decision method and
discuss its correctness guarantees, time and space complexity, and
possible optimizations.

\subsection{Invariant Creation}

\label{sub:Invariant-Creation}

A \textit{decision invariant} (or simply invariant) will be defined
as an inequality of the following form:
\[
f_{1}\left(stat_{1}\right)<f_{2}\left(stat_{2}\right),
\]
where $stat_{1},stat_{2}\in STAT$ are sets of the monitored statistic
values and $f_{1},f_{2}:\, STAT\rightarrow\mathbb{R}$ are arbitrary functions.

We are interested in finding a single invariant for each \textit{building
block} of the evaluation plan in current use. A building block is
defined as the most primitive, indivisible part of a plan. An evaluation
plan can then be seen as a collection of building blocks. For instance, 
the plan for detecting a sequence of three events
of types A, B and C, which we discussed in Example 1, is formed by
the following three blocks:
\begin{enumerate}
\item ``Accept an event of type C'';
\item ``Scan the history for events of type B matching the accepted C'';
\item ``Scan the history for events of type A matching the accepted C and
B''.
\end{enumerate}
In general, in an order-based plan, each step in the selected order will be 
considered a block, whereas for tree-based plans a block is equivalent to 
an internal node. 

We know that the specific plan from the above example was chosen because the plan generation
algorithm $\mathcal{A}$ sorts the event types according to their
arrival rates. If, for instance, the rate of B exceeded that of A,
the second block would have been ``Scan the history for events of
type A matching the accepted C'' and the third would also have changed
accordingly. In other words, the second block of the plan is so defined
because, during the run of $\mathcal{A}$, the condition $rate_{B}<rate_{A}$
was at some point checked, and the result of this check was positive.
Following the terminology defined above, in this example $STAT$ consists
of all valid arrival rate values and $f_{1},f_{2}$ are trivial functions,
i.e., $f_{1}\left(x\right)=f_{2}\left(x\right)=x$.

We will denote any condition (over the measured statistic values)
whose verification has led the algorithm to include some building
block in the final plan as a \textit{deciding condition}. Obviously,
no generic method exists to distinguish between a deciding condition
and a regular one. This process is to be applied separately on any
particular algorithm $\mathcal{A}$ based on its semantics. In our
example, assume that the arrival rates are sorted using a simple min-sort
algorithm, selecting the smallest remaining one at each iteration.
Then, any direct comparison between two arrival rates will be considered
a deciding condition, as opposed to any other condition which may
or may not be a part of this algorithm's implementation.

When $\mathcal{A}$ is invoked on a given input, locations can be
marked in the algorithm's execution flow where the deciding conditions
are verified. We will call any actual verification of a deciding condition
a \textit{block-building comparison (BBC)}. For instance, assume that
we start executing our min-sort algorithm and a deciding condition
$rate_{C}<rate_{A}$ is verified. Further assume that $rate_{C}$
is smaller than $rate_{A}$. Then, this verification is a BBC associated
with the building block ``Accept an event of type C first'', because, unless
this deciding condition holds, the block will not be included in the
final plan. This will also be the case if $rate_{C}<rate_{B}$ is subsequently verified
and $rate_{C}$ is smaller. If $rate_{B}$ is smaller, the opposite
condition, $rate_{B}<rate_{C}$,/ becomes a BBC associated with a block ``Accept
an event of type B first''. Overall, $\left(n-1\right)$ BBCs take place during the
first min-sort iteration, $\left(n-2\right)$ during the second iteration,
and so forth.

In general, for each building block $b$ of any evaluation plan, we can
determine a \textit{deciding condition set (DCS)}. A DCS of $b$ consists
of all deciding conditions that were actually checked and satisfied
by BBCs belonging to $b$ as explained above. Note that, by definition,
the intersection of two DCSs is always empty. In our example, assuming
that the blocks listed above are denoted as $b_{1},b_{2},b_{3}$,
the DCSs are as follows:\begin{align*}
&DCS_{1}=\left\{ rate_{C}<rate_{B},rate_{C}<rate_{A}\right\} ,\\
&DCS_{2}=\left\{ rate_{B}<rate_{A}\right\} ,\\
&DCS_{3}=\varnothing.
\end{align*}%
As long as the above conditions hold, no other evaluation plan can
be returned by $\mathcal{A}$. On the other hand, if any of the conditions
is violated, the outcome of $\mathcal{A}$ will result in generating
a different plan. If we define the decision function $\mathcal{D}$
as a conjunction of the deciding condition sets, we will recognize
situations in which the current plan becomes sub-optimal with high
precision and confidence.

However, verifying all deciding conditions for all building blocks
is very inefficient. In our simple example, the total number of such
conditions is quadratic in the number of event types participating
in the pattern. For more complicated plan categories and generation
algorithms, this dependency may grow to a high-degree polynomial or
even become exponential. Since the adaptation decision is made during
every iteration of Algorithm \ref{alg:Detection-adaptation-loop-in},
the overhead may not only decrease the system throughput, but also
negatively affect the response time.

To overcome this problem, we will constrain the number of conditions
to be verified by $\mathcal{D}$ to one per building block. For each
deciding condition set $DCS_{i}$, we will determine the tightest
condition, that is, the one that was closest to being violated during
plan generation. This tightest condition will be selected as an \textit{invariant}
of the building block $b_{i}$. In other words, we may alternatively define an invariant
as a deciding condition selected for actual verification by $\mathcal{D}$
out of a DCS. More formally, given a set,
\[
DCS_{i}=\left\{ c_{1},\cdots,c_{m}\right\} ;\, c_{k}=\left(f_{k,1}\left(stat_{k,1}\right)<f_{k,2}\left(stat_{k,2}\right)\right),
\]
we will select a condition minimizing the expression
\[
f_{k,2}\left(stat_{k,2}\right)-f_{k,1}\left(stat_{k,1}\right),
\]
as an invariant of the building block $b_{i}$.

In the example above, the invariant for $DCS_{i}$ is $rate_{C}<rate_{B}$,
since we know that $rate_{B}<rate_{A}$, and therefore $rate_{B}-rate_{C}<rate_{A}-rate_{C}$.
It is clear that $rate_{B}$ is a tighter bound for the value of $rate_{C}$
than $rate_{A}$.

To summarize, the process of invariant creation proceeds as follows.
During the run of $\mathcal{A}$ on the current set of statistics
$Stat$, we closely monitor its execution. Whenever a block-building
comparison is detected for some block $b$, we add the corresponding
deciding condition to the DCS of $b$. After the completion of $\mathcal{A}$,
the tightest condition of each DCS is extracted and added to the invariant
list.

\begin{figure*}
	\centering
	\subfloat[]{\includegraphics[width=.3\linewidth]{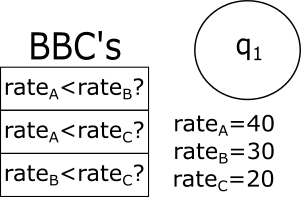}\label{fig:invariants1}}\quad\quad
	\subfloat[]{\includegraphics[width=.4\linewidth]{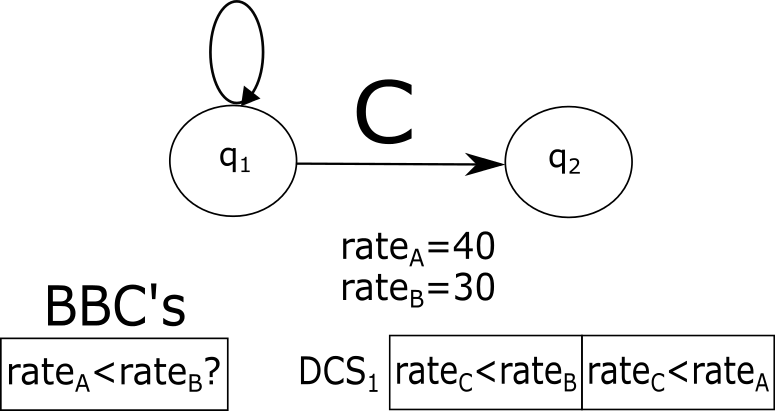}\label{fig:invariants2}}\quad\quad
	\subfloat[]{\includegraphics[width=.4\linewidth]{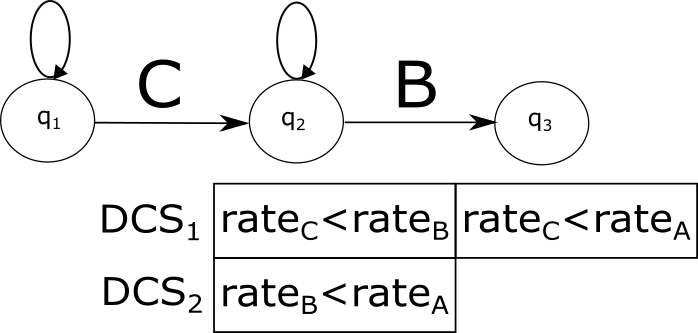}\label{fig:invariants3}}\quad\quad
	\subfloat[]{\includegraphics[width=.4\linewidth]{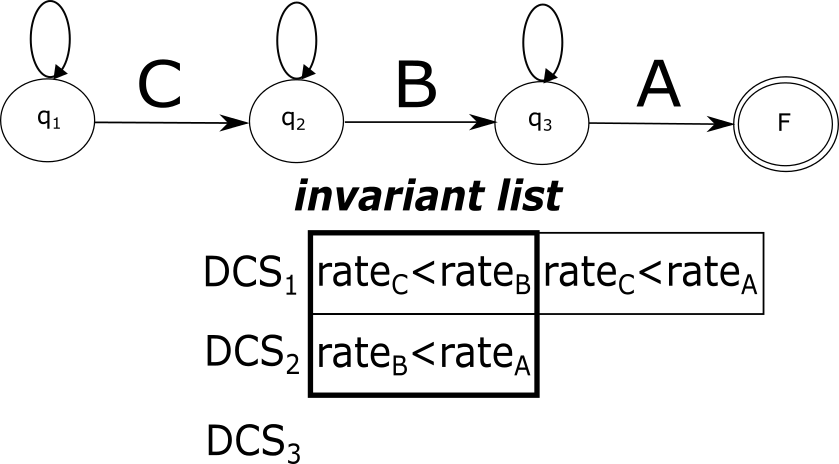}\label{fig:invariants4}}\quad
    \caption{Invariant creation for pattern \textit{SEQ(A,B,C)} from Example 1: \protect\subref{fig:invariants1} selecting the first event type in the detection order; \protect\subref{fig:invariants2} \textit{C} is set as the first event type, and selection of the second event type is in process; \protect\subref{fig:invariants3} \textit{B} is set as the second type, and only a single event type remains for the third position; \protect\subref{fig:invariants4} the evaluation plan and the invariant set are finalized.}
	\label{fig:invariant-creation-example}
\end{figure*}

Figure \ref{fig:invariant-creation-example} demonstrates the invariant
creation process applied on the pattern from Example 1 and the rate-sorting
algorithm $\mathcal{A}$ discussed above. Each subfigure depicts a
different stage in the plan generation and presents the DCSs and the
BBCs involved at this stage.

As discussed above, this generic method has to be adapted to any specific
implementation of $\mathcal{A}$. This is trivially done for any $\mathcal{A}$
which constructs the solution plan in a step-by-step manner, selecting
and appending one building block at a time. However, for algorithms
incorporating other approaches, such as dynamic programming, it is
more challenging to attribute a block-building comparison to a single
block of the plan. In Section \ref{sec:Applications-of-the}, we will
exemplify this process on two algorithms taken from the previous work
in the field and discuss its applicability on broader algorithm categories.

\subsection{Invariant Verification and Adaptation}

\label{sub:Invariant-Verification}

During the execution of the detection-adaptation loop (Algorithm \ref{alg:Detection-adaptation-loop-in}),
$\mathcal{D}$ traverses the list of invariants built as described
above. It returns $true$ if a violated invariant was found (according
to the current statistic estimates) and $false$ otherwise. This list
is sorted according to the order of the respective building blocks
in the evaluation plan. In Example 1, first the invariant $rate_{C}<rate_{B}$
will be verified, followed by $rate_{B}<rate_{A}$. The reason is
that an invariant implicitly assumes the correctness
of the preceding invariants (e.g., $rate_{B}<rate_{A}$ assumes that
$rate_{C}<rate_{B}$ holds; otherwise, it should have been changed
to $rate_{C}<rate_{A}$). For tree-based plans, the verification
proceeds in a bottom-up order. For example, for the tree plan displayed
in Figure \ref{fig:zstream-example}\subref{fig:zstream-abc-left},
the order is $\left(A,B\right)\rightarrow\left(A,B,C\right)$.

If a violation of an invariant is detected, $\mathcal{A}$ is invoked
to create a new evaluation plan. In this case, the currently used
invariants are invalidated and a new list is created following the
process described above. Subsequent verifications performed by $\mathcal{D}$
are then based on the new invariants.

Assuming that any invariant can be verified in constant time and memory,
the complexity of $\mathcal{D}$ using the invariant-based method
is $O\left(B\right)$, where $B$ is the number of the building blocks
in an evaluation plan. This number is bounded by the pattern size
(the number of event types participating in a pattern) for both order-based
and tree-based plans. To guarantee this result, an application of
the invariant-based method on a specific implementation of $\mathcal{A}$
has to ensure that the verification of a single invariant is a constant-time
operation, as we exemplify in Section \ref{sec:Applications-of-the}.

\subsection{Correctness Guarantees and the K-invariant Method}

\label{sub:Correctness-Guarantees}

We will now formally prove that the invariant-based method presented
above guarantees that no false positive detections will occur during
the detection-adaptation loop.

\newtheorem{thrm}{Theorem}
\begin{thrm}

Let $\mathcal{D}$ be a reoptimizing decision function implemented
according to the invariant-based method. Let $\mathcal{A}$ be a deterministic
plan generation algorithm in use and let $p$ be the currently employed
plan. Then, if at some point during execution $\mathcal{D}$ returns
$true$, the subsequent invocation of $\mathcal{A}$ will return a
plan $p'$, such that $p'\neq p$.

\end{thrm}

By definition, if $\mathcal{D}$ returns $true$, then there is at
least one invariant whose verification failed, i.e., its deciding
condition does not hold anymore. Let $c$ be the first such condition,
and let $b_{i}$ be the building block such that $c\in DCS_{i}$ (recall
that there is only one such $b_{i}$). Then, by determinism of $\mathcal{A}$
and by the ordering defined on the invariants, the new run of $\mathcal{A}$
will be identical to the one that produced $p$ until the block-building
comparison that checks $c$. At that point, by definition of the block-building
comparison, the negative result of validating $c$ will cause $\mathcal{A}$
to reject $b_{i}$ as the current building block and select a different
one, thus producing a plan $p'$, which is different from $p$. $\blacksquare$

Since we assume $\mathcal{A}$ to always produce the optimal solution,
the above result can be extended.

\newtheorem{corr}{Corollary}
\begin{corr}

Let $\mathcal{D}$ be an invariant-based reoptimizing decision function
and let $\mathcal{A}$ be a deterministic plan generation algorithm
in use. Then, if at some point during execution $\mathcal{D}$ returns
$true$, the subsequent invocation of $\mathcal{A}$ will return a
plan that is more efficient than the currently employed one.

\end{corr}

Note that the opposite direction of Theorem 1 does not hold. It is
still possible that a more efficient evaluation plan can be deployed,
yet this opportunity will not be detected by $\mathcal{D}$ because
we only pick a single condition from each deciding condition set (see
Section \ref{sub:Dynamic-Programming-Algorithm} for an example).
If we were to include the whole union of the above sets in the invariant
set, even stronger guarantees could be achieved, as stated in the
following theorem.

\begin{thrm}

Let $\mathcal{D}$ be a reoptimizing decision function implemented
according to the invariant-based method, \textbf{with all conditions
from all DCSs} included in the invariant set. Let $\mathcal{A}$ be
a deterministic plan generation algorithm in use and let $p$ be the
currently employed plan. Then, if \textbf{and only if} at some point
during the execution $\mathcal{D}$ returns $true$, the subsequent invocation
of $\mathcal{A}$ will return a plan $p'$, such that $p'\neq p$.

\end{thrm}

The first direction follows immediately from Theorem 1. For the second
direction, let $p'\neq p$ and let $b_{i}\in p,b_{i}'\in p'$ be the
first building blocks that differ in $p$ and $p'$. By $\mathcal{A}$'s
determinism, there exist $f_{1},f_{2},stat_{1},stat_{2}$ s. t.\begin{align*}
&\left(f_{1}\left(stat_{1}\right)<f_{2}\left(stat_{2}\right)\right)\in DCS_{i} \\
&\left(f_{2}\left(stat_{2}\right)<f_{1}\left(stat_{1}\right)\right)\in DCS_{i}',
\end{align*}%
as otherwise there would be no way for $\mathcal{A}$ to deterministically
choose between $b_{i}$ and $b_{i}'$. Since $p'$ was created by
$\mathcal{A}$ using the currently estimated statistic values, we
can deduce that $f_{2}\left(stat_{2}\right)<f_{1}\left(stat_{1}\right)$ holds.
Consequently, $f_{1}\left(stat_{1}\right)<f_{2}\left(stat_{2}\right)$ does
not hold. By the assumption that all deciding conditions are included
in the invariant set, $\mathcal{D}$ will necessarily detect this
violation, which completes the proof. $\blacksquare$

The above result shows that greater precision can be gained if we
do not limit the number of monitored invariants per building block.
However, as discussed above, validating all deciding conditions may
drastically increase the adaptation overhead.

The tradeoff between performance and precision can be controlled by
introducing a new parameter $K$, defined as the maximal number of
conditions from a deciding set to select as invariants. We will refer
to the method using a specific value of $K$ as the \textit{K-invariant
method}, as opposed to the \textit{basic invariant method} discussed
above. Note that the 1-invariant method is equivalent to the basic
one. The K-invariant method becomes more accurate and more time-consuming
for higher values of $K$. The total number of the invariants in this
case is at most $K\cdot\left(B-1\right)$.

\subsection{Distance-Based Invariants}

\label{sub:Distance-Based-Invariants}

By Corollary 1, it is guaranteed that a new, better evaluation plan
will be produced following an invariant violation. However, the magnitude
of its improvement over the old plan is not known. Consider a scenario
in which two event types in a pattern have very close arrival rates.
Further assume that there are slight oscillations in the rates, causing
the event types to swap positions periodically when ordered according
to this statistic. If an invariant is defined comparing the arrival
rates of these two types, then $\mathcal{D}$ will discover these
minor changes and two evaluation plans with little to no difference
in performance will be repeatedly produced and deployed. Although
not a ``false positive'' by definition, the overhead implied by
this situation may exceed any benefit of using an adaptive platform.

To overcome this problem, we will introduce the notion of the minimal
distance $d$, defined as the smallest relative difference between
the two sides of the inequality required for an invariant to be considered
as violated. That is, given a deciding condition
\[
f_{k,1}\left(stat_{k,1}\right)<f_{k,2}\left(stat_{k,2}\right),
\]
we will construct the invariant to be verified by $\mathcal{D}$ as
follows:
\[
\left(1+d\right)\cdot f_{k,1}\left(stat_{k,1}\right)<f_{k,2}\left(stat_{k,2}\right).
\]
The experimental study in Section \ref{sec:Experimental-Evaluation}
demonstrates that a correctly chosen $d$ leads to a significant performance
improvement over the basic technique. However, finding a sufficiently
good $d$ is a difficult task, as it depends on the data, the type
of statistics, the invariant expression, and the frequency 
and magnitude of the runtime changes. We identify the following directions for solving this problem:
\begin{enumerate}
\item Parameter scanning: empirically checking a range of candidate values
to find the one resulting in the best performance. This method is
the simplest, but often infeasible in real-life scenarios.
\item Data analysis methods: calculating the distance by applying a heuristic
rule on the currently available statistics can provide a good estimate
in some cases. For instance, we can monitor the initial execution
of the plan generation algorithm and set $d$ as the average obtained
relative difference between the sides of a deciding condition or,
more formally: 
\[
d=AVG\left(\frac{\left|\left(f_{k,2}\left(stat_{k,2}\right)-f_{k,1}\left(stat_{k,1}\right)\right)\right|}{min\left(f_{k,1}\left(stat_{k,1}\right),f_{k,2}\left(stat_{k,2}\right)\right)}\right).
\]
The effectiveness of this approach depends on the distribution and the 
runtime behavior of the statistical values. Specifically, false positives may 
occur when the values are very close and the changes are frequent. Still, 
we expect it to perform reasonably well in the common case. This technique 
can also be utilized to produce a starting point for parameter scanning.
\item Meta-adaptive methods: dynamically tuning $d$ on-the-fly to adapt it 
to the current stream statistics. This might be the most accurate and 
reliable solution. We start with some initial value, possibly obtained using 
the above techniques. Then, as invariants are violated and
new plans are computed, modify $d$ to prevent repeated reoptimization
attempts when the observed gain in plan quality is low.  An even higher 
precision can be achieved by additionally utilizing fine-grained per-invariant 
distances. This advanced research direction is a subject for our future work.
\end{enumerate}
We implement and experimentally evaluate the first two approaches
in Section \ref{sec:Experimental-Evaluation}.

\subsection{Tightest Conditions Selection Strategy}

\label{sub:Tightest-Invariant-Selection}

In Section \ref{sub:Invariant-Creation} we explained that, given
a DCS for a block $b$, the condition to be included in the invariant
set is the one with the smallest difference between the sides of the
inequality (according to the currently observed values of the statistics).
The intention of this approach is to pick a condition most likely
to be violated later. This, however, is merely a heuristic. In many
cases, there may be no correlation at all between the difference of
the currently observed values and the probability of the new values
to violate the inequality. Hence, this selection strategy may result
in suboptimal invariant selection.

However, sometimes the information regarding the expected variance
of a data property is either given in advance or can be calculated
to some degree of precision and even approximated on-the-fly \cite{BabcockDMO03}.
In these cases, a possible optimization would be to explicitly calculate
the violation probability of every deciding condition and use it as
a metric for selecting an invariant from a deciding condition set.

\section{Applications of the Invariant-Based Method}

\label{sec:Applications-of-the}

In Section \ref{sec:Invariant-Based-Method-for}, we presented a generic
method for defining a reoptimizing decision function $\mathcal{D}$
as a list of invariants. As we have seen, additional steps are required
in order to apply this method to a specific choice of the evaluation
plan structure and the plan generation algorithm. Namely, the following
should be strictly defined: 1)what is considered a building block
in a plan; 2)what is considered a block-building comparison in $\mathcal{A}$;
3)how we associate a BBC with a building block. Additionally, efficient
verification of the invariants must be ensured. In this section, we
will exemplify this process on two plan-algorithm combinations taken
from previous works in the field. The experimental study in Section
\ref{sec:Experimental-Evaluation} will also be conducted on these
adapted algorithms. We also discuss how the presented techniques can
be generalized to several classes of algorithms.

\subsection{Greedy Algorithm for Order-Based Plans}

\label{sub:Greedy-Algorithm-for}

The greedy heuristic algorithm based on cardinalities and predicate
selectivities was first described in \cite{Swami89} for creating
left-deep tree plans for join queries. The algorithm supports all operators described in Section
\ref{sub:Events-and-Patterns} and their arbitrary composition. Its
basic form, which we describe shortly, only targets conjunction and
sequence patterns of arbitrary complexity. Support for other operators
and their composition is implemented by either activating transformation
rules on the input pattern or applying post-processing steps on the
generated plan (e.g., to augment it with negated events). As these
additional operations do not affect the application of the invariant-based
method, we do not describe them here. The reader is referred to \cite{KolchinskySK17Long}
for more details.

The algorithm proceeds iteratively, selecting at each step the event
type which is expected to minimize the overall number of partial matches
(subsets of valid pattern matches) to be kept in memory. At the beginning,
the event type with the lowest arrival rate (multiplied by the selectivities
of any predicates possibly defined solely on this event type) is chosen.
At each subsequent step $i;\, i>1$, the event type to be selected
is the one minimizing the expression
\[
\prod_{j=1}^{i}r_{p_{j}}\cdot\prod_{j,k\leq i}sel_{p_{j},p_{k}},
\]
where $r_{x}$ stands for the arrival rate of the $x^{th}$ event
type in a pattern, $sel_{x,y}$ is the selectivity of the predicate
defined between the $x^{th}$ and the $y^{th}$ event types (equals
to 1 if no predicate is defined), $p_{1},\cdots,p_{i-1}$ are the
event types selected during previous steps, and $p_{i}$ is the candidate
event type for the current step. Since a large part of this expression
is constant when selecting $p_{i}$, it is sufficient to find an event
type, out of those still not included in the plan, minimizing the
expression
\[
r_{p_{i}}\cdot sel_{p_{i},p_{i}}\cdot\prod_{k<i}sel_{p_{k},p_{i}}.
\]

\begin{algorithm}
\caption{Greedy algorithm for order-based plan generation (basic form) \label{alg:Greedy-algorithm-for}}

Input: event types $e_{1},\cdots,e_{n}$, arrival rates $r_{1},\cdots,r_{n}$,
inter-event predicate selectivities $sel_{1,1},\cdots,sel_{n,n}$

Output: order-based evaluation plan $E=e_{p_{1}},e_{p_{2}},\cdots,e_{p_{n}}$

\hfill

$E\Leftarrow\varnothing$

$p_{1}=argmin_{j}\left\{ r_{j}\cdot sel_{j,j}\right\}$

add $e_{p_{1}}$ to $E$

for $i$ from 2 to $n$:

\quad $p_{i}=argmin_{j\notin E}\left\{ r_{j}\cdot sel_{j,j}\cdot\prod_{k<i}sel_{p_{k},j}\right\} $

\quad add $e_{p_{i}}$ to $E$

return $E$
\end{algorithm}

Algorithm \ref{alg:Greedy-algorithm-for} depicts the plan generation
process. When all selectivities satisfy $sel_{x,y}=1$, i.e., no predicates
are defined for the pattern, this algorithm simply sorts the events
in an ascending order of their arrival rates.

We will define a building block for order-based evaluation plans produced
by Algorithm \ref{alg:Greedy-algorithm-for} as a single directive
of processing an event type in a specific position of a plan. That
is, a building block is an expression of the form ``Process the event
type $e_{j}$ at $i^{th}$ position in a plan''. Obviously, a full
plan output by the algorithm will contain exactly $n$ blocks, while
a total of $O\left(n^{2}\right)$ blocks will be considered during
the run. Deciding conditions created for such a building block will
be of the following form:
\[
r_{j}\cdot sel_{j,j}\cdot\prod_{k<i}sel_{p_{k},j}<r_{j'}\cdot sel_{j',j'}\cdot\prod_{k<i}sel_{p_{k},j'}.
\]
Here, $e_{j'},j'\neq j$ is an event type which was considered to
occupy $i^{th}$ position at some point but eventually $e_{j}$ was
selected. Note that, while in the worst case the products may contain
up to $n-1$ multiplicands, in most cases the number of the predicates
defined over the events in a pattern is significantly lower than $n^{2}$.
Therefore, invariant verification will be executed in near-constant
time.

\subsection{Dynamic Programming Algorithm for Tree-Based Plans}

\label{sub:Dynamic-Programming-Algorithm}

\begin{algorithm}
\caption{ZStream algorithm for tree-based plan generation\label{alg:ZStream-algorithm-for}}

Input: event types $e_{1},\cdots,e_{n}$, arrival rates $r_{1},\cdots,r_{n}$,
inter-event predicate selectivities $sel_{1,1},\cdots,sel_{n,n}$

Output: a tree-based evaluation plan $T$

\hfill

$subtrees\Leftarrow$ new two-dimensional matrix of size $n\times n$

for $i$ from 1 to $n$:

\quad$subtrees[i][1].cardinality=subtrees[i][1].cost=r_{i}$

for $i$ from 2 to $n$:

\quad for $j$ from 1 to $n-i+1$:

\quad\quad for $k$ from $j+1$ to $j+i$:

\quad\quad \quad$new\_cardinality=Card($

\quad\quad \quad\quad\quad\quad$subtrees[k-j][j].cardinality,$

\quad\quad \quad\quad\quad\quad$subtrees[i-(k-j)][k].cardinality)$

\quad\quad \quad$new\_cost=subtrees[k-j][j].cost\,+$

\quad\quad \quad\quad$+\, subtrees[i-(k-j)][k].cost+new\_cardinality$

\quad\quad \quad if $new\_cost<subtrees[i][j].cost:$

\quad\quad \quad\quad$subtrees[i][j].tree=new\_tree($

\quad\quad \quad\quad\quad$subtrees[k-j][j],subtrees[i-(k-j)][k])$

\quad\quad \quad\quad$subtrees[i][j].cardinality=new\_cardinality$

\quad\quad \quad\quad$subtrees[i][j].cost=new\_cost$

return $subtrees[n][1].tree$
\end{algorithm}

The authors of ZStream \cite{MeiM09} introduced an efficient algorithm
for producing tree-based plans based on dynamic programming (Algorithm
\ref{alg:ZStream-algorithm-for}). The algorithm consists of $n-1$
steps, where during the $i^{th}$ step the tree-based plans for all
subsets of the pattern of size $i+1$ are calculated (for the trees
of size 1, the only possible tree containing the lone leaf is assumed).
During this calculation, previously memoized results for the two subtrees
of each tree are used. To calculate the cost of a tree $T$ with the
subtrees $L$ and $R$, the following formula is used:
\[
Cost\left(T\right)=\begin{cases}
r_{i} & T\: is\: a\: leaf\\
Cost\left(L\right)+Cost\left(R\right)+Card\left(L,R\right) & otherwise,
\end{cases}
\]
where $Card\left(L,R\right)$ is the cardinality (the expected number
of partial matches reaching the root) of $T$, whose calculation depends
on the operator applied by the root. For example, the cardinality
of a conjunction node is defined as the product of the cardinalities
of its operands multiplied by the total selectivity of the conditions
between the events in $L$ and the events in $R$. That is,
\[
Card\left(T\right)=Card\left(L\right)\times Card\left(R\right)\times SEL\left(L,R\right),
\]
where $SEL\left(L,R\right)$ is a product of all predicate selectivities
$sel_{i,j}:\, i\in L,j\in R$. Leaf cardinalities are defined as the
arrival rates of the respective event types. The reader is referred
to \cite{MeiM09} for more details.

To apply the invariant-based adaptation method on this algorithm,
we will define each internal node of a tree-based plan as a building
block. This way, up to $O\left(n^{3}\right)$ blocks will be formed
during the run of Algorithm \ref{alg:ZStream-algorithm-for}, with
only $O\left(n\right)$ included in the resulting plan.

A comparison between the costs of two trees will be considered a block-building
comparison for the root of the less expensive tree. The deciding conditions
for this algorithm will be thus defined simply as $Cost\left(T_{1}\right)<Cost\left(T_{2}\right)$,
where $T_{1},T_{2}$ are the two compared trees. These comparisons
are invoked at each step during the search for the cheapest tree over
a given subset of events. For $k$ events, the number of candidate
trees is $C_{k-1}=\frac{\left(2k-2\right)!}{\left(k-1\right)!k!}$,
where $C_{m}$ is the $m^{th}$ Catalan number. Therefore, picking
only one comparison as an invariant and dismissing the rest of the
candidates may create a problem of false negatives, and K-invariant
method is recommended instead (see discussion in Section \ref{sub:Correctness-Guarantees}).

The obvious problem with the above definition is that tree cost calculation
is a recursive function, which contradicts our constant-time invariant
verification assumption. We will eliminate this recursion by utilizing
the following observation. In Algorithm \ref{alg:ZStream-algorithm-for},
all block-building comparisons are performed on pairs of trees defined
over the same set of event types. By invariant definition, one of
these trees is always a subtree of a plan currently being in use.
Recall that invariants on tree-based plans are always verified in
the direction from leaves to the root. Hence, if any change was detected
in one of the statistics affecting the subtrees of the two compared
trees, it would be noticed during verification of earlier invariants.
Thus, it is safe to represent the cost of a subtree in an invariant
as a constant whose value is initialized to the cost of that subtree
during invariant creation (i.e., plan construction).

\subsection{General Applicability of the Invariant-Based Method}

\label{sub:General-Applicability-of}

The approaches described in Sections \ref{sub:Greedy-Algorithm-for}
and \ref{sub:Dynamic-Programming-Algorithm} only cover two special
cases. Here, we generalize the presented methodologies to apply the
invariant-based method to any greedy or dynamic programming algorithm.
We also discuss the applicability of our method to other algorithm categories.

A generalized variation of the technique illustrated in Section \ref{sub:Greedy-Algorithm-for}
can be utilized for any greedy plan generation algorithm. To that
end, a part of a plan constructed during a single greedy iteration
should be defined as a building block. Additionally, a conjunction
of all conditions evaluated to select a specific block is to be defined
as a block-building comparison associated with this block. Since most
greedy algorithms require constant time and space for a single step,
the complexity requirements for the invariant verification will be
satisfied.

Using similar observations, we can generalize the approach described
in Section \ref{sub:Dynamic-Programming-Algorithm} to any dynamic 
programming algorithm. A subplan memoized by the algorithm
will correspond to a building block. A comparison between two subplans
will serve as a BBC for the block that was selected during the initial
run.

In general, the invariant-based method can be similarly adapted to
any algorithm that constructs a plan in a deterministic, bottom-up
manner, or otherwise includes a notion of a ``building block''.
To the best of our knowledge, the majority of the proposed solutions
share this property.

In contrast, algorithms based on randomized local search (adapted
to CEP in \cite{ArxivJoin}) cannot be used in conjunction with the
invariant-based method. Rather than building a plan step-by-step,
these algorithms start with a complete initial solution and randomly
modify it to create an improved version \cite{AartsL97} until some
stopping condition is satisfied.

\section{Experimental Evaluation}

\label{sec:Experimental-Evaluation}

In this section, the results of our experimental evaluation are presented.
The objectives of this empirical study were twofold. First, we wanted
to assess the overall system performance achieved by our approach
and the computational overhead implied by its adaptation process as
compared to the existing strategies for ACEP systems, outlined in
Section \ref{sec:Introduction}. Our second goal was to explore how
changes in the parameters of our method and of the data characteristics
impact the above metrics.

\subsection{Experimental Setup}

\label{sub:Experimental-Setup}

\begin{figure}
	\centering
	\subfloat[]{\includegraphics[width=.5\linewidth]{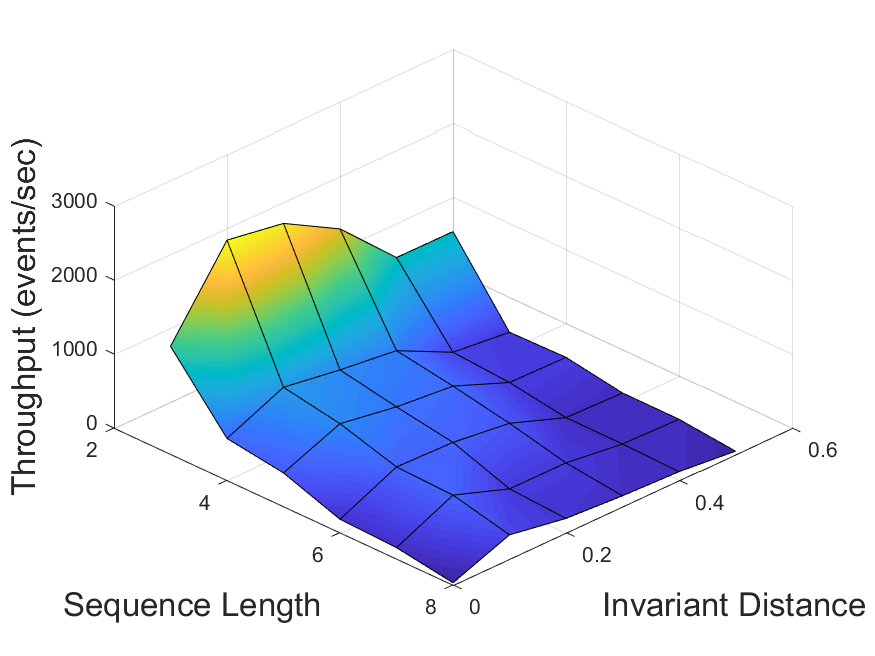}\label{fig:traffic-greedy}}
	\subfloat[]{\includegraphics[width=.5\linewidth]{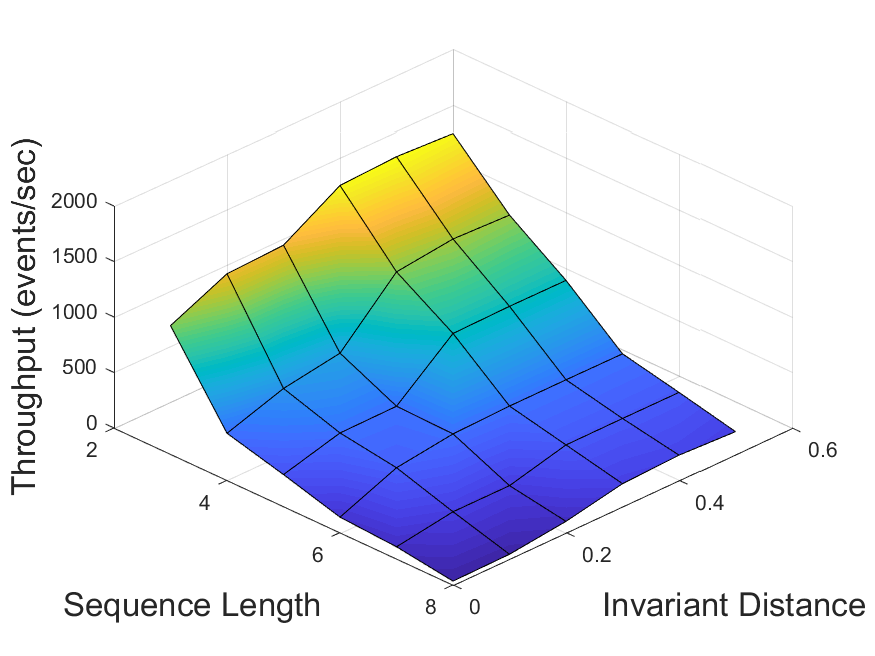}\label{fig:traffic-zstream}}\hfill
	\subfloat[]{\includegraphics[width=.5\linewidth]{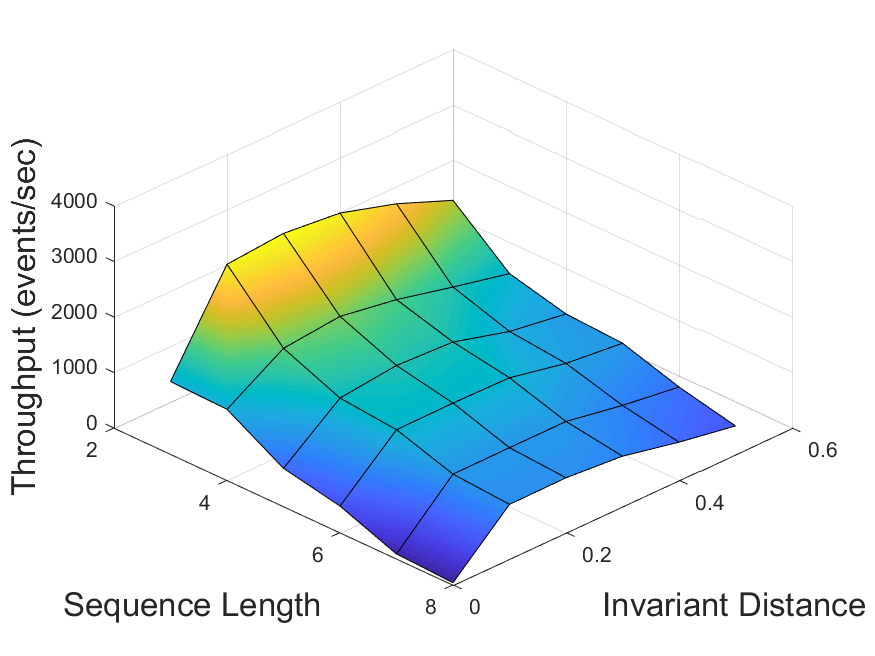}\label{fig:stocks-greedy}}
	\subfloat[]{\includegraphics[width=.5\linewidth]{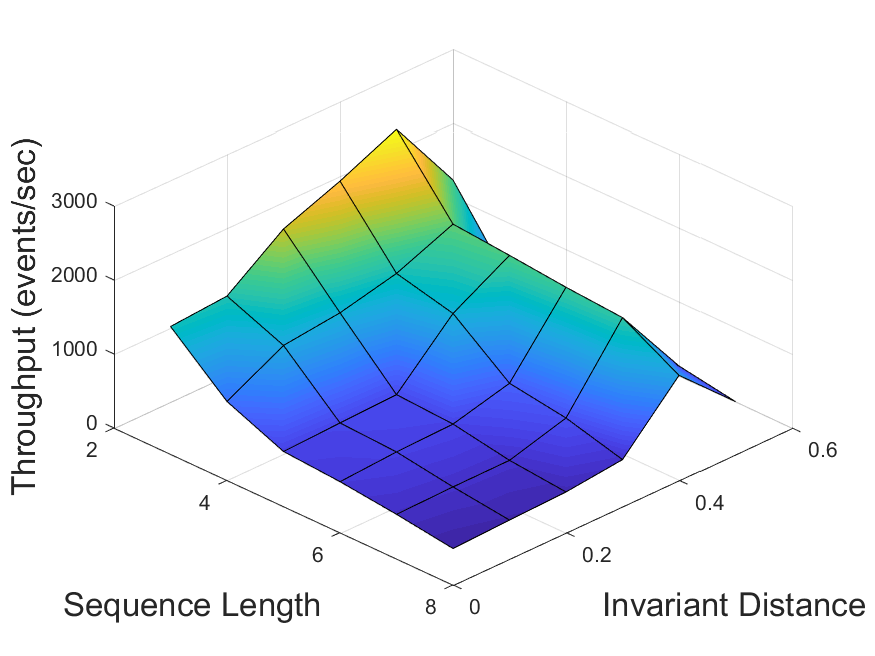}\label{fig:stocks-zstream}}
    \caption{Throughput of the invariant-based method for different dataset-algorithm pairs as a function of the pattern size and the invariant distance \textit{d}: \protect\subref{fig:traffic-greedy} traffic dataset / greedy algorithm; \protect\subref{fig:traffic-zstream} traffic dataset / ZStream algorithm; \protect\subref{fig:stocks-greedy} stocks dataset / greedy algorithm; \protect\subref{fig:stocks-zstream} stocks dataset / ZStream algorithm.}
	\label{fig:minimal-distance}
\end{figure}

\begin{figure*}
	\centering
	\subfloat[]{\includegraphics[width=.5\linewidth]{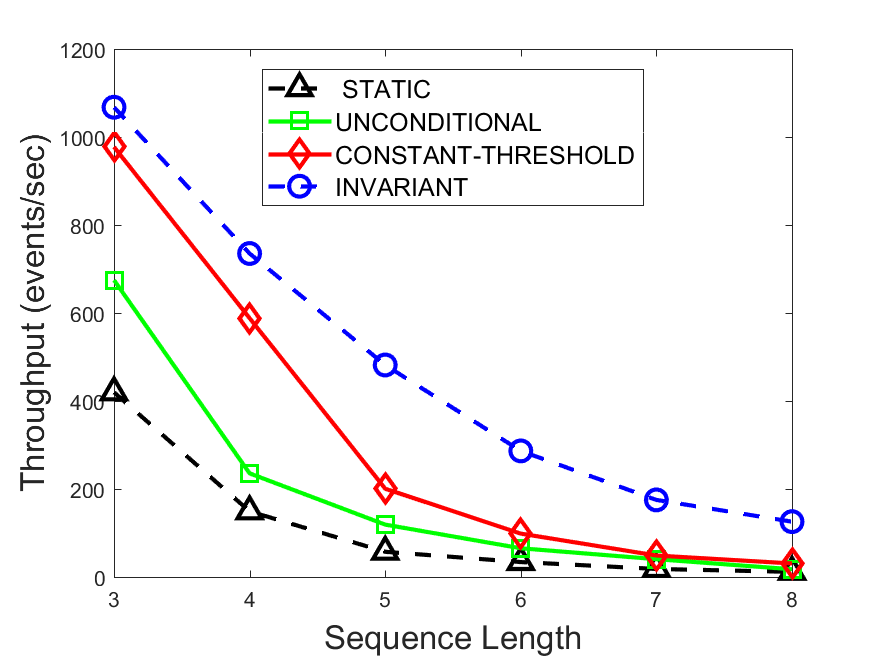}\label{fig:throughput}}
	\subfloat[]{\includegraphics[width=.5\linewidth]{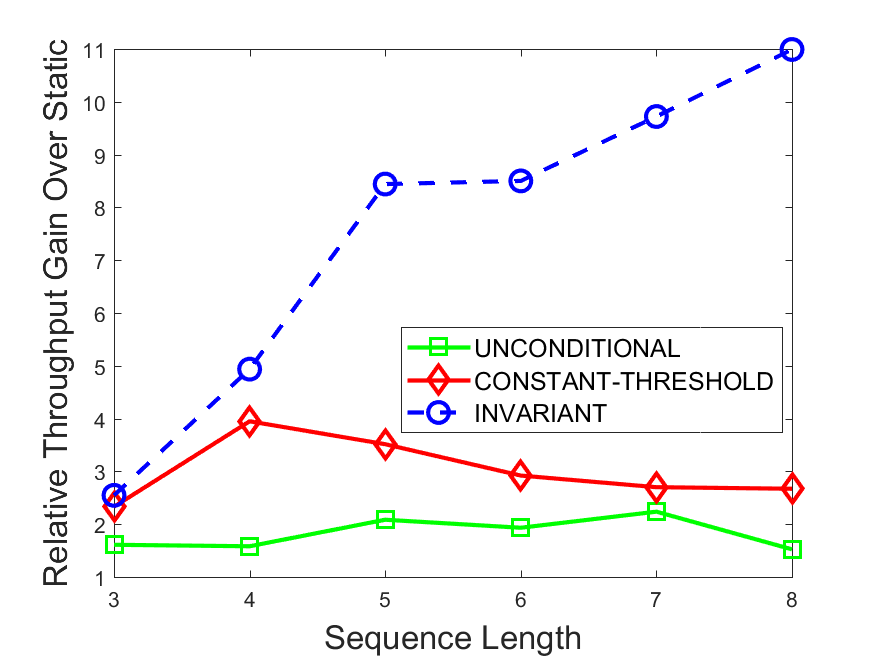}\label{fig:relative-throughput}}\hfill
	\subfloat[]{\includegraphics[width=.5\linewidth]{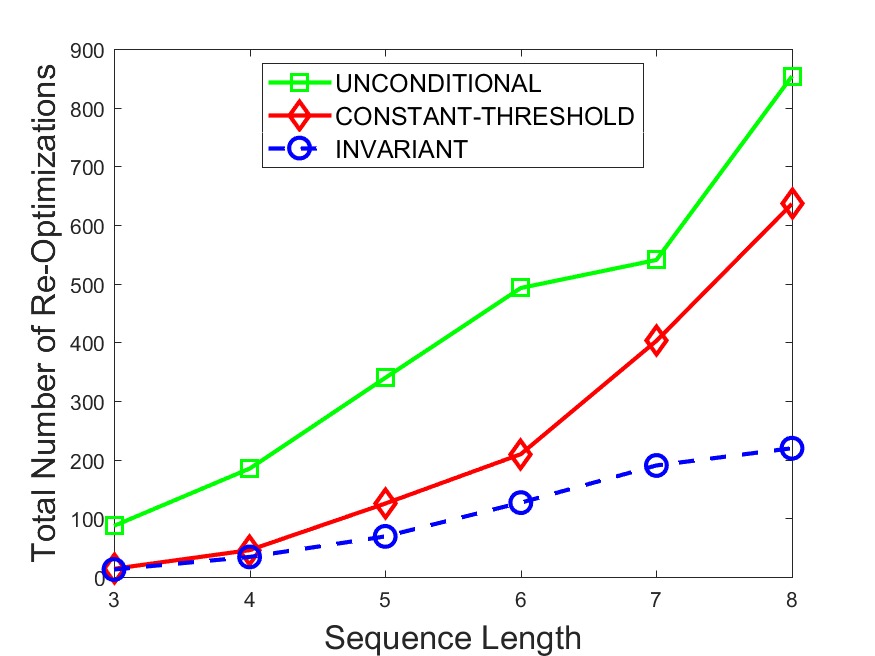}\label{fig:plan-switch}}
	\subfloat[]{\includegraphics[width=.5\linewidth]{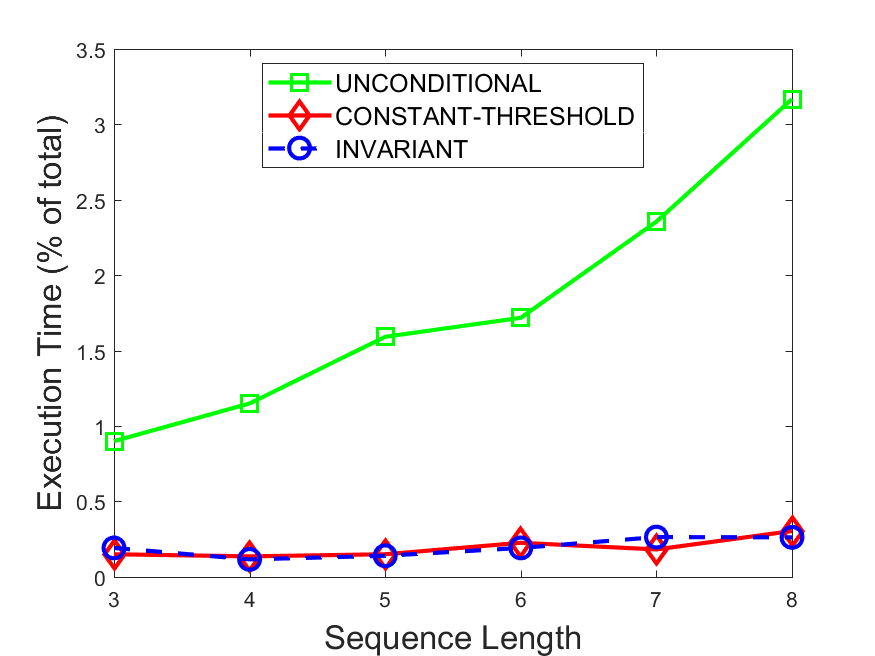}\label{fig:overhead}}
    \caption{Comparison of the adaptation methods applied on the traffic dataset in conjunction with the greedy algorithm: \protect\subref{fig:throughput} throughput (higher is better); \protect\subref{fig:relative-throughput} relative throughput gain over the non-adaptive method (higher is better); \protect\subref{fig:plan-switch} total number of plan reoptimizations; \protect\subref{fig:overhead} computational overhead (lower is better).}
	\label{fig:traffic-greedy}
\end{figure*}

\begin{figure*}
	\centering
	\subfloat[]{\includegraphics[width=.5\linewidth]{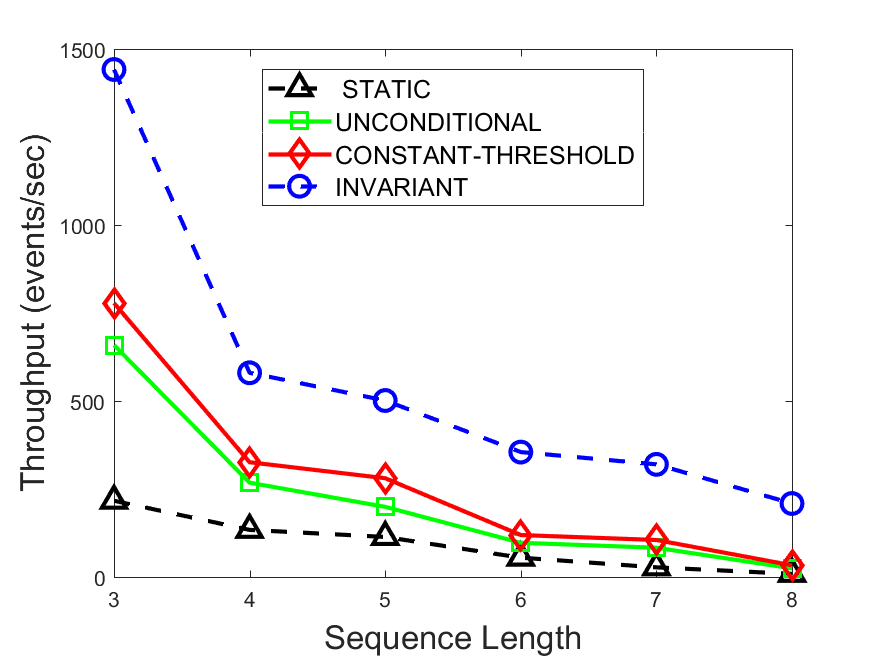}\label{fig:throughput}}
	\subfloat[]{\includegraphics[width=.5\linewidth]{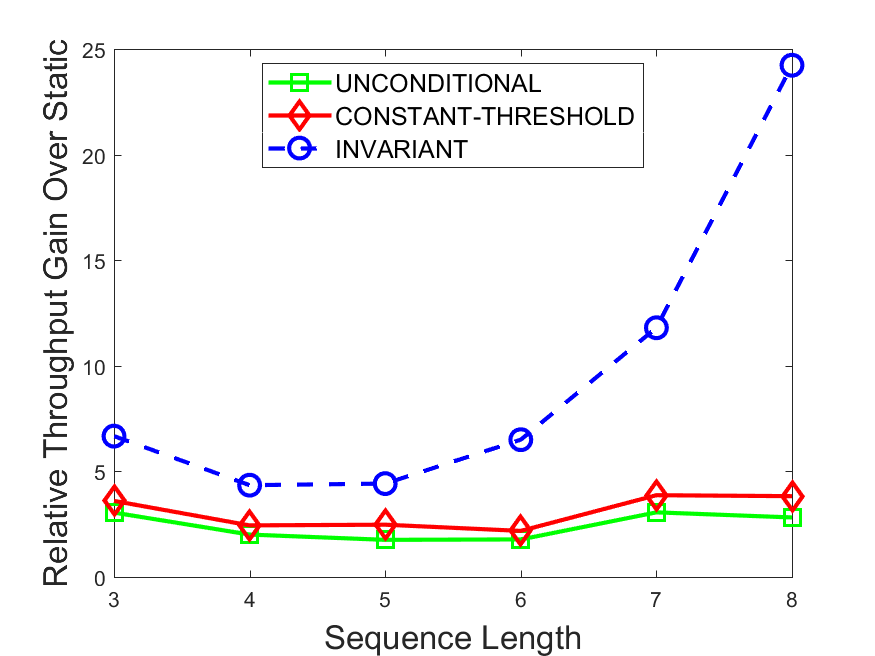}\label{fig:relative-throughput}}\hfill
	\subfloat[]{\includegraphics[width=.5\linewidth]{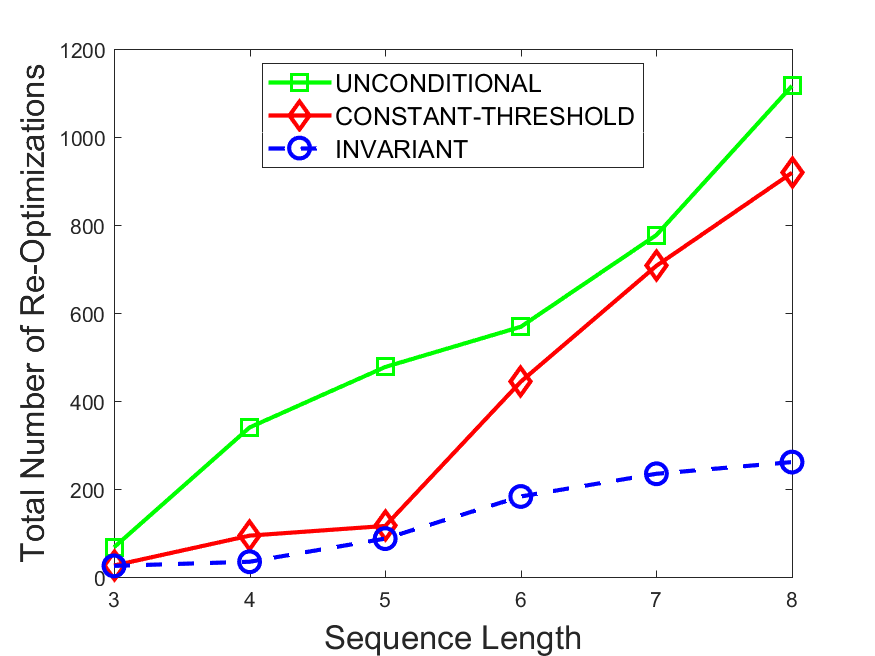}\label{fig:plan-switch}}
	\subfloat[]{\includegraphics[width=.5\linewidth]{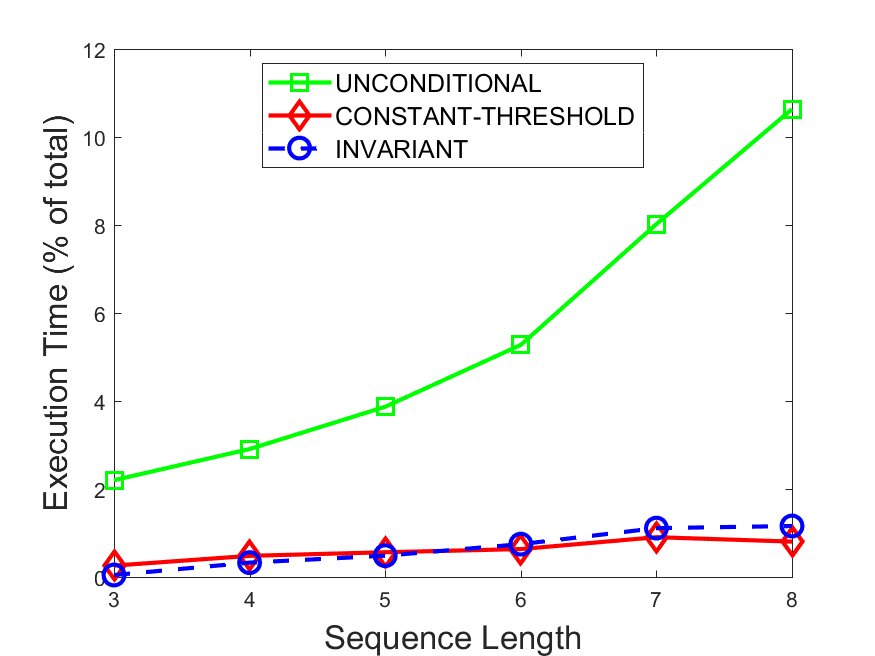}\label{fig:overhead}}
    \caption{Comparison of the adaptation methods applied on the traffic dataset in conjunction with ZStream algorithm: \protect\subref{fig:throughput} throughput (higher is better); \protect\subref{fig:relative-throughput} relative throughput gain over the non-adaptive method (higher is better); \protect\subref{fig:plan-switch} total number of plan reoptimizations; \protect\subref{fig:overhead} computational overhead (lower is better).}
	\label{fig:traffic-zstream}
\end{figure*}

We implemented the two CEP models described in Section \ref{sec:Applications-of-the},
the lazy NFA \cite{KolchinskySS15} with the greedy order-based algorithm
\cite{Swami89} and the ZStream model with tree-based dynamic programming
algorithm \cite{MeiM09}. We also added support for three adaptation
methods (i.e., implementations of $\mathcal{D}$): 1) the unconditional
reoptimization method from \cite{KolchinskySS15}; 2) the constant-threshold
method from \cite{MeiM09}; 3) the invariant-based method. To accurately
estimate the event arrival rates and predicate selectivities on-the-fly,
we utilized the algorithm presented in \cite{DatarGIM02} for maintaining
statistics over sliding window.

Since the plan generation algorithms used during this study create
plans optimized for maximal throughput, we choose throughput as a
main performance metric, reflecting the effectiveness of the above
algorithms in the presence of changes in the input. We believe that
similar results could be obtained for algorithms targeting any other
optimization goal, such as minimizing latency or communication cost.

Two real-world datasets were used in the experiments. For each
of them, we created 5 sets of patterns containing different operators
(Section \ref{sub:Events-and-Patterns}), as follows: (1)sequences;
(2)sequences with an event under negation; (3)conjunctions; (4)sequences
with an event under Kleene closure; (5)composite patterns, consisting
of a disjunction of three shorter sequences. Each set contained 6
patterns of sizes varying from 3 to 8. The details are specified below
for each dataset. Our main results presented in this section are averaged
over all pattern sets unless otherwise stated. We provide the full description
of the specific results obtained for each set in Appendix \ref{sec:Additional-Experimental-Results}.

The first dataset contains the vehicle traffic sensor data, provided
by City of Aarhus, Denmark \cite{AliGM15} and collected over a period
of 4 months from 449 observation points, with 13,577,132 primitive
events overall. Each event represents an observation of traffic at
the given point. The attributes of an event include, among others,
the point ID, the average observed speed, and the total number of
observed vehicles during the last 5 minutes. The arrival rates and
selectivities for this dataset were highly skewed and stable, with
few on-the-fly changes. However, the changes that did occur were mostly
very extreme. The patterns for this dataset were motivated by normal
driving behavior, where the average speed tends to decrease with the
increase in the number of vehicles on the road. We requested to detect
the violations of this model, i.e., combinations (sequences, conjunctions,
etc., depending on the operator involved) of three or more observations
with either an increase or a decline in both the number of vehicles
and the average speed.

The second dataset was taken from the NASDAQ stock market historical
records \cite{EODData}. Each record in this dataset represents a
single update to the price of a stock, spanning a 1-year period and
covering over 2100 stock identifiers with prices updated on a per
minute basis. Our input stream contained 80,509,033 primitive events,
each consisting of a stock identifier, a timestamp, and a current
price. For each stock identifier, a separate event type was defined.
In addition, we preprocessed the data to include the difference between
the current and the previous price. Contrary to the traffic dataset,
low skew in data statistics was observed, with the initial values
nearly identical for all event types. The changes were highly frequent,
but mostly minor. The patterns to evaluate were then defined as combinations
of different stock identifiers (types), with the predefined price
differences (e.g., for a conjunction pattern $AND\left(A,B,C\right)$
we require $A.diff<B.diff<C.diff$).

All models and algorithms under examination were implemented in Java.
All experiments were run on a machine with 2.20 Ghz CPU and 16.0 GB
RAM.

\subsection{Experimental Results}

\label{sub:Experimental-Results}

\begin{figure*}
	\centering
	\subfloat[]{\includegraphics[width=.5\linewidth]{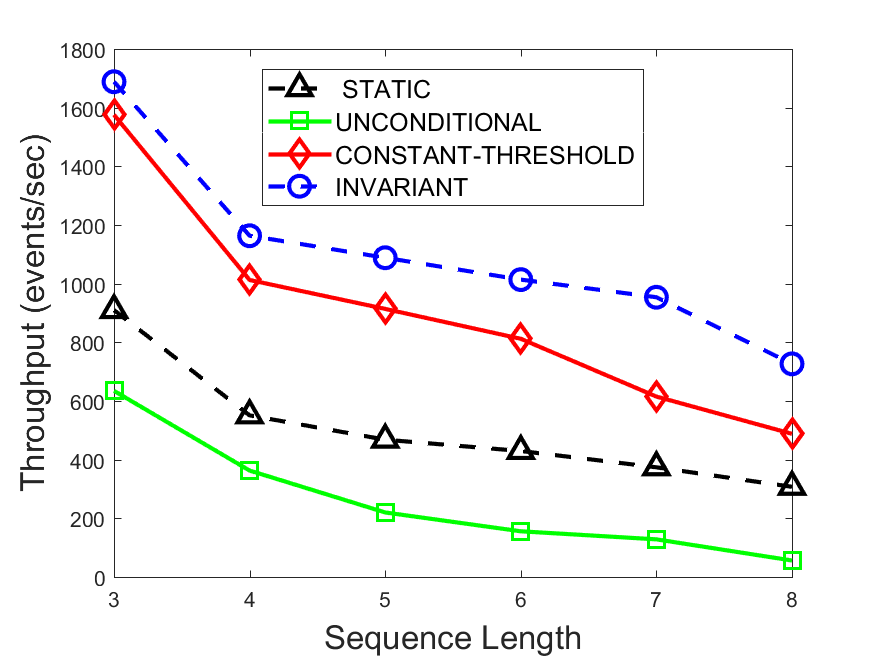}\label{fig:throughput}}
	\subfloat[]{\includegraphics[width=.5\linewidth]{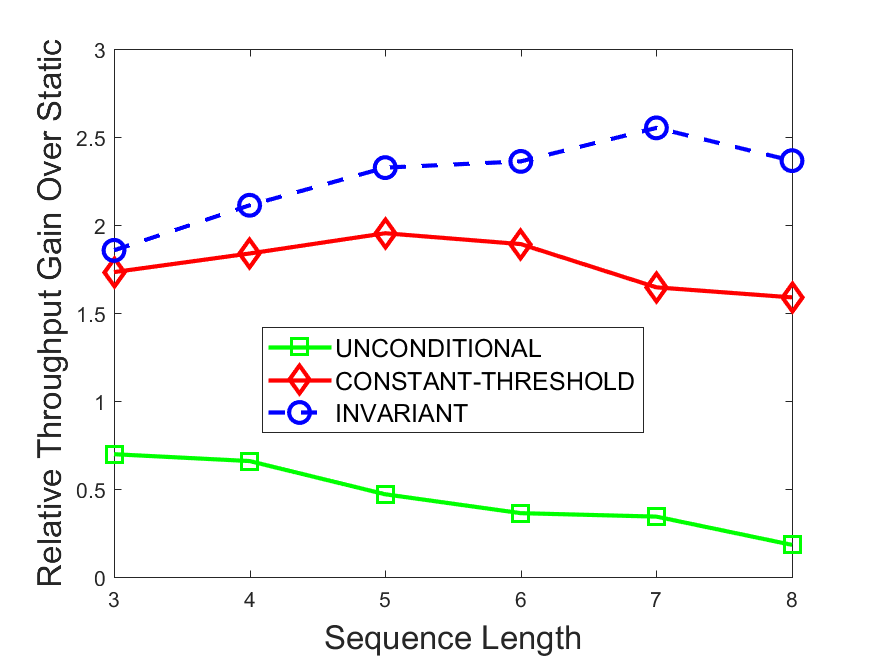}\label{fig:relative-throughput}}\hfill
	\subfloat[]{\includegraphics[width=.5\linewidth]{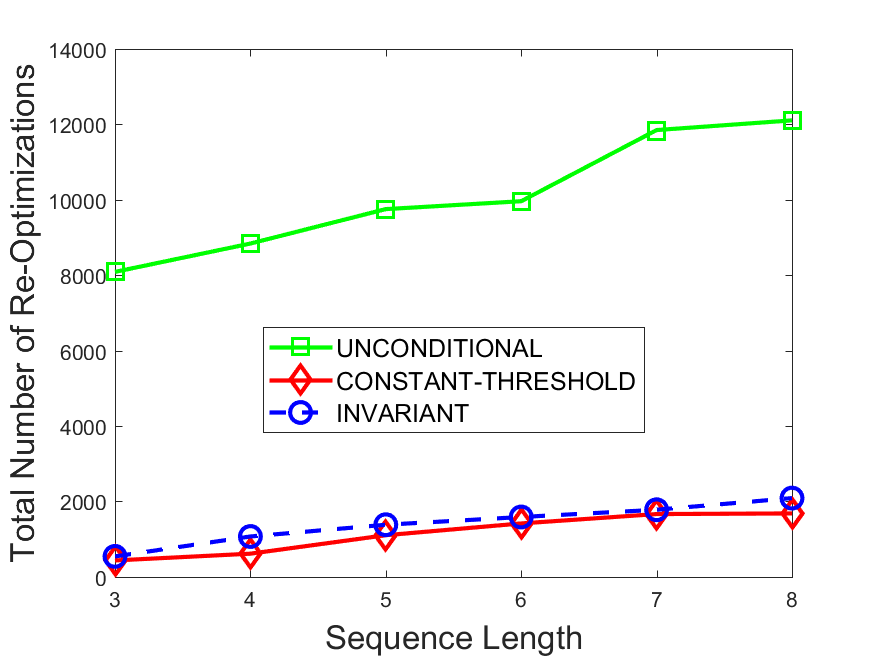}\label{fig:plan-switch}}
	\subfloat[]{\includegraphics[width=.5\linewidth]{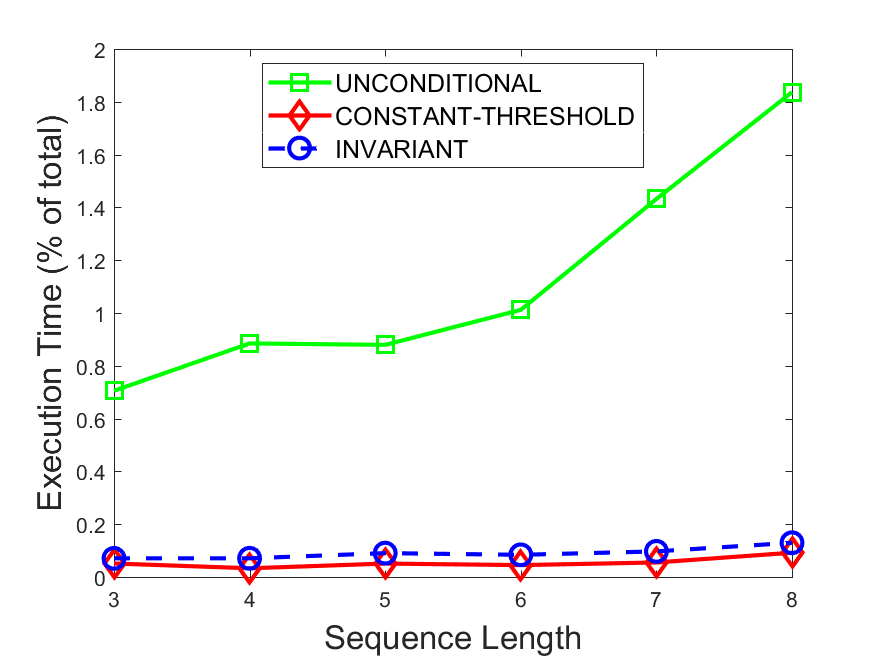}\label{fig:overhead}}
    \caption{Comparison of the adaptation methods applied on the stocks dataset in conjunction with the greedy algorithm: \protect\subref{fig:throughput} throughput (higher is better); \protect\subref{fig:relative-throughput} relative throughput gain over the non-adaptive method (higher is better); \protect\subref{fig:plan-switch} total number of plan reoptimizations; \protect\subref{fig:overhead} computational overhead (lower is better).}
	\label{fig:stocks-greedy}
\end{figure*}

\begin{figure*}
	\centering
	\subfloat[]{\includegraphics[width=.5\linewidth]{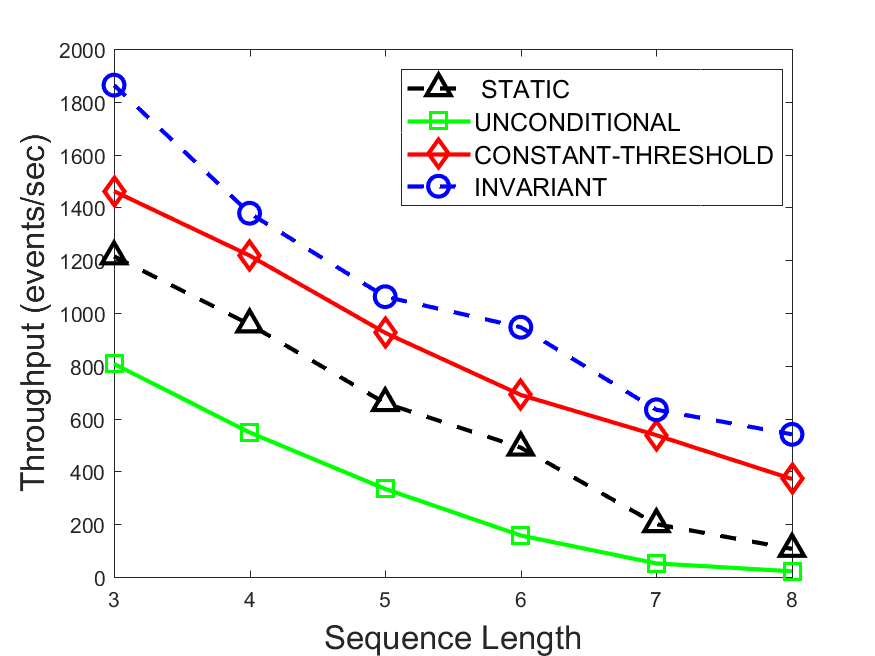}\label{fig:throughput}}
	\subfloat[]{\includegraphics[width=.5\linewidth]{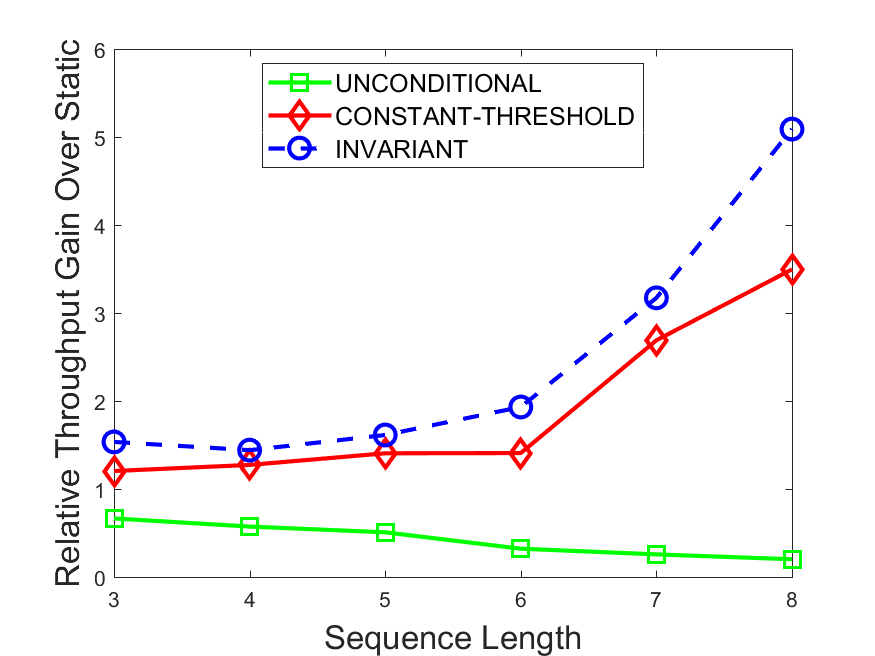}\label{fig:relative-throughput}}\hfill
	\subfloat[]{\includegraphics[width=.5\linewidth]{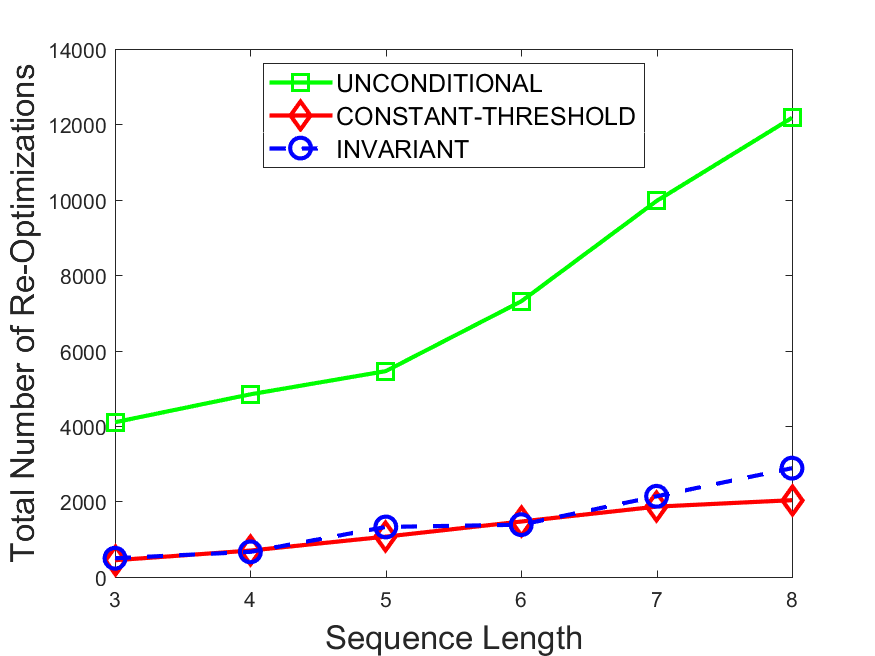}\label{fig:plan-switch}}
	\subfloat[]{\includegraphics[width=.5\linewidth]{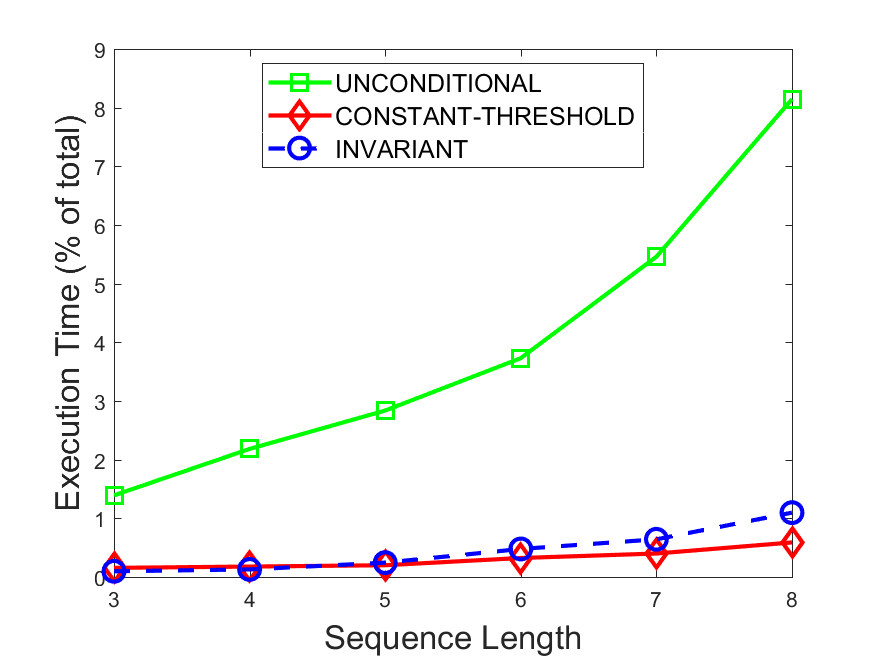}\label{fig:overhead}}
    \caption{Comparison of the adaptation methods applied on the stocks dataset in conjunction with ZStream algorithm: \protect\subref{fig:throughput} throughput (higher is better); \protect\subref{fig:relative-throughput} relative throughput gain over the non-adaptive method (higher is better); \protect\subref{fig:plan-switch} total number of plan reoptimizations; \protect\subref{fig:overhead} computational overhead (lower is better).}
	\label{fig:stocks-zstream}
\end{figure*}

In our first experiment, we evaluated the performance of the invariant-based
method for different values of the invariant distance $d$ (Section
\ref{sub:Distance-Based-Invariants}). In this experiment, only the 
sequence pattern sets were used. For each of the four possible
dataset-algorithm combinations, the system throughput was measured
as a function of the tested pattern size and of $d$, with its values
ranging from 0 (which corresponds to the basic method) to 0.5. The
results are displayed in Figure \ref{fig:minimal-distance}. It can
be observed that in each scenario there exists an optimal value $d_{opt}$,
which depends on the data and the algorithm in use, consistently outperforming
the other values for all pattern sizes. For distances higher than
$d_{opt}$, too many changes in the statistics are undetected, while
the lower values trigger unnecessary adaptations. Overall, the throughput
achieved by using invariants with distance $d_{opt}$ is 2 to 25 times
higher than that of the basic method ($d=0$).

Then, we validated the average relative difference method described
in Section \ref{sub:Distance-Based-Invariants} by comparing its output
value $d_{avg}$ to $d_{opt}$ (obtained via parameter scanning as
described above) for each scenario. The results are summarized in
Table \ref{tab:Quality-of-distance}.

For the traffic dataset, the computed values were considerably close
to the optimal ones for patterns of length 6 and above, with precision
reaching at least 87\% (for ZStream algorithm and pattern length 7)
and as high as 92\% (Greedy algorithm, length 8). For the stocks dataset,
the achieved accuracy was only 31-44\%. We can thus conclude that
the tested method does not function well in presence of low data skew,
matching our expectations from Section \ref{sub:Distance-Based-Invariants}.
This highlights the need for developing better solutions, which is
the goal of our future work.

It can also be observed that for all dataset-algorithm combinations
the prevision of the average relative difference method increases
with pattern size. We estimate that the scalability of this method
would further increase for even larger patterns.

\begin{table}
\caption{Quality of distance estimates obtained by the average relative difference
method\label{tab:Quality-of-distance}}

\begin{tabular}{|c|c|c|c|c|c|}
\hline 
Dataset & Algorithm & Pattern Size & $d_{avg}$ & $d_{opt}$ & $min\left(\frac{d_{avg}}{d_{opt}},\frac{d_{opt}}{d_{avg}}\right)$\tabularnewline
\hline 
\hline 
Traffic & Greedy & 4 & 0.1695 & 0.1 & 0.59\tabularnewline
\hline 
Traffic & Greedy & 5 & 0.1211 & 0.1 & 0.826\tabularnewline
\hline 
Traffic & Greedy & 6 & 0.1163 & 0.1 & 0.86\tabularnewline
\hline 
Traffic & Greedy & 7 & 0.0909 & 0.1 & 0.909\tabularnewline
\hline 
Traffic & Greedy & 8 & 0.0921 & 0.1 & 0.921\tabularnewline
\hline 
Traffic & ZStream & 4 & 0.9153 & 0.4 & 0.437\tabularnewline
\hline 
Traffic & ZStream & 5 & 0.6514 & 0.4 & 0.614\tabularnewline
\hline 
Traffic & ZStream & 6 & 0.4551 & 0.4 & 0.879\tabularnewline
\hline 
Traffic & ZStream & 7 & 0.4581 & 0.4 & 0.873\tabularnewline
\hline 
Traffic & ZStream & 8 & 0.4464 & 0.4 & 0.896\tabularnewline
\hline 
Stocks & Greedy & 4 & 0.0556 & 0.2 & 0.278\tabularnewline
\hline 
Stocks & Greedy & 5 & 0.0268 & 0.2 & 0.134\tabularnewline
\hline 
Stocks & Greedy & 6 & 0.0661 & 0.2 & 0.331\tabularnewline
\hline 
Stocks & Greedy & 7 & 0.0818 & 0.2 & 0.409\tabularnewline
\hline 
Stocks & Greedy & 8 & 0.0866 & 0.2 & 0.443\tabularnewline
\hline 
Stocks & ZStream & 4 & 0.095 & 0.4 & 0.24\tabularnewline
\hline 
Stocks & ZStream & 5 & 0.102 & 0.4 & 0.255\tabularnewline
\hline 
Stocks & ZStream & 6 & 0.1231 & 0.4 & 0.308\tabularnewline
\hline 
Stocks & ZStream & 7 & 0.1563 & 0.4 & 0.391\tabularnewline
\hline 
Stocks & ZStream & 8 & 0.1206 & 0.4 & 0.304\tabularnewline
\hline 
\end{tabular}
\end{table}

Next, we performed an experimental comparison of all previously described
adaptation methods. The comparison was executed separately for each
dataset-algorithm combination. For the invariant-based method, the
$d_{opt}$ values obtained during the previous experiment were used.
For the constant-threshold method, an optimal threshold $t_{opt}$
was empirically found for each of the above combinations using a similar
series of runs.

Figures \ref{fig:traffic-greedy}-\ref{fig:stocks-zstream} summarize
the results. Each figure represents the measurements for a particular
dataset-algorithm combination and contains four graphs, presenting
different statistics as a function of the pattern size. The first
graph presents the throughput achieved using each of the adaptation
methods. Here, we have also included the ``static'' method in our
study, where no adaptation is supported and the dataset is processed
using a single, predefined plan. The second graph is a different way
of viewing the previous one, comparing the adaptation methods by the
relative speedup they achieve over the ``static plan'' approach.
The third graph depicts the total number of reoptimizations (actual
plan replacements) recorded during each run. Finally, we report the
computational overhead of each method, that is, a percentage of the
total execution time spent on executions of $\mathcal{D}$ and $\mathcal{A}$
(i.e., checking whether a reoptimization is necessary and computing
new plans).

The throughput comparison demonstrates the superiority of the invariant-based
method over its alternatives for all scenarios. Its biggest performance
gain is achieved in the traffic scenario, characterized by high skew
and major statistic shifts (Figures \ref{fig:traffic-greedy}-\ref{fig:traffic-zstream}).
This gain reaches its peak for larger patterns, with the maximal recorded
performance of more than 6 times that of the second-best constant-threshold
method: the greater the discrepancy between the data characteristics,
the more difficult it is to find a single threshold to accurately
monitor all the changes. Since this discrepancy may only increase
as more statistic values are added to the monitored set, we expect
the superiority of this method to keep growing with the pattern size
beyond the values we experimented with. Figures \ref{fig:traffic-greedy}\subref{fig:relative-throughput}-\ref{fig:traffic-zstream}\subref{fig:relative-throughput}
provide a clear illustration of the above phenomenon and of the invariant-based
method scalability. Note also that, for larger pattern sizes, the
constant-threshold method nearly converges to the unconditional one
due to the increasing number of false positives it produces.

For the stocks dataset (Figures \ref{fig:stocks-greedy}-\ref{fig:stocks-zstream}),
the throughput measurements for the constant-threshold and the invariant-based
methods are considerably closer. Due to the near-uniformity of the
statistic values and of their variances, finding a single $t_{opt}$
is sufficient to recognize most important changes. Hence, the precision
of the constant-threshold method is very high on this input. Nevertheless,
the invariant-based method achieves a performance speedup for this
dataset as well (albeit only about 30-60\%) without adding significant
overhead. Also, for the same reason, the static plan performs reasonably
well in this scenario, decidedly outperforming the unconditional method.
The latter suffers from extreme over-adapting to the numerous small-scale
statistic shifts.

The total number of reoptimizations performed in each scenario (Figures
\ref{fig:traffic-greedy}\subref{fig:plan-switch}-\ref{fig:stocks-zstream}\subref{fig:plan-switch})
backs up and augments the above results. The invariant-based method
consistently requires few plan replacements while also achieving the
best throughput. The extremely high numbers produced by the unconditional
strategy lead to its poor performance. For the traffic dataset, the
constant-threshold method tends to approach these numbers for larger
patterns. This can either be a sign of multiple false positives or
over-adapting. For the stocks dataset, this method becomes more similar
to the invariant-based one, executing nearly identical reoptimizations.

Figures \ref{fig:traffic-greedy}\subref{fig:overhead}-\ref{fig:stocks-zstream}\subref{fig:overhead}
present the computational overhead of the compared approaches. Here,
the same behavior is observed for all dataset-algorithm combinations.
While the invariant-based and the constant-threshold methods consume
negligible system resources, unconditional reoptimization results
in up to 11\% of the running time devoted to the adaptation process.

As evident by the experiments with stock market data (Figures \ref{fig:stocks-greedy}-\ref{fig:stocks-zstream}),
smaller number of reoptimizations and lower computational overhead
do not necessarily result in better overall system performance. On
this dataset, the invariant-based method achieves the highest throughput
despite a slightly higher overhead as compared to the second-best
constant-threshold method. This can be attributed to the false negatives
of the latter, that is, cases in which it missed a reoptimization
opportunity and kept using an old plan despite a better one being
available.

In all experiments, the gains of the invariant-based method were considerably
higher for ZStream algorithm than for the greedy one. There are two
reasons for this result. First, the more complex structure of the
tree-based plans makes it more difficult to capture the dependencies
between plan components without fine-grained invariants. Second, as
this algorithm is more computationally expensive, the penalty for
a redundant reoptimization is higher. Following these observations,
we believe that the invariant-based method is capable of achieving
even larger benefit for more advanced and precise (and hence more
complex) plan generation algorithms. Utilizing this method will thus
encourage the adoption of such algorithms by CEP engines.

\section{Related Work}

\label{sec:Related-Work}

Complex event processing is an increasingly active research field
\cite{CugolaM12}. The origins of CEP systems can be traced to older
data stream managements systems (DSMSs), including Aurora/Borealis
\cite{AbadiABCHLMRRTXZ05}, Stream \cite{ArasuBBCDIMSW16}, TelegraphCQ
\cite{ChandrasekaranDFHHKMRRS03}, and NiagaraCQ \cite{ChenDTW00}.
This was followed by the emergence of a broad variety of solutions
for detecting occurrences of situations of interest, as opposed to
generic data, including frameworks such as SASE/SASE+ \cite{WuDR06,AgrawalDGI2008,ZhangDI2014},
CEDR \cite{BargaGAH07}, Cayuga \cite{DemersGHRW06}, T-Rex \cite{CugolaGM12}
and Amit \cite{AdiE04}. Esper \cite{Esper10} and IBM System S \cite{AminiABEKSPV06}
are examples of widely used commercial CEP providers.

Many CEP approaches incorporate NFAs as their primary evaluation structure
\cite{WuDR06,DemersGHRW06,CugolaGM12}. Various extensions to this
model were developed, such as AFA \cite{ChandramouliGM10} and lazy
NFA \cite{KolchinskySS15}. ZStream \cite{MeiM09} utilizes tree-based
detection plans for the representation of event patterns. Event processing
networks \cite{EtzionN10} is another conceptual model, presenting
a pattern as a network of simple agents.

Multiple works have addressed the broad range of CEP optimization
opportunities arising when the statistical characteristics of the
primitive events are taken into account. In \cite{AkdereMCT08} ``plan-based
evaluation'' is described, where the arrival rates of events are
exploited to reduce network communication costs. The authors of NextCEP
\cite{Schultz-MollerMP09} propose a framework for pattern rewriting
in which operator properties are utilized to assign a cost to every
candidate evaluation plan. Then, a search algorithm (either greedy
or dynamic) is applied to select the lowest cost detection scheme.
ZStream \cite{MeiM09} applies a set of algebraic rule-based transformations
on a given pattern, and then reorders the operators to minimize the
cost of a plan.

Adaptive query processing (AQP) is the widely studied problem of adapting
a query plan to the unstable data characteristics \cite{DeshpandeIR07}.
Multiple solutions consider traditional data-bases \cite{AcostaVLCR11,IvesHW04,BabuBD05,MarklRSLPC04,KabraD98,StillgerLMK01}.
The mid-query reoptimization mechanism \cite{KabraD98}, one of the
first to possess adaptive properties, collects statistics at the predefined
checkpoints and compares them to the past estimates. If severe deviation
is observed, the remainder of the data is processed using a new plan.
The methods described in \cite{BabuBD05} and \cite{MarklRSLPC04}
are the closest in spirit to our work. Rather than executing reoptimization
on a periodic basis or upon a constant change, the authors compute
an individual range for each monitored value within which the current
plan is considered close-to-optimal.

The field of stream processing has developed adaptive techniques of
its own. A-Greedy \cite{BabuMMNW04} is an algorithm for adaptive
ordering of pipelined filters, providing strong theoretical guarantees.
Similarly to our method, it detects violations of invariants defined
on the filter drop probabilities. The authors of \cite{LiuIL16} describe
``incremental reoptimization,'' where the optimizer constantly attempts
to locate a better plan using efficient search and pruning techniques.
Eddy \cite{AvnurH00,BizarroBDW05,MaddenSHR02} presents stateless
routing operators, redirecting incoming tuples to query operators
according to a predefined routing policy. This system discovers execution
routes on-the-fly in a per-tuple manner. Query Mesh \cite{NehmeWLRB13}
is a middle-ground approach, maintaining a set of plans and using
a classifier to select a plan for each data item. Large DSMSs have
also incorporated adaptive mechanisms \cite{TatbulCZCS03,BabuW04}.

The majority of the proposed CEP techniques are deprived from adaptivity
considerations \cite{FlourisGDGKM17}. The two notable exceptions,
ZStream \cite{MeiM09} and tree-based NFA \cite{KolchinskySS15} were
covered in detail above. Additional works labeled as 'adaptive' refer
to on-the-fly switching between several detection algorithms \cite{SadoghiJ14,YiLW16}
or dynamic rule mining \cite{CoffiMM12,LeeYHJ15}.

\section{Conclusions and Future Work}

\label{sec:Conclusions-and-Future}

In this paper, we discussed the problem of efficient adaptation of
a CEP system to on-the-fly changes in the statistical properties of
the data. A new method was presented to avoid redundant reoptimizations
of the pattern evaluation plan by periodically verifying a small set
of simple conditions defined on the monitored data characteristics,
such as the arrival rates and the predicate selectivities. We proved
that validating this set of conditions will only fail if a better
evaluation plan is available. We applied our method on two real-life
algorithms for plan generation and experimentally demonstrated the
achieved performance gain.

In addition to the research of distance estimators (Section \ref{sub:Distance-Based-Invariants}),
one area of interest that was not yet addressed by the existing approaches
is the multi-pattern adaptive CEP, where the system is given a set
of patterns possibly containing common subexpressions. In this case,
the detection process typically follows a single global plan that
exploits sharing opportunities, rather than executing multiple individual
plans in parallel. While our method can be trivially applied to multi-pattern
systems with no sharing, substantially more sophisticated optimization
techniques are required for the general case. We intend to target
this research direction in our future work.

\bibliographystyle{plain}
\bibliography{references}

\appendix

\section{Additional Experimental Results}

\label{sec:Additional-Experimental-Results}

This appendix extends the experimental results discussed in Section
\ref{sub:Experimental-Results} by organizing and presenting them
by pattern type.

\begin{figure*}
	\centering
	\subfloat[]{\includegraphics[width=.25\linewidth]{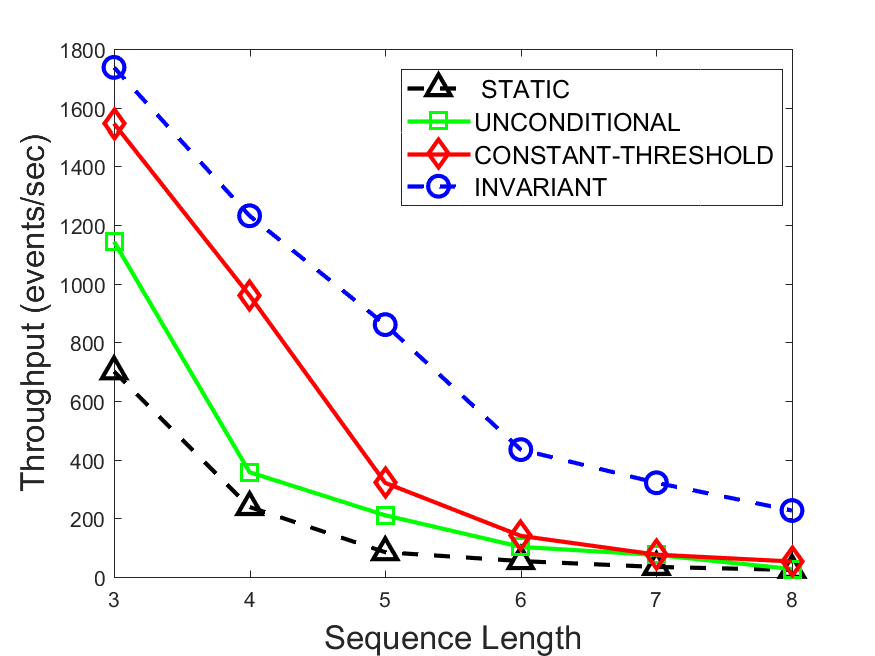}\label{fig:throughput}}
	\subfloat[]{\includegraphics[width=.25\linewidth]{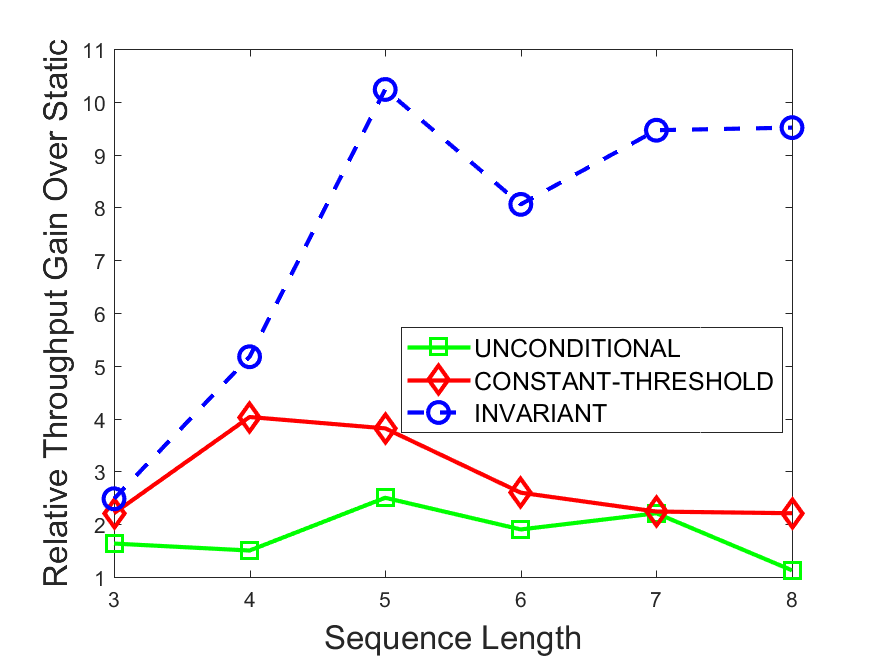}\label{fig:relative-throughput}}
	\subfloat[]{\includegraphics[width=.25\linewidth]{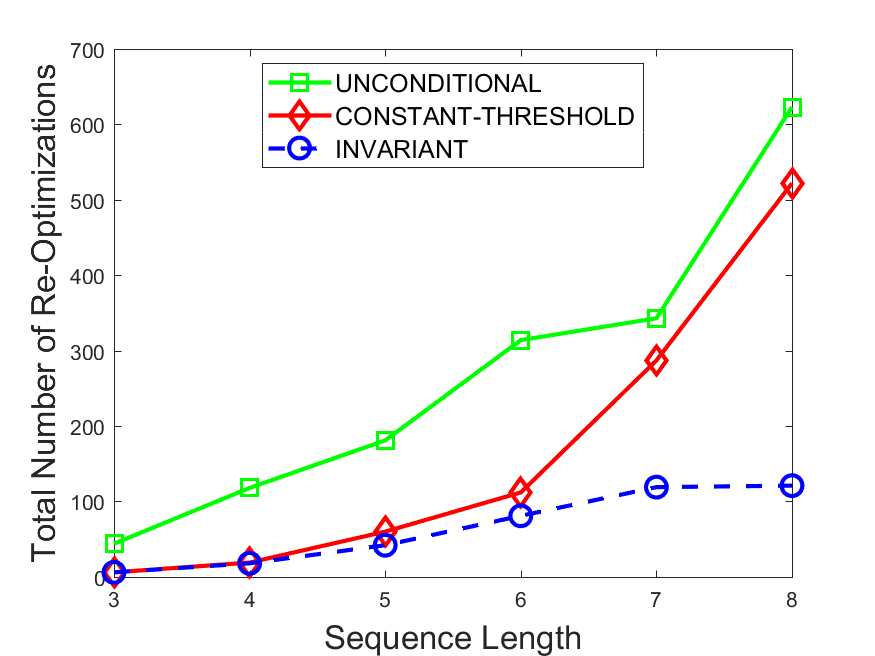}\label{fig:plan-switch}}
	\subfloat[]{\includegraphics[width=.25\linewidth]{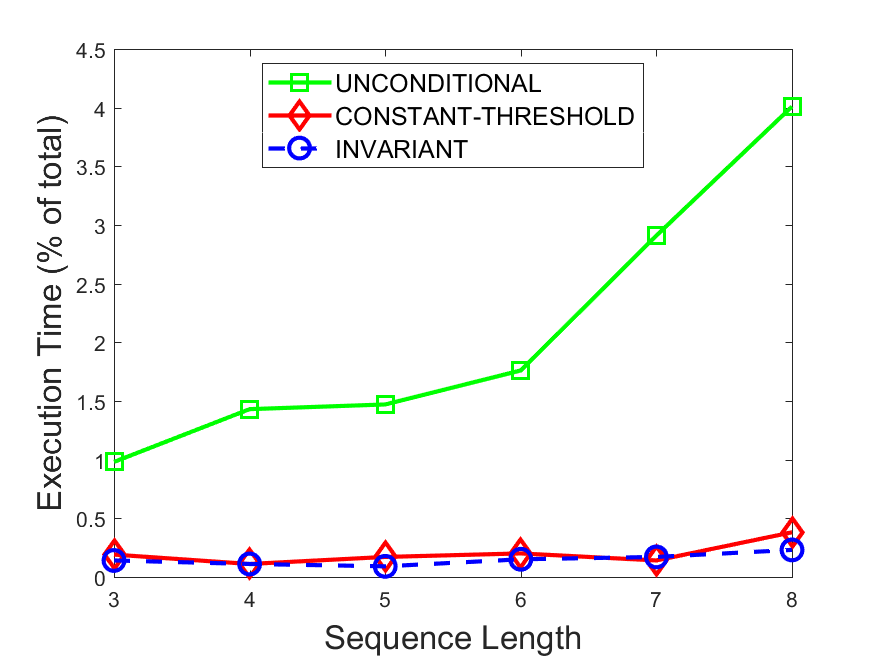}\label{fig:overhead}}
    \caption{Comparison of the adaptation methods applied on the traffic dataset in conjunction with the greedy algorithm (sequence patterns): \protect\subref{fig:throughput} throughput (higher is better); \protect\subref{fig:relative-throughput} relative throughput gain over the non-adaptive method (higher is better); \protect\subref{fig:plan-switch} total number of plan reoptimizations; \protect\subref{fig:overhead} computational overhead (lower is better).}
	\label{fig:seq-traffic-greedy}
\end{figure*}

\begin{figure*}
	\centering
	\subfloat[]{\includegraphics[width=.25\linewidth]{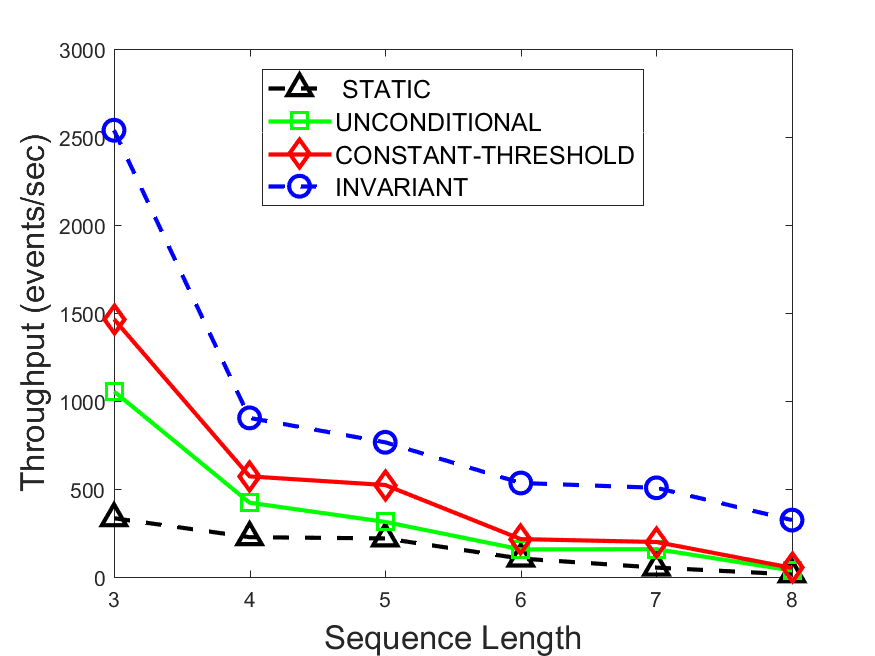}\label{fig:throughput}}
	\subfloat[]{\includegraphics[width=.25\linewidth]{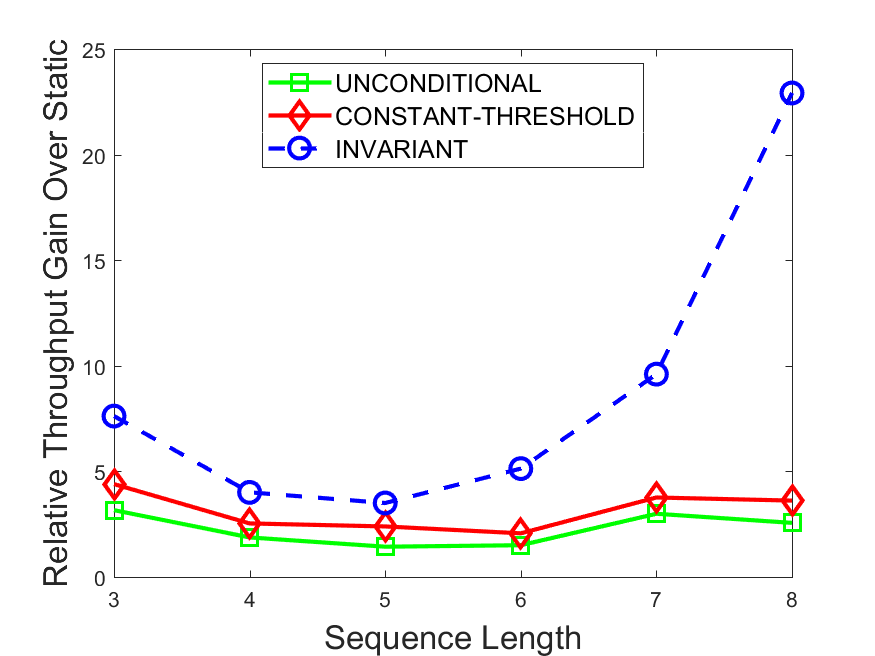}\label{fig:relative-throughput}}
	\subfloat[]{\includegraphics[width=.25\linewidth]{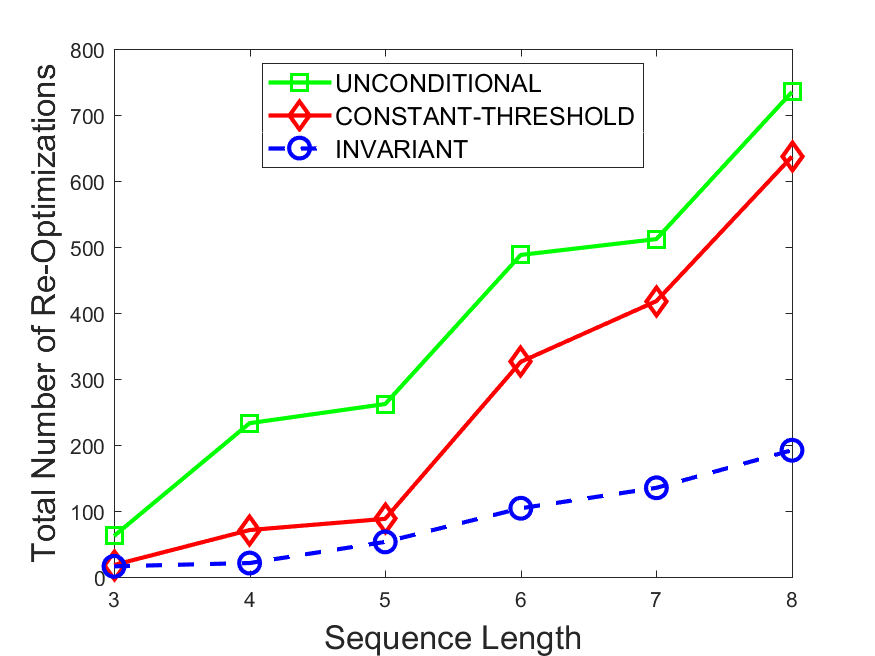}\label{fig:plan-switch}}
	\subfloat[]{\includegraphics[width=.25\linewidth]{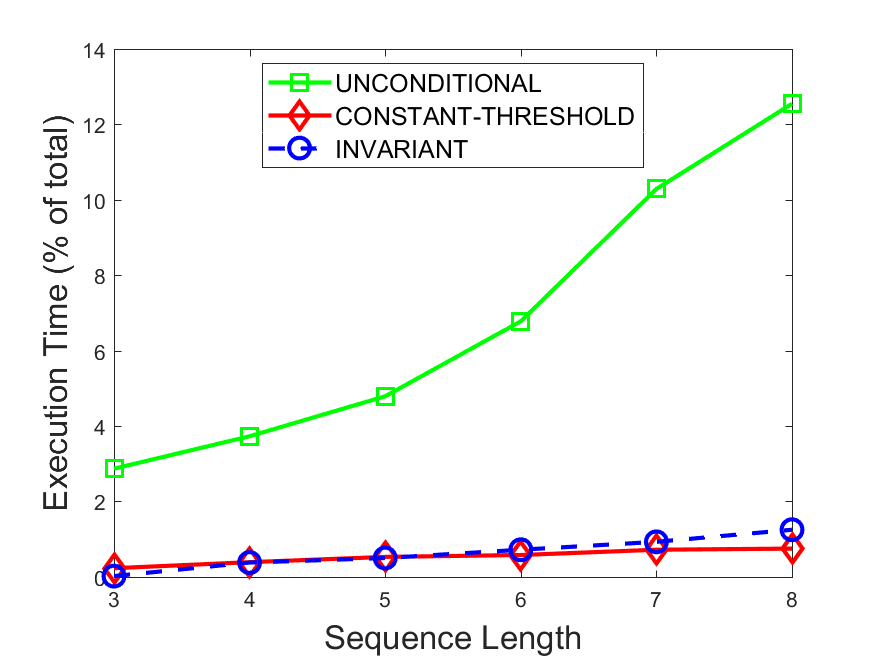}\label{fig:overhead}}
    \caption{Comparison of the adaptation methods applied on the traffic dataset in conjunction with ZStream algorithm (sequence patterns): \protect\subref{fig:throughput} throughput (higher is better); \protect\subref{fig:relative-throughput} relative throughput gain over the non-adaptive method (higher is better); \protect\subref{fig:plan-switch} total number of plan reoptimizations; \protect\subref{fig:overhead} computational overhead (lower is better).}
	\label{fig:seq-traffic-zstream}
\end{figure*}

\begin{figure*}
	\centering
	\subfloat[]{\includegraphics[width=.25\linewidth]{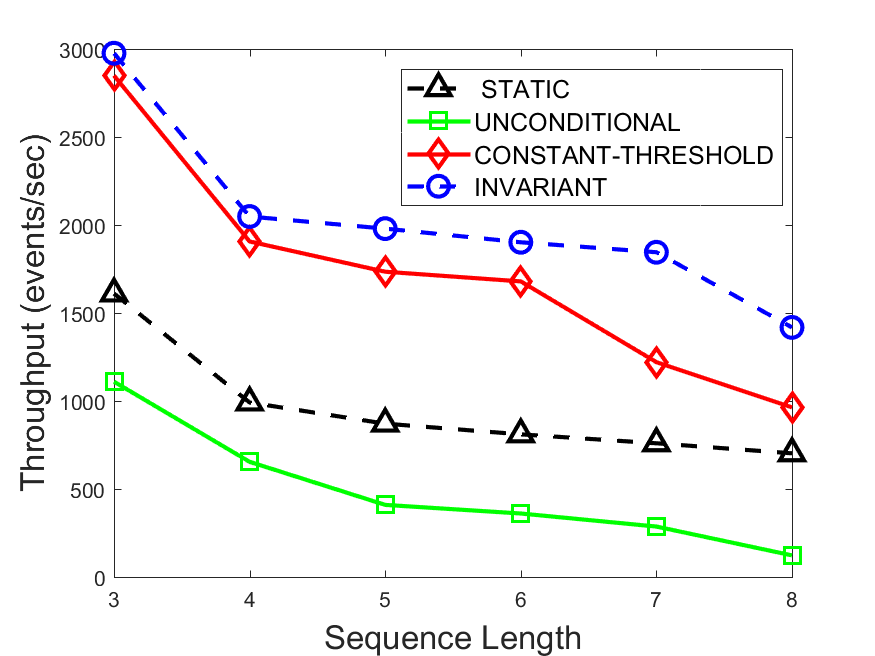}\label{fig:throughput}}
	\subfloat[]{\includegraphics[width=.25\linewidth]{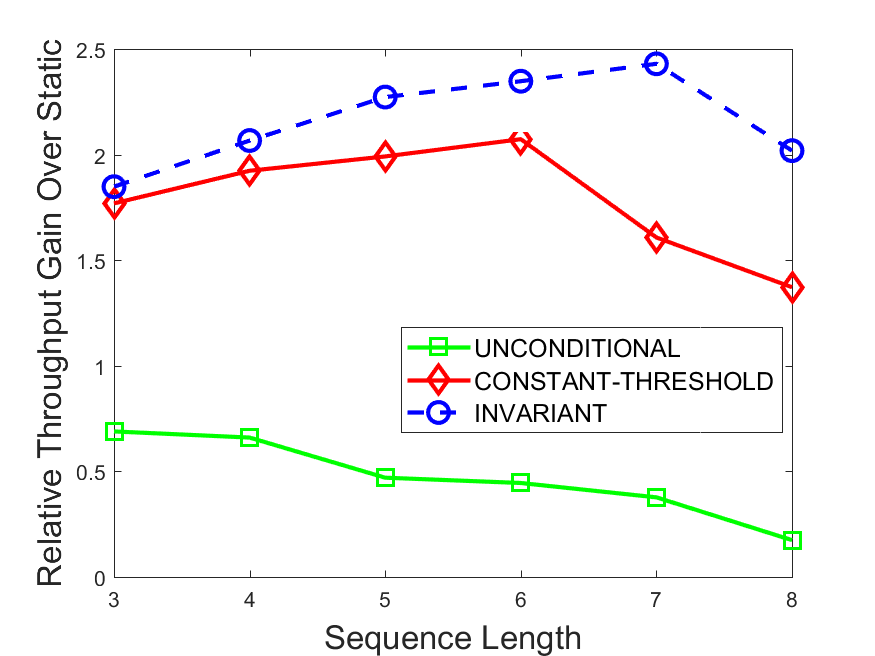}\label{fig:relative-throughput}}
	\subfloat[]{\includegraphics[width=.25\linewidth]{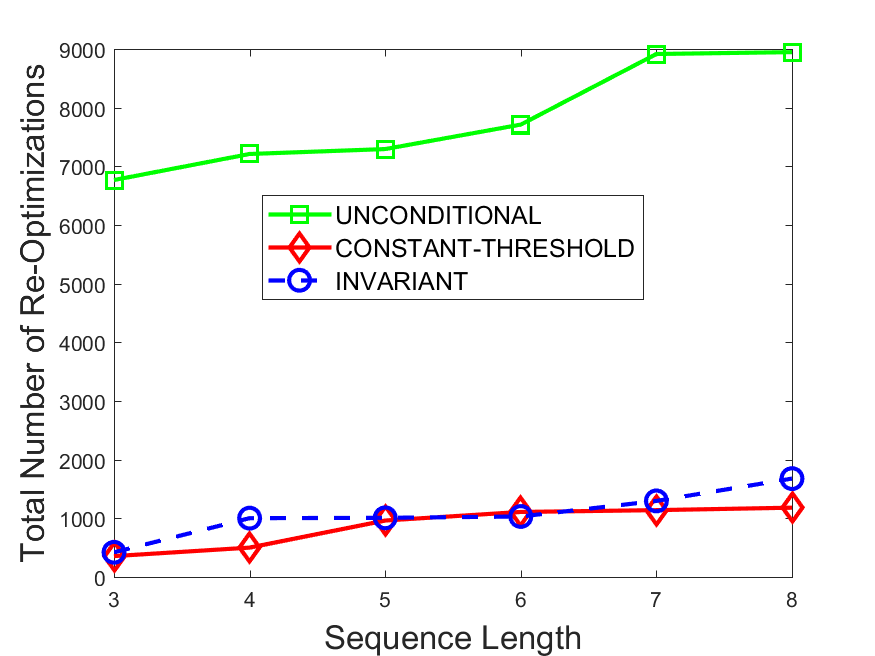}\label{fig:plan-switch}}
	\subfloat[]{\includegraphics[width=.25\linewidth]{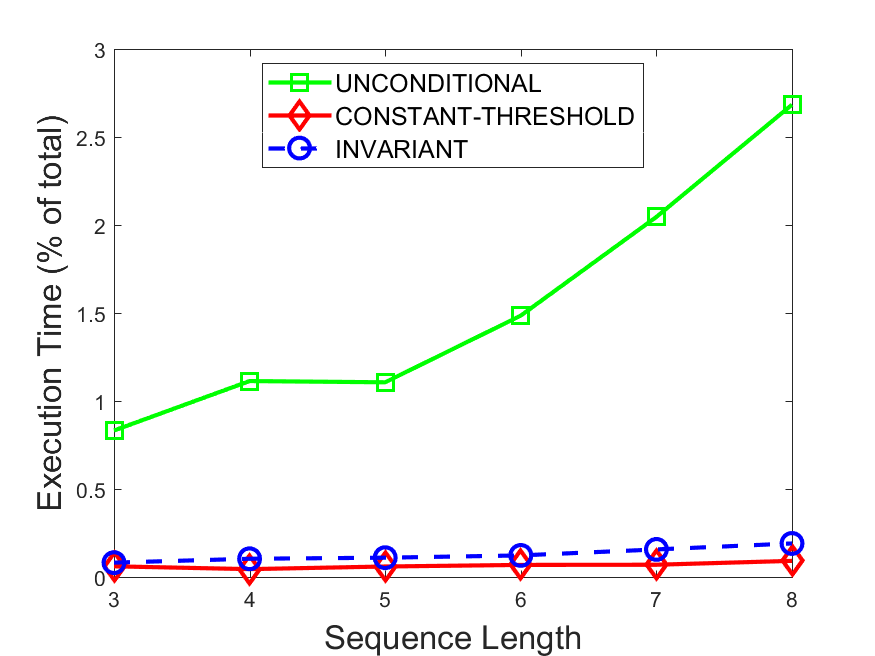}\label{fig:overhead}}
    \caption{Comparison of the adaptation methods applied on the stocks dataset in conjunction with the greedy algorithm (sequence patterns): \protect\subref{fig:throughput} throughput (higher is better); \protect\subref{fig:relative-throughput} relative throughput gain over the non-adaptive method (higher is better); \protect\subref{fig:plan-switch} total number of plan reoptimizations; \protect\subref{fig:overhead} computational overhead (lower is better).}
	\label{fig:seq-stocks-greedy}
\end{figure*}

\begin{figure*}
	\centering
	\subfloat[]{\includegraphics[width=.25\linewidth]{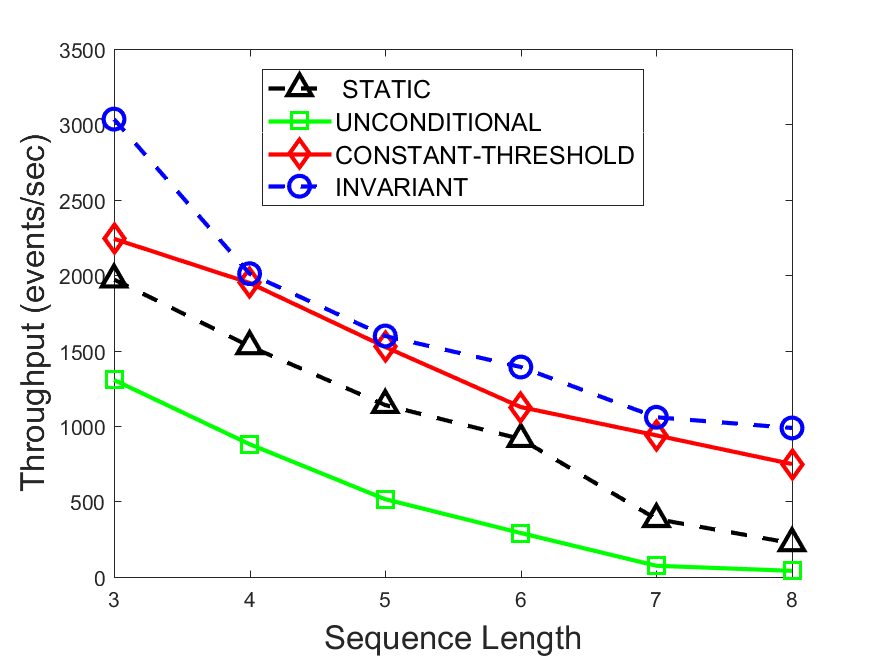}\label{fig:throughput}}
	\subfloat[]{\includegraphics[width=.25\linewidth]{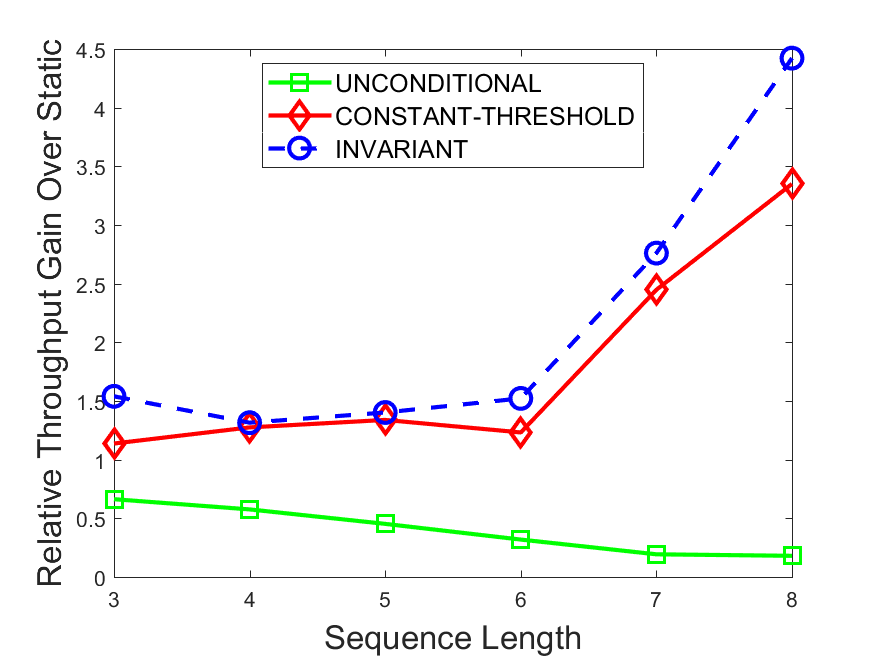}\label{fig:relative-throughput}}
	\subfloat[]{\includegraphics[width=.25\linewidth]{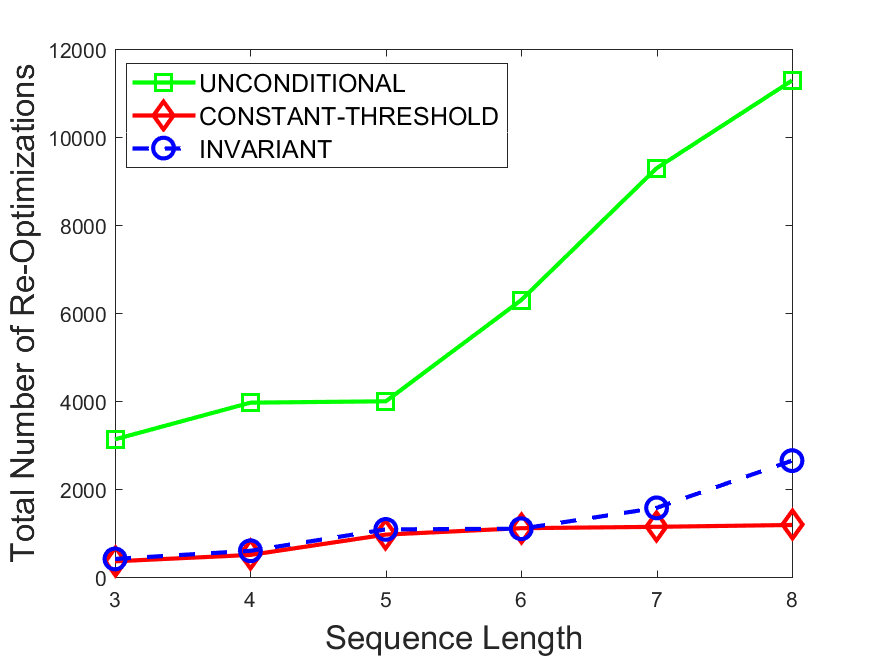}\label{fig:plan-switch}}
	\subfloat[]{\includegraphics[width=.25\linewidth]{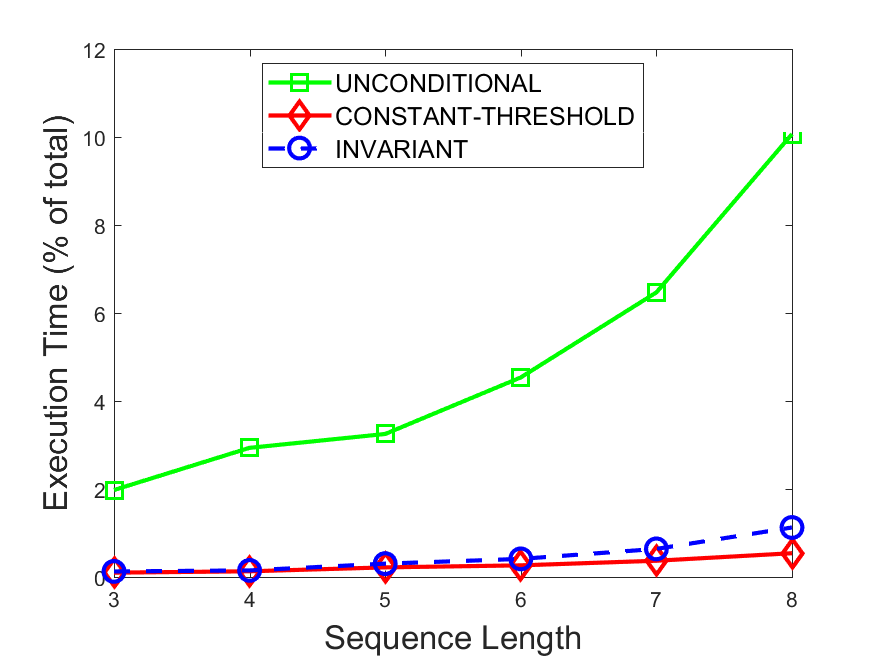}\label{fig:overhead}}
    \caption{Comparison of the adaptation methods applied on the stocks dataset in conjunction with ZStream algorithm (sequence patterns): \protect\subref{fig:throughput} throughput (higher is better); \protect\subref{fig:relative-throughput} relative throughput gain over the non-adaptive method (higher is better); \protect\subref{fig:plan-switch} total number of plan reoptimizations; \protect\subref{fig:overhead} computational overhead (lower is better).}
	\label{fig:seq-stocks-zstream}
\end{figure*}

\begin{figure*}
	\centering
	\subfloat[]{\includegraphics[width=.25\linewidth]{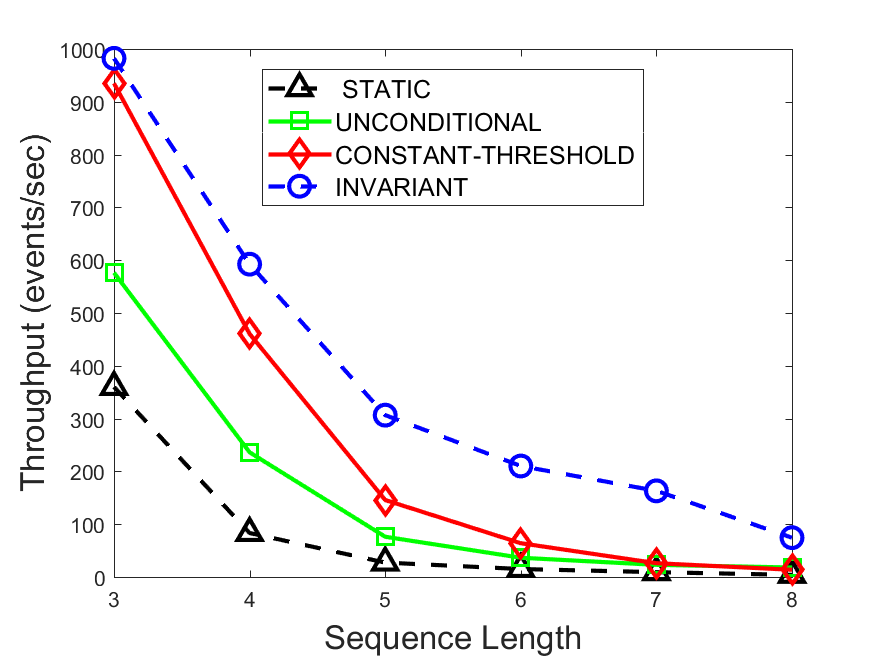}\label{fig:throughput}}
	\subfloat[]{\includegraphics[width=.25\linewidth]{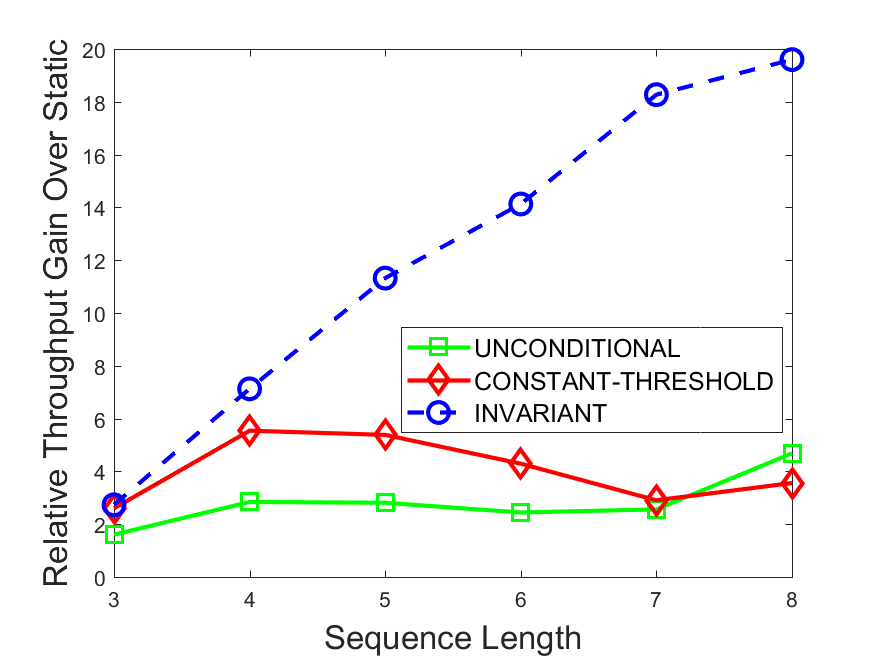}\label{fig:relative-throughput}}
	\subfloat[]{\includegraphics[width=.25\linewidth]{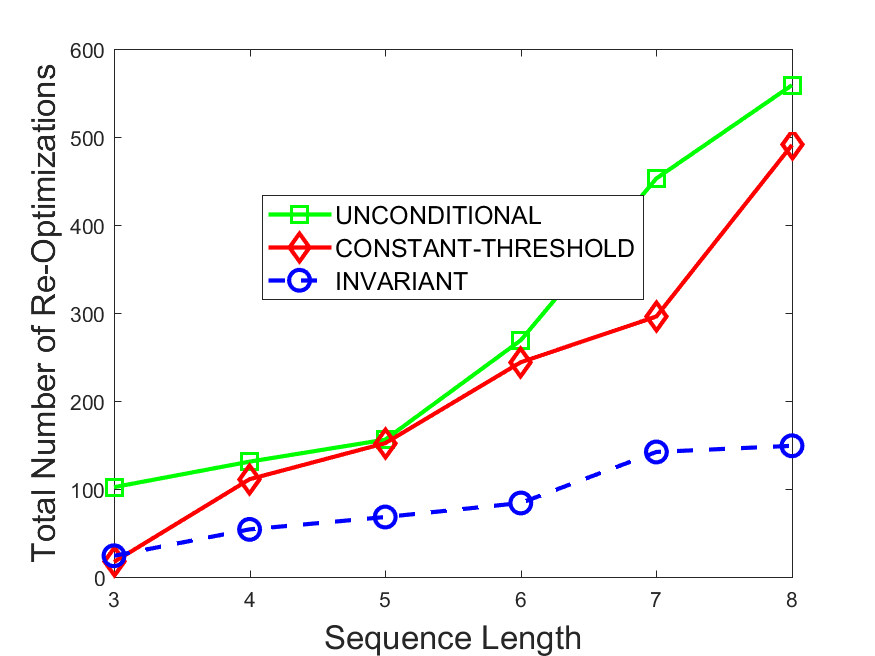}\label{fig:plan-switch}}
	\subfloat[]{\includegraphics[width=.25\linewidth]{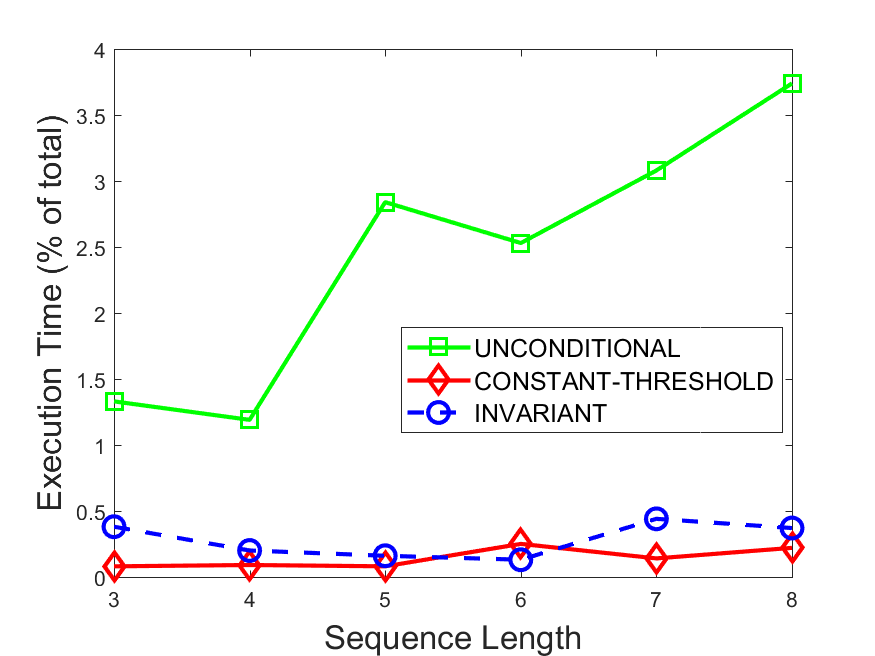}\label{fig:overhead}}
    \caption{Comparison of the adaptation methods applied on the traffic dataset in conjunction with the greedy algorithm (conjunction patterns): \protect\subref{fig:throughput} throughput (higher is better); \protect\subref{fig:relative-throughput} relative throughput gain over the non-adaptive method (higher is better); \protect\subref{fig:plan-switch} total number of plan reoptimizations; \protect\subref{fig:overhead} computational overhead (lower is better).}
	\label{fig:con-traffic-greedy}
\end{figure*}

\begin{figure*}
	\centering
	\subfloat[]{\includegraphics[width=.25\linewidth]{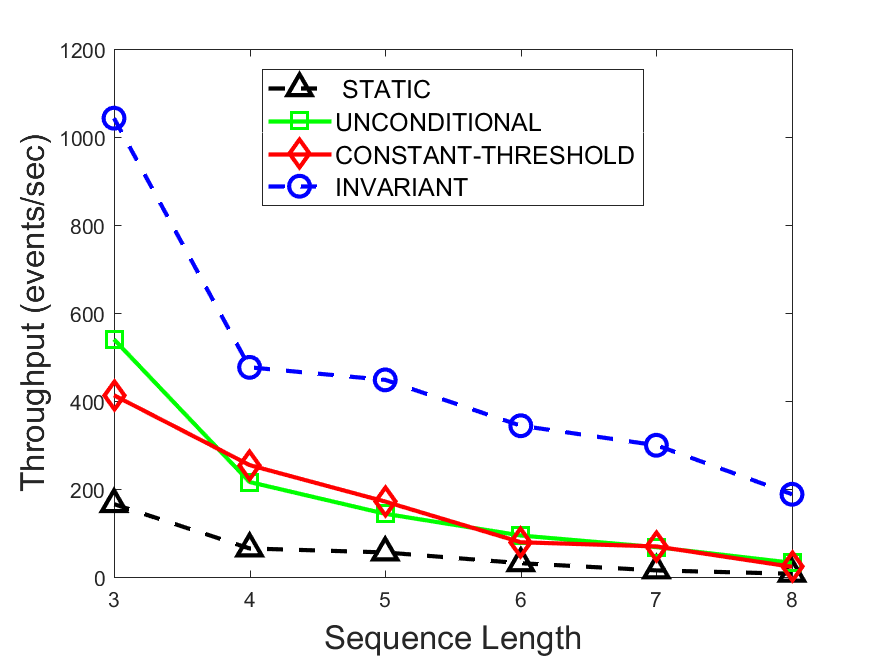}\label{fig:throughput}}
	\subfloat[]{\includegraphics[width=.25\linewidth]{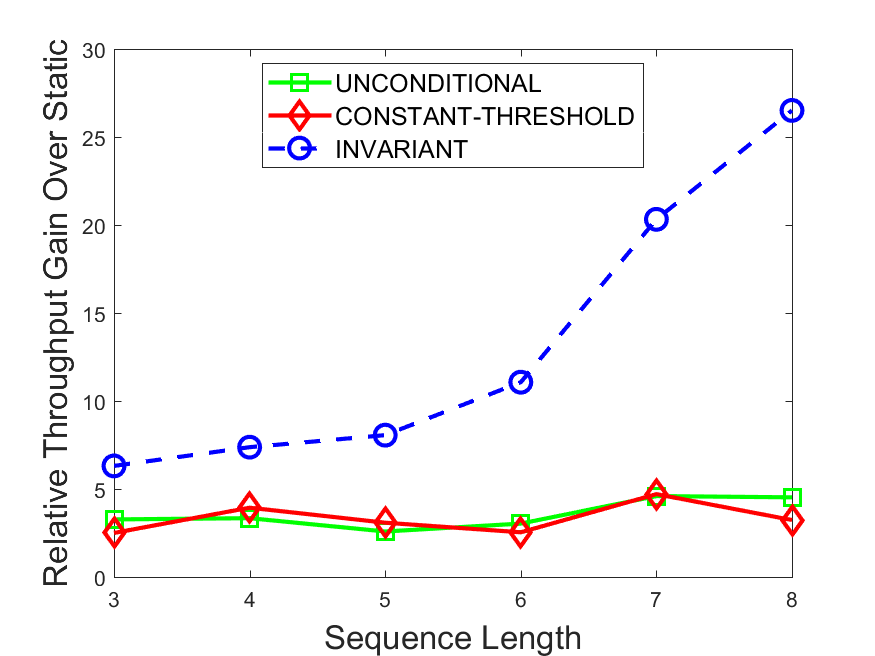}\label{fig:relative-throughput}}
	\subfloat[]{\includegraphics[width=.25\linewidth]{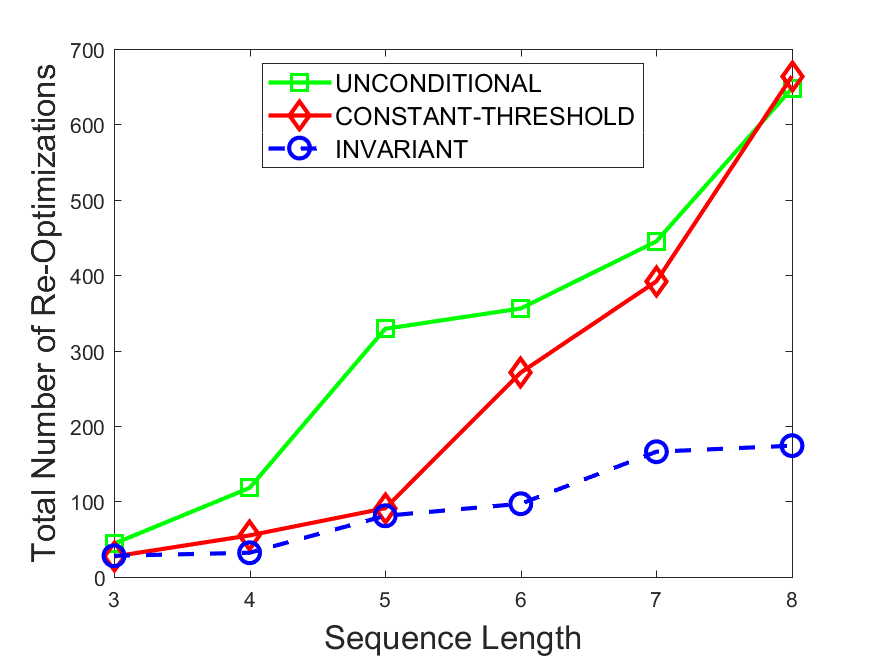}\label{fig:plan-switch}}
	\subfloat[]{\includegraphics[width=.25\linewidth]{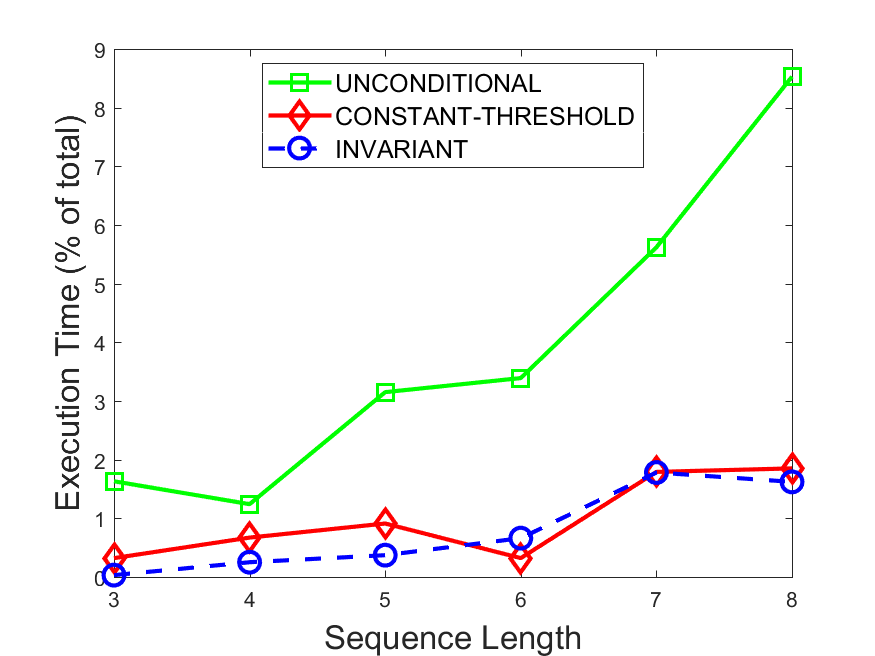}\label{fig:overhead}}
    \caption{Comparison of the adaptation methods applied on the traffic dataset in conjunction with ZStream algorithm (conjunction patterns): \protect\subref{fig:throughput} throughput (higher is better); \protect\subref{fig:relative-throughput} relative throughput gain over the non-adaptive method (higher is better); \protect\subref{fig:plan-switch} total number of plan reoptimizations; \protect\subref{fig:overhead} computational overhead (lower is better).}
	\label{fig:con-traffic-zstream}
\end{figure*}

\begin{figure*}
	\centering
	\subfloat[]{\includegraphics[width=.25\linewidth]{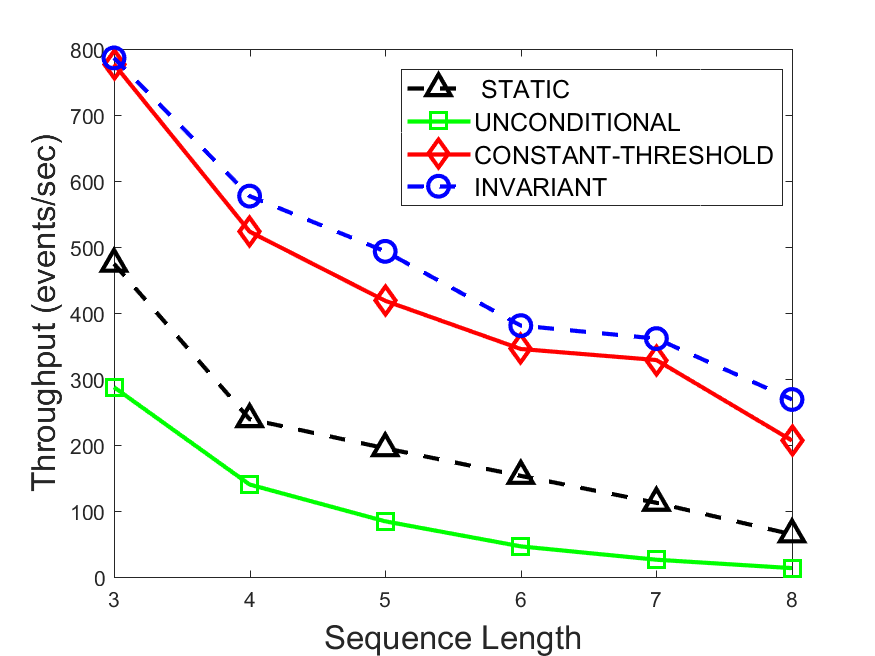}\label{fig:throughput}}
	\subfloat[]{\includegraphics[width=.25\linewidth]{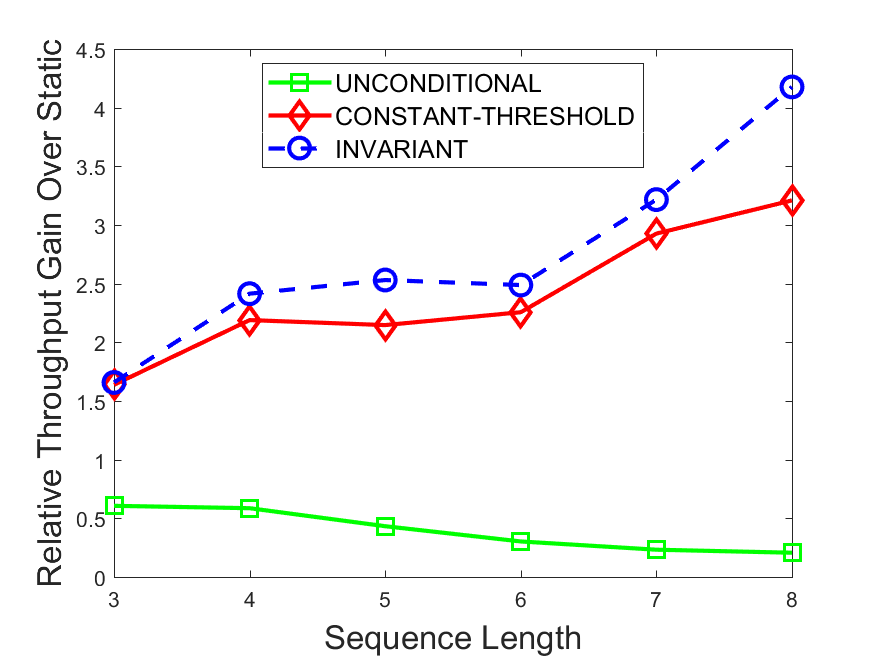}\label{fig:relative-throughput}}
	\subfloat[]{\includegraphics[width=.25\linewidth]{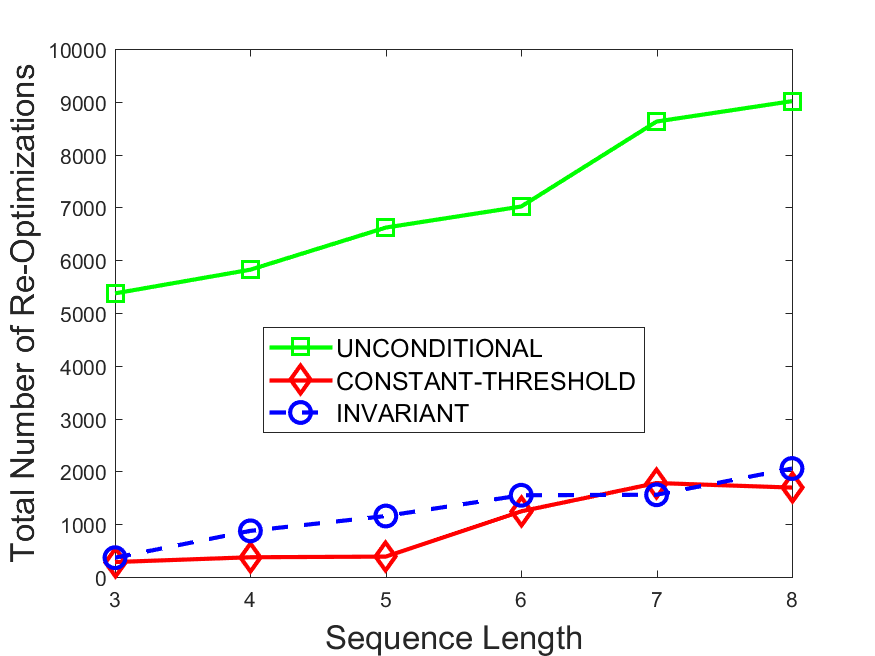}\label{fig:plan-switch}}
	\subfloat[]{\includegraphics[width=.25\linewidth]{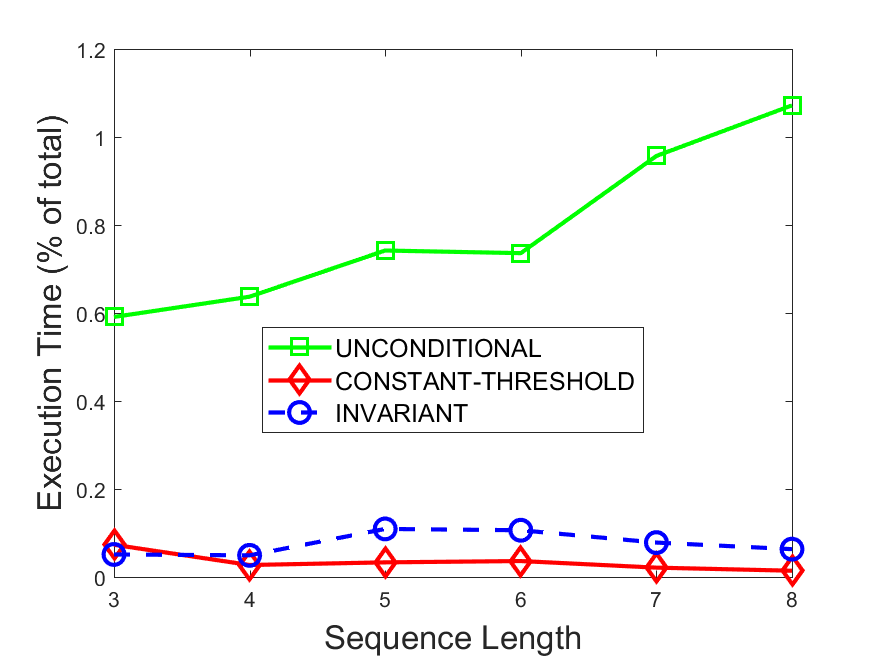}\label{fig:overhead}}
    \caption{Comparison of the adaptation methods applied on the stocks dataset in conjunction with the greedy algorithm (conjunction patterns): \protect\subref{fig:throughput} throughput (higher is better); \protect\subref{fig:relative-throughput} relative throughput gain over the non-adaptive method (higher is better); \protect\subref{fig:plan-switch} total number of plan reoptimizations; \protect\subref{fig:overhead} computational overhead (lower is better).}
	\label{fig:con-stocks-greedy}
\end{figure*}

\begin{figure*}
	\centering
	\subfloat[]{\includegraphics[width=.25\linewidth]{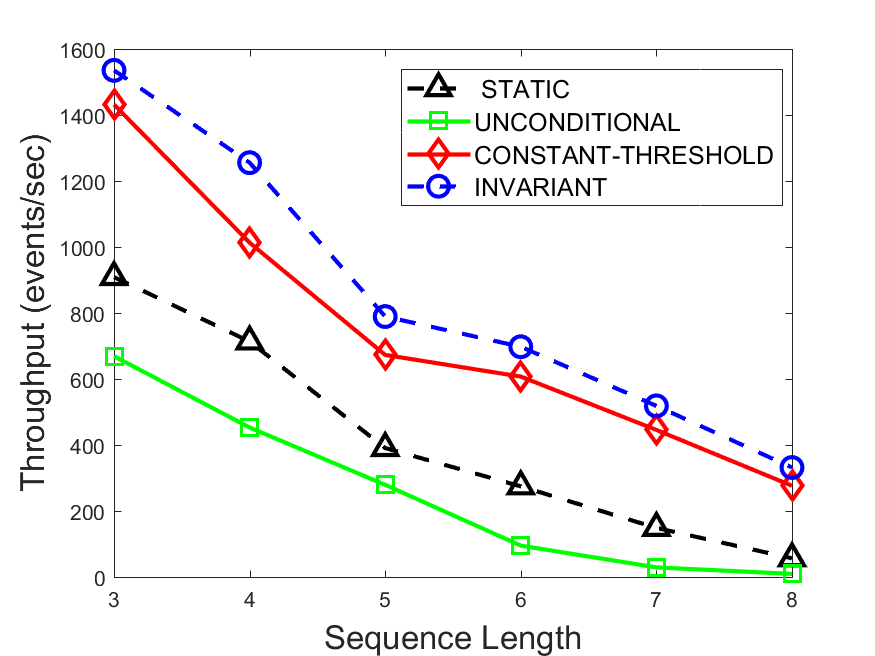}\label{fig:throughput}}
	\subfloat[]{\includegraphics[width=.25\linewidth]{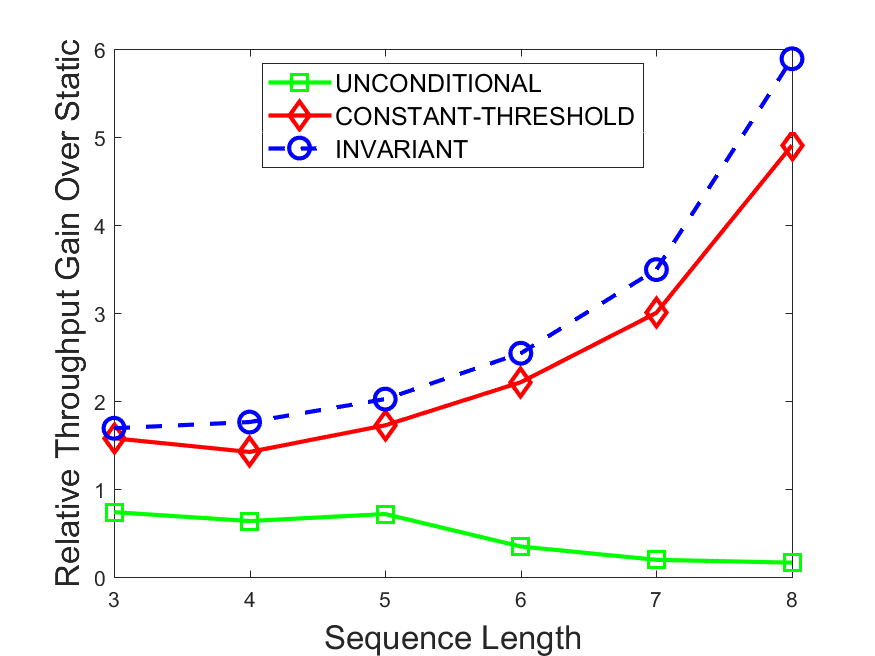}\label{fig:relative-throughput}}
	\subfloat[]{\includegraphics[width=.25\linewidth]{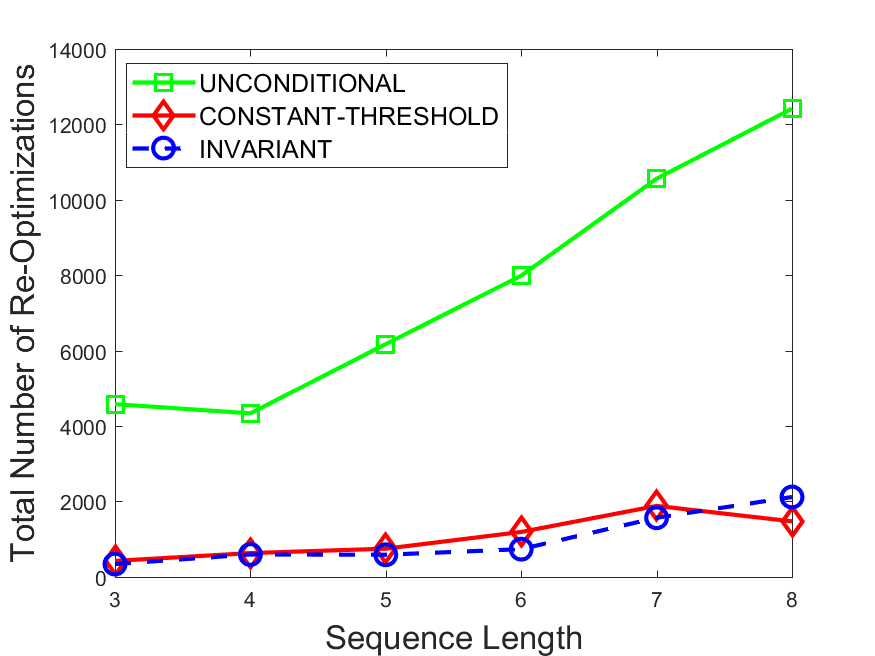}\label{fig:plan-switch}}
	\subfloat[]{\includegraphics[width=.25\linewidth]{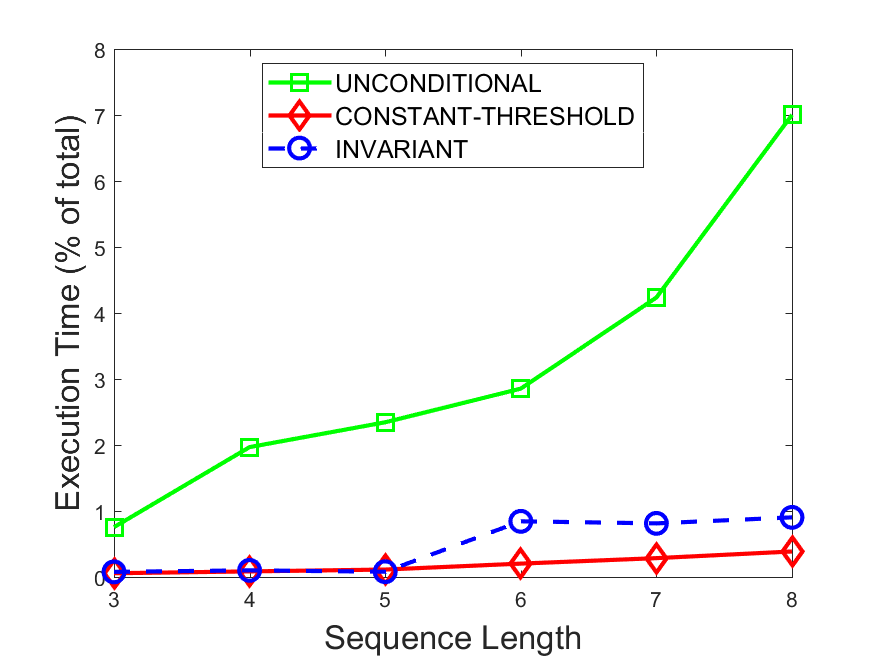}\label{fig:overhead}}
    \caption{Comparison of the adaptation methods applied on the stocks dataset in conjunction with ZStream algorithm (conjunction patterns): \protect\subref{fig:throughput} throughput (higher is better); \protect\subref{fig:relative-throughput} relative throughput gain over the non-adaptive method (higher is better); \protect\subref{fig:plan-switch} total number of plan reoptimizations; \protect\subref{fig:overhead} computational overhead (lower is better).}
	\label{fig:con-stocks-zstream}
\end{figure*}

\begin{figure*}
	\centering
	\subfloat[]{\includegraphics[width=.25\linewidth]{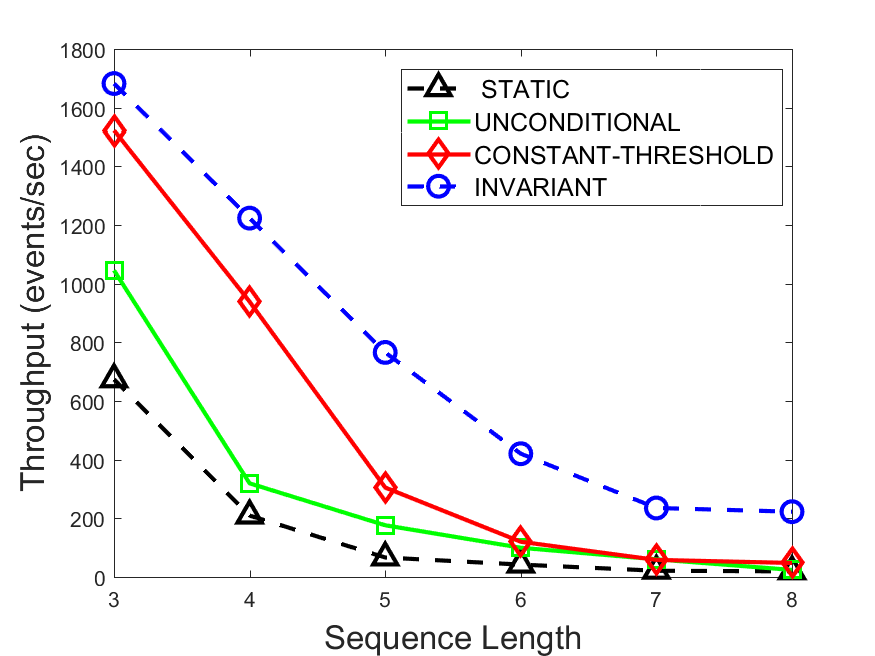}\label{fig:throughput}}
	\subfloat[]{\includegraphics[width=.25\linewidth]{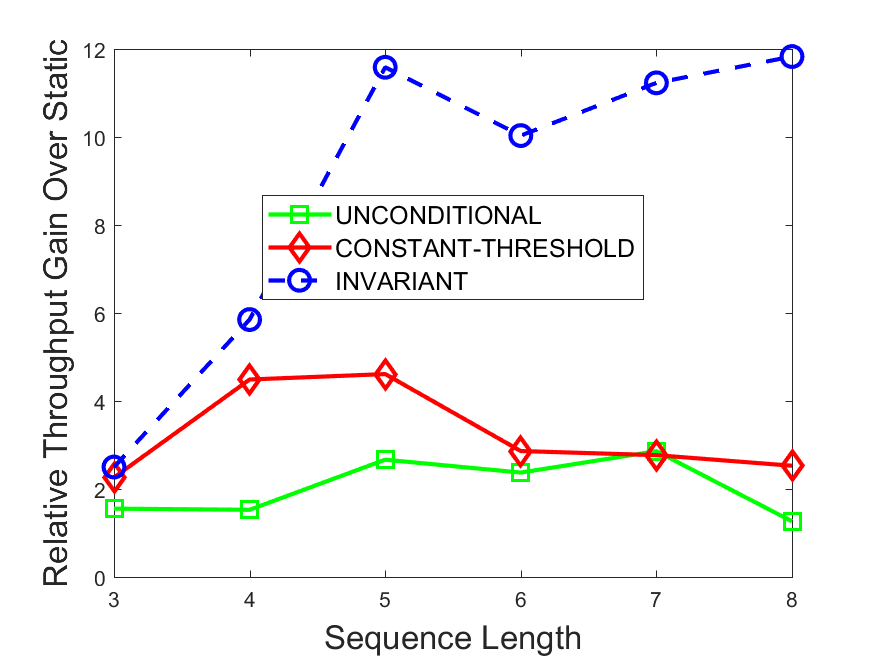}\label{fig:relative-throughput}}
	\subfloat[]{\includegraphics[width=.25\linewidth]{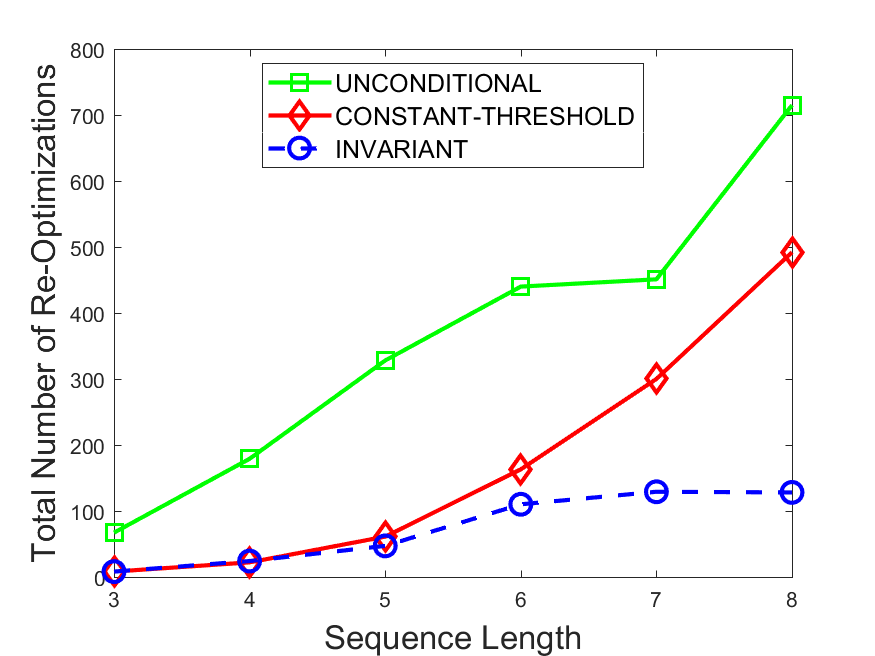}\label{fig:plan-switch}}
	\subfloat[]{\includegraphics[width=.25\linewidth]{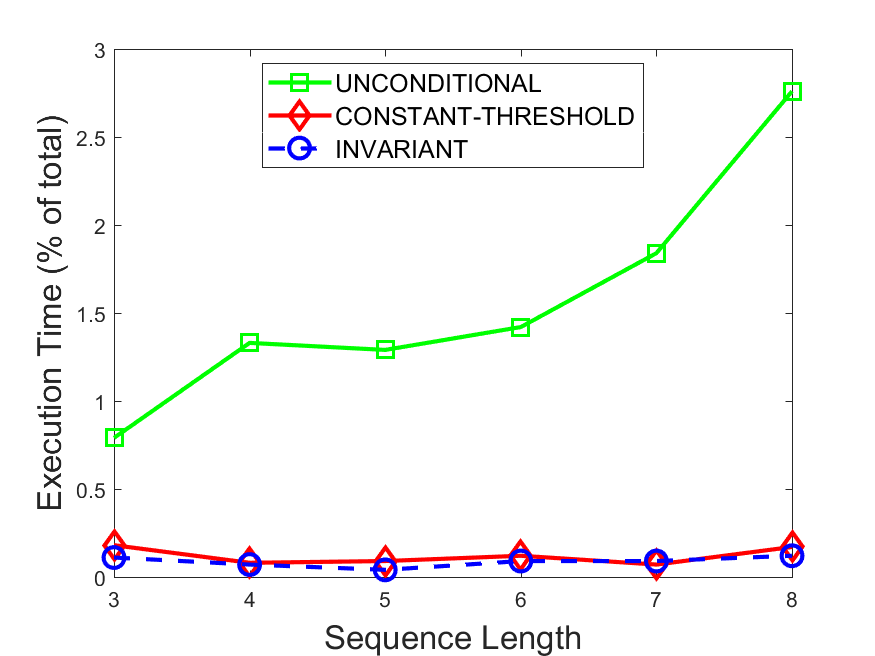}\label{fig:overhead}}
    \caption{Comparison of the adaptation methods applied on the traffic dataset in conjunction with the greedy algorithm (negation patterns): \protect\subref{fig:throughput} throughput (higher is better); \protect\subref{fig:relative-throughput} relative throughput gain over the non-adaptive method (higher is better); \protect\subref{fig:plan-switch} total number of plan reoptimizations; \protect\subref{fig:overhead} computational overhead (lower is better).}
	\label{fig:neg-traffic-greedy}
\end{figure*}

\begin{figure*}
	\centering
	\subfloat[]{\includegraphics[width=.25\linewidth]{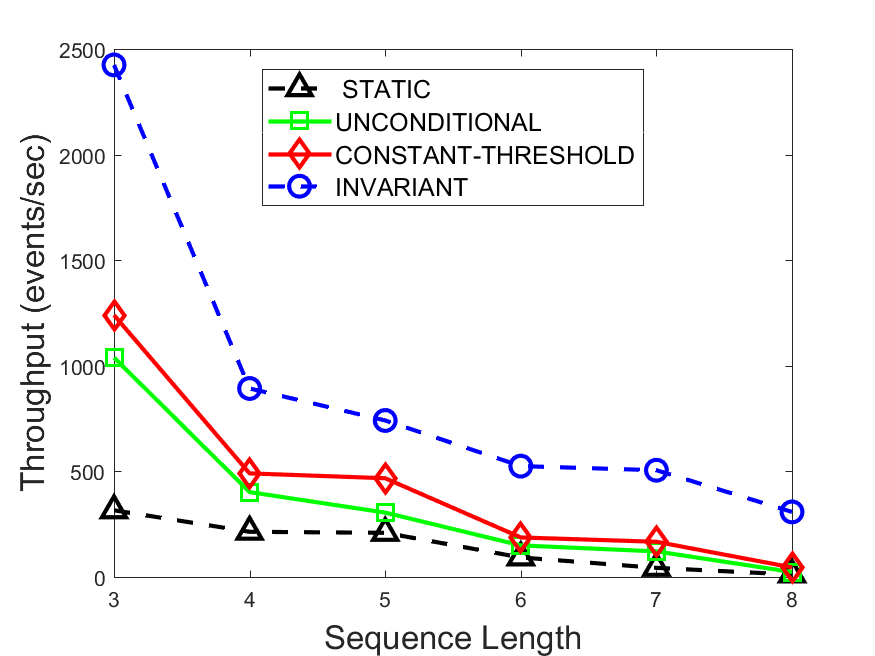}\label{fig:throughput}}
	\subfloat[]{\includegraphics[width=.25\linewidth]{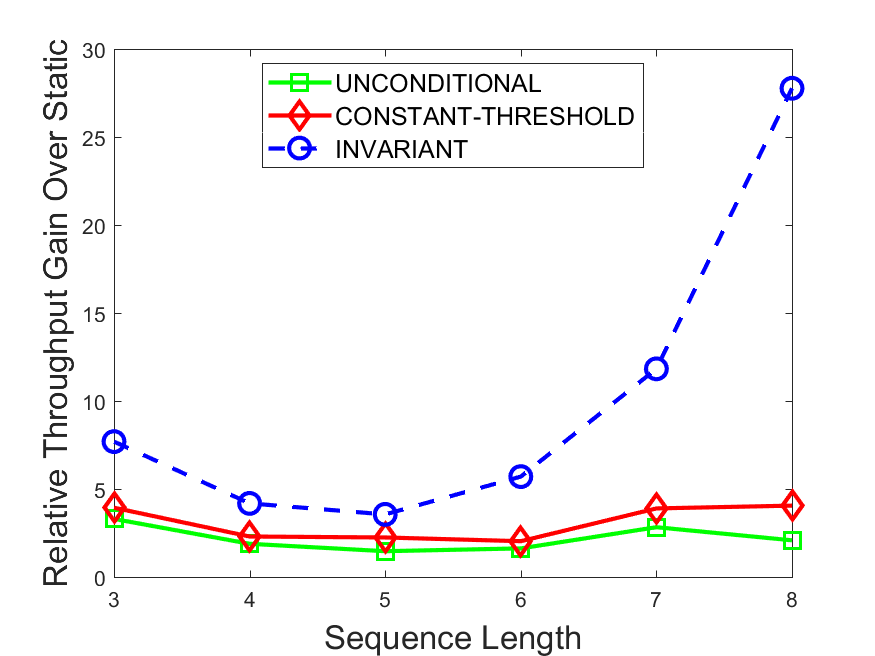}\label{fig:relative-throughput}}
	\subfloat[]{\includegraphics[width=.25\linewidth]{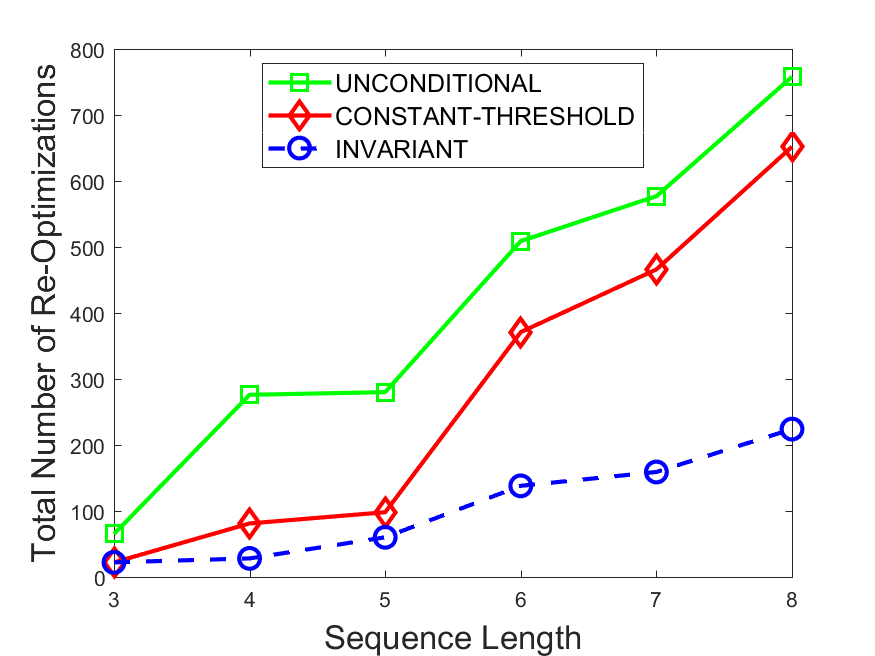}\label{fig:plan-switch}}
	\subfloat[]{\includegraphics[width=.25\linewidth]{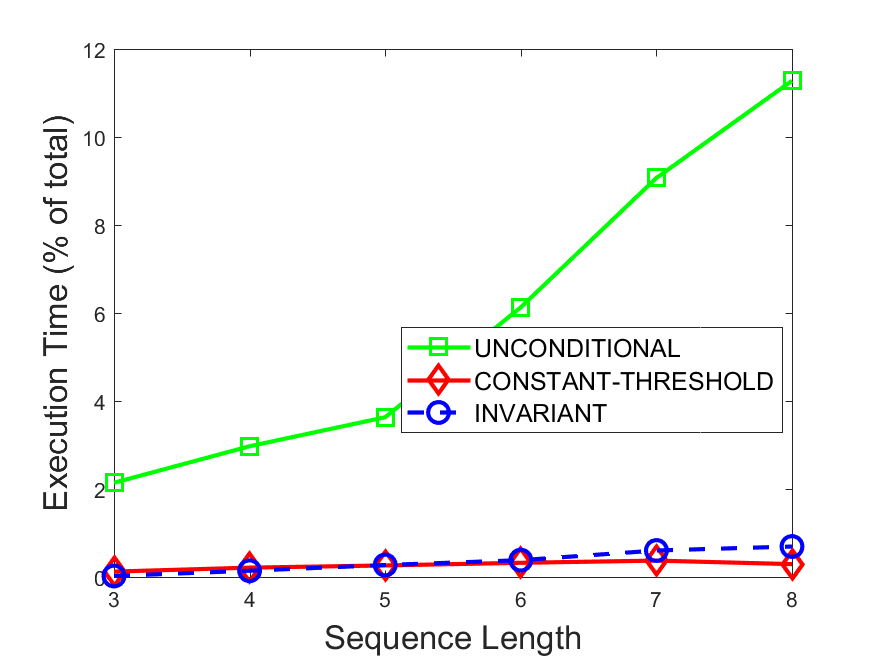}\label{fig:overhead}}
    \caption{Comparison of the adaptation methods applied on the traffic dataset in conjunction with ZStream algorithm (negation patterns): \protect\subref{fig:throughput} throughput (higher is better); \protect\subref{fig:relative-throughput} relative throughput gain over the non-adaptive method (higher is better); \protect\subref{fig:plan-switch} total number of plan reoptimizations; \protect\subref{fig:overhead} computational overhead (lower is better).}
	\label{fig:neg-traffic-zstream}
\end{figure*}

\begin{figure*}
	\centering
	\subfloat[]{\includegraphics[width=.25\linewidth]{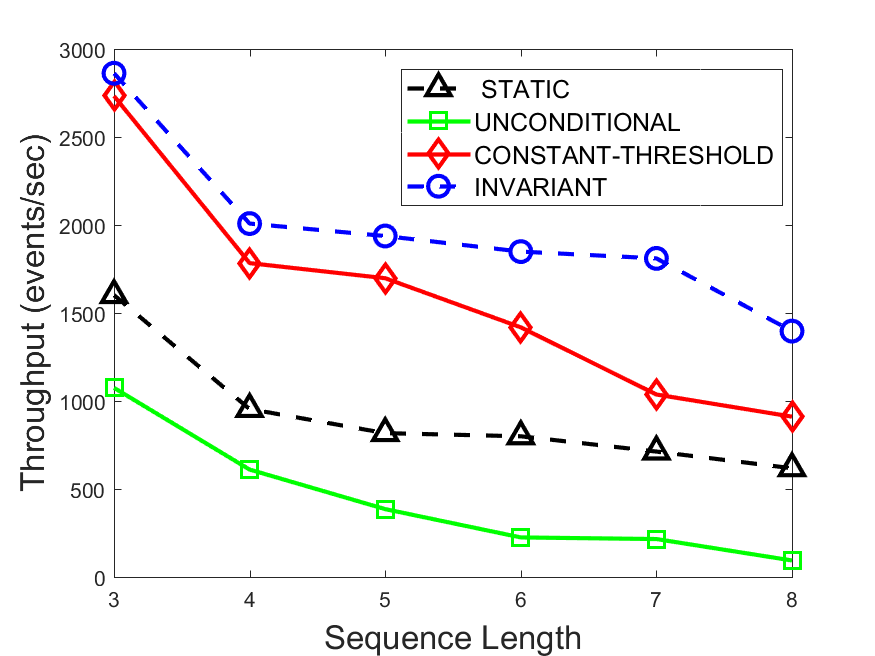}\label{fig:throughput}}
	\subfloat[]{\includegraphics[width=.25\linewidth]{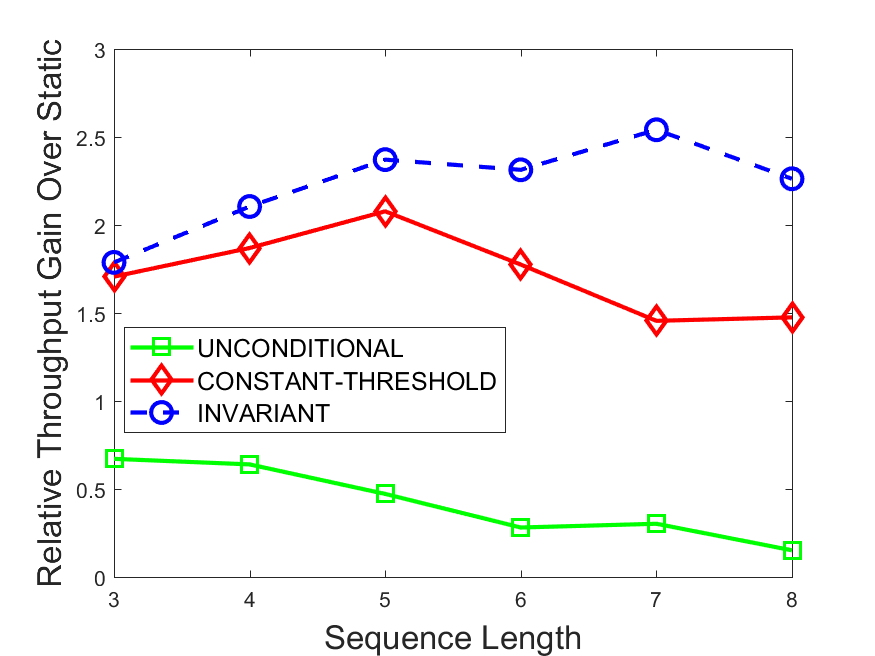}\label{fig:relative-throughput}}
	\subfloat[]{\includegraphics[width=.25\linewidth]{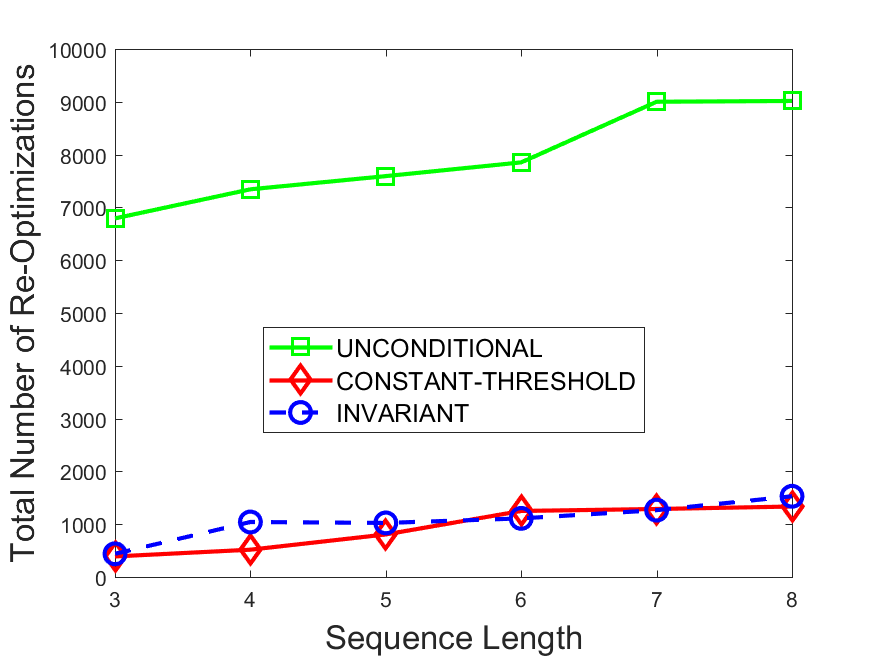}\label{fig:plan-switch}}
	\subfloat[]{\includegraphics[width=.25\linewidth]{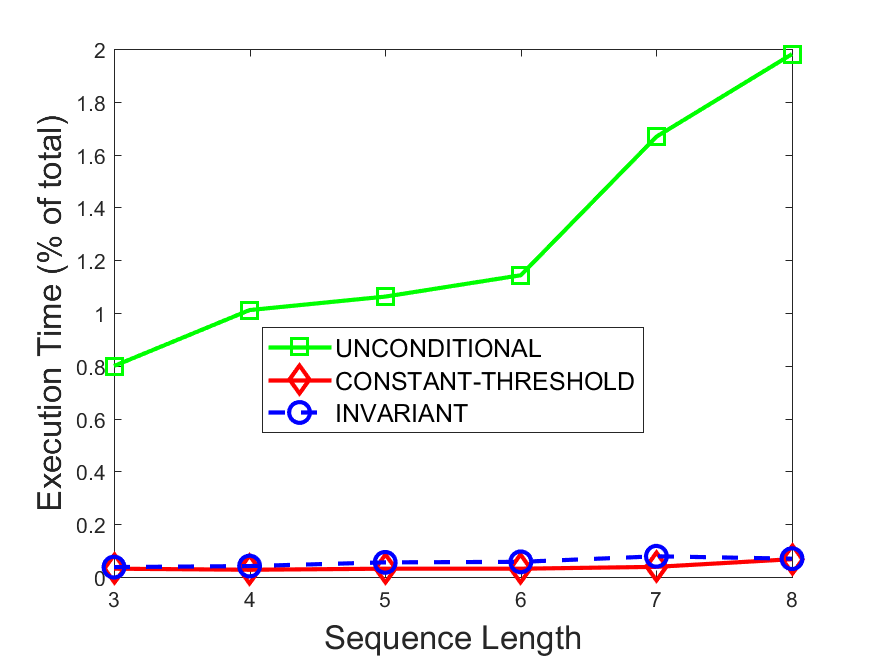}\label{fig:overhead}}
    \caption{Comparison of the adaptation methods applied on the stocks dataset in conjunction with the greedy algorithm (negation patterns): \protect\subref{fig:throughput} throughput (higher is better); \protect\subref{fig:relative-throughput} relative throughput gain over the non-adaptive method (higher is better); \protect\subref{fig:plan-switch} total number of plan reoptimizations; \protect\subref{fig:overhead} computational overhead (lower is better).}
	\label{fig:neg-stocks-greedy}
\end{figure*}

\begin{figure*}
	\centering
	\subfloat[]{\includegraphics[width=.25\linewidth]{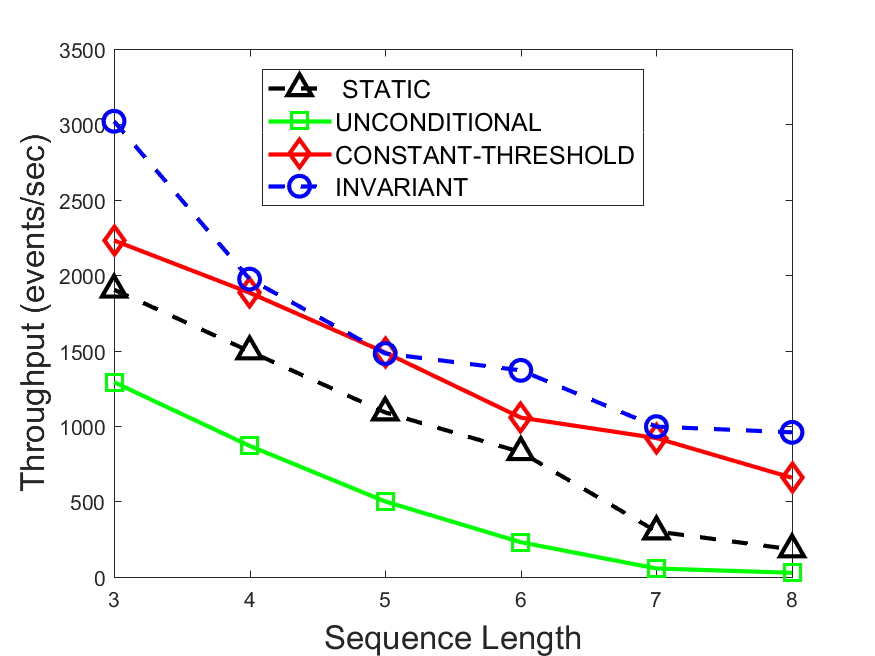}\label{fig:throughput}}
	\subfloat[]{\includegraphics[width=.25\linewidth]{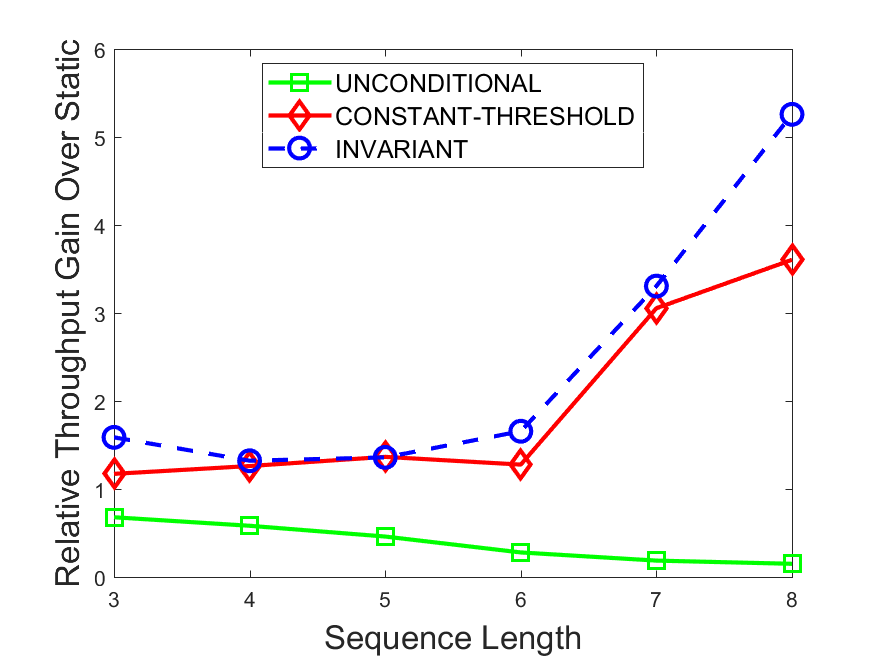}\label{fig:relative-throughput}}
	\subfloat[]{\includegraphics[width=.25\linewidth]{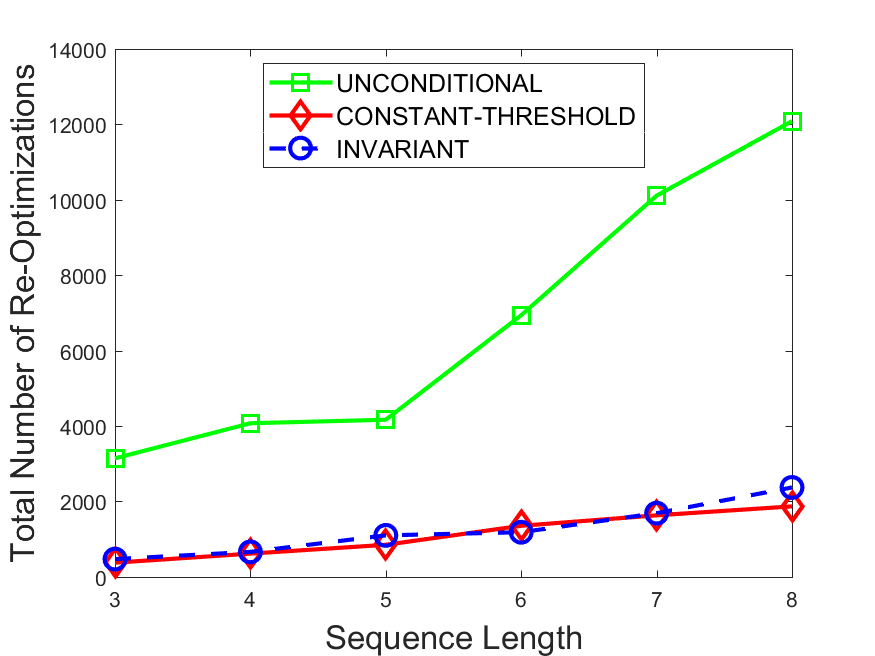}\label{fig:plan-switch}}
	\subfloat[]{\includegraphics[width=.25\linewidth]{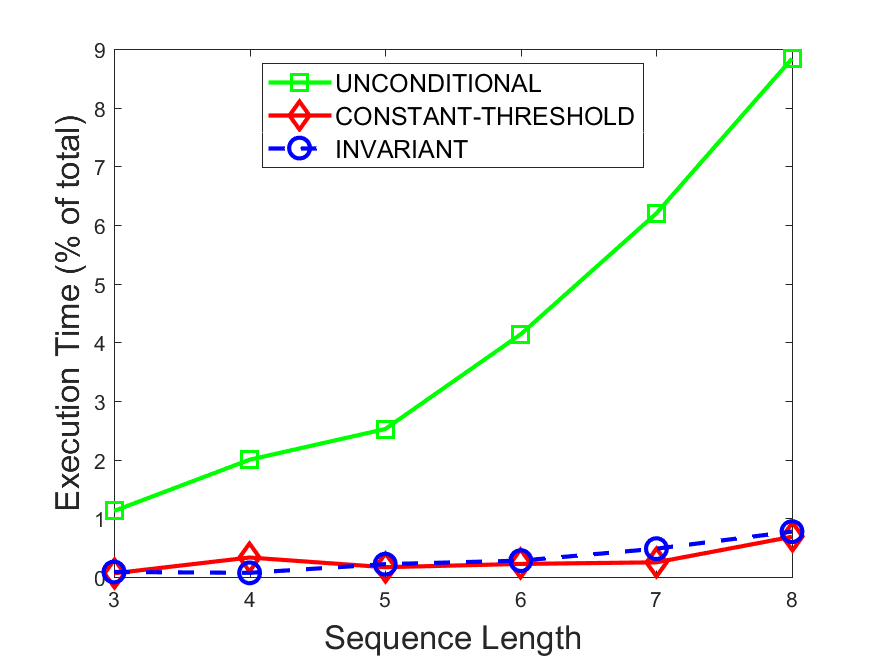}\label{fig:overhead}}
    \caption{Comparison of the adaptation methods applied on the stocks dataset in conjunction with ZStream algorithm (negation patterns): \protect\subref{fig:throughput} throughput (higher is better); \protect\subref{fig:relative-throughput} relative throughput gain over the non-adaptive method (higher is better); \protect\subref{fig:plan-switch} total number of plan reoptimizations; \protect\subref{fig:overhead} computational overhead (lower is better).}
	\label{fig:neg-stocks-zstream}
\end{figure*}

\begin{figure*}
	\centering
	\subfloat[]{\includegraphics[width=.25\linewidth]{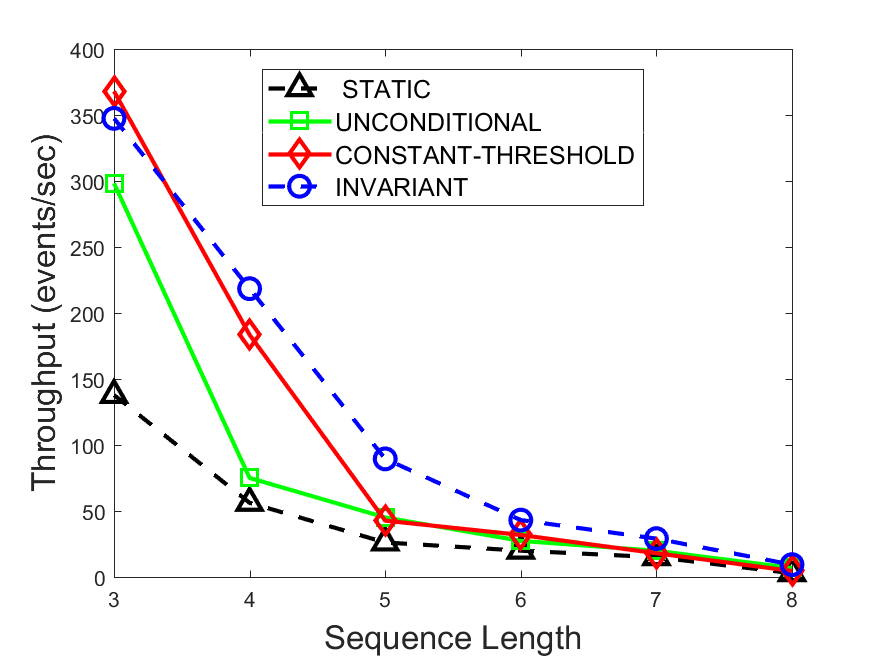}\label{fig:throughput}}
	\subfloat[]{\includegraphics[width=.25\linewidth]{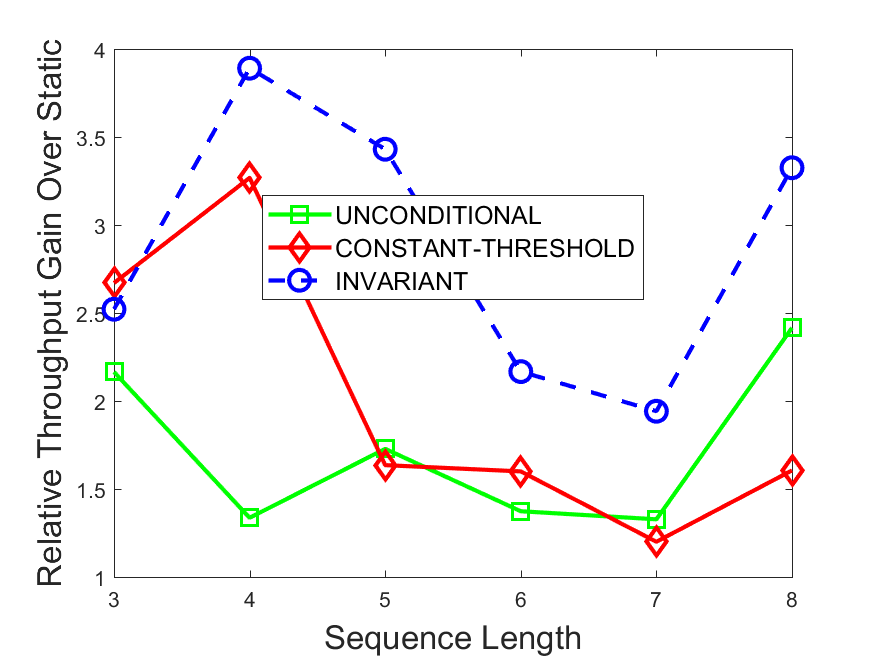}\label{fig:relative-throughput}}
	\subfloat[]{\includegraphics[width=.25\linewidth]{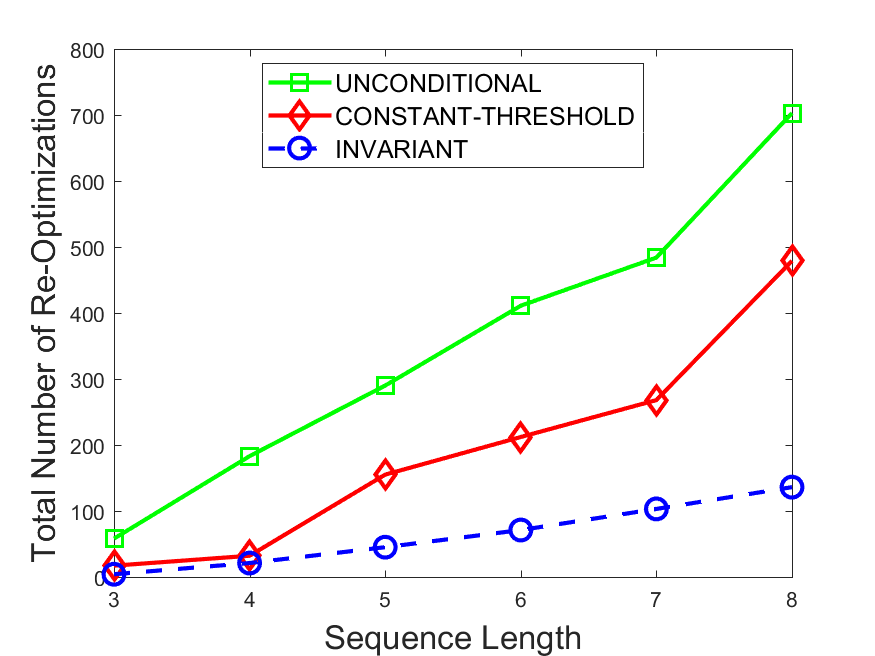}\label{fig:plan-switch}}
	\subfloat[]{\includegraphics[width=.25\linewidth]{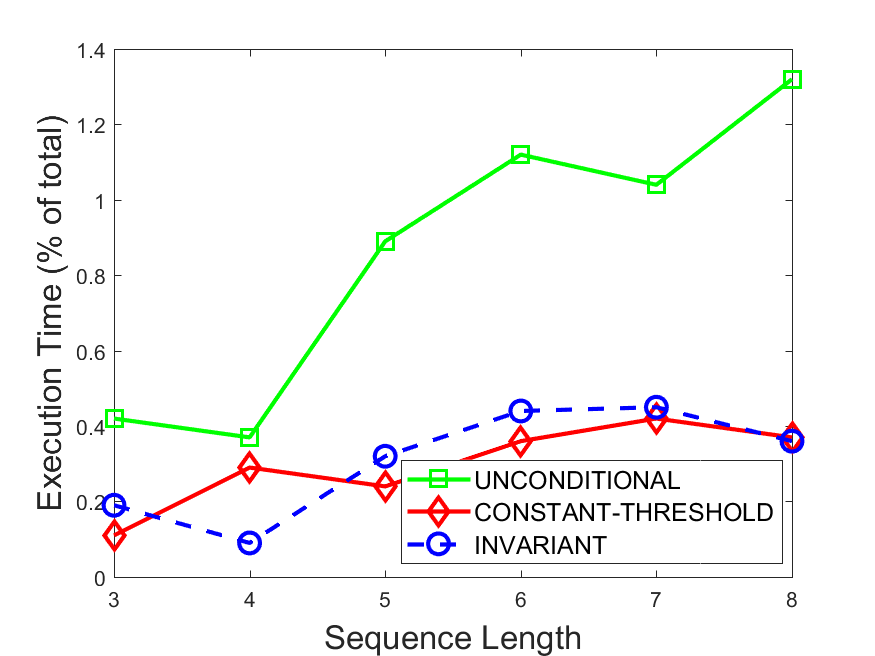}\label{fig:overhead}}
    \caption{Comparison of the adaptation methods applied on the traffic dataset in conjunction with the greedy algorithm (Kleene closure patterns): \protect\subref{fig:throughput} throughput (higher is better); \protect\subref{fig:relative-throughput} relative throughput gain over the non-adaptive method (higher is better); \protect\subref{fig:plan-switch} total number of plan reoptimizations; \protect\subref{fig:overhead} computational overhead (lower is better).}
	\label{fig:it-traffic-greedy}
\end{figure*}

\begin{figure*}
	\centering
	\subfloat[]{\includegraphics[width=.25\linewidth]{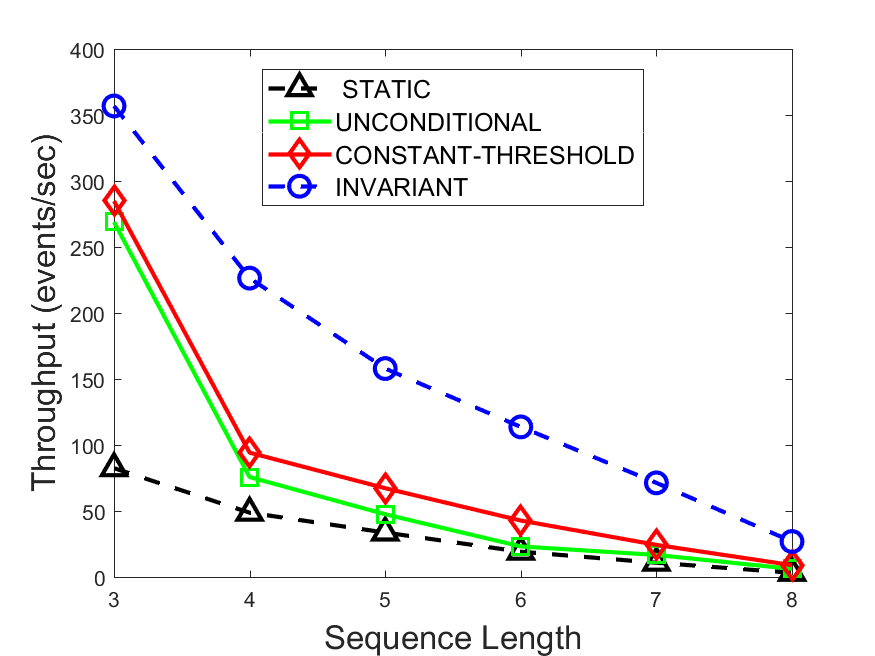}\label{fig:throughput}}
	\subfloat[]{\includegraphics[width=.25\linewidth]{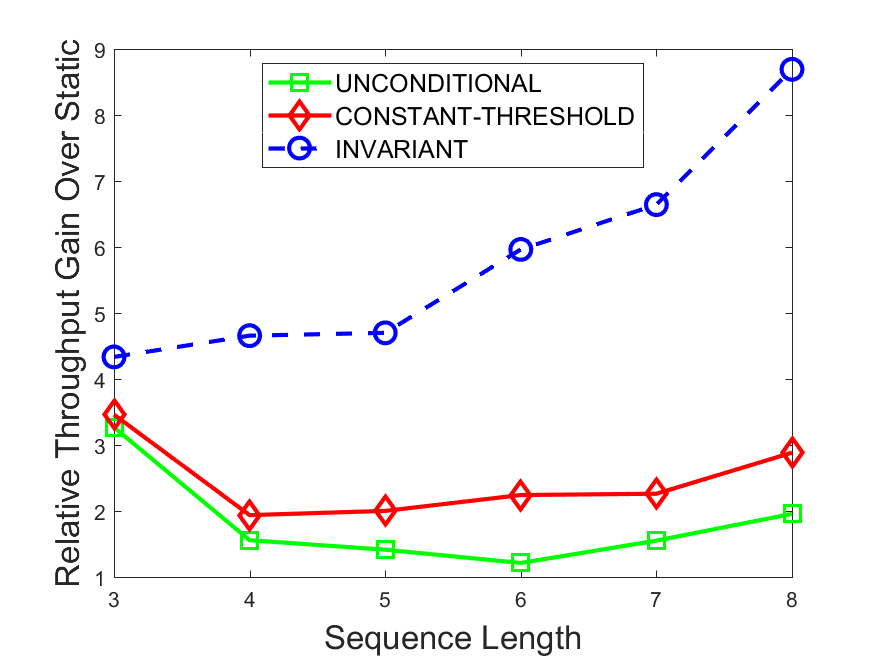}\label{fig:relative-throughput}}
	\subfloat[]{\includegraphics[width=.25\linewidth]{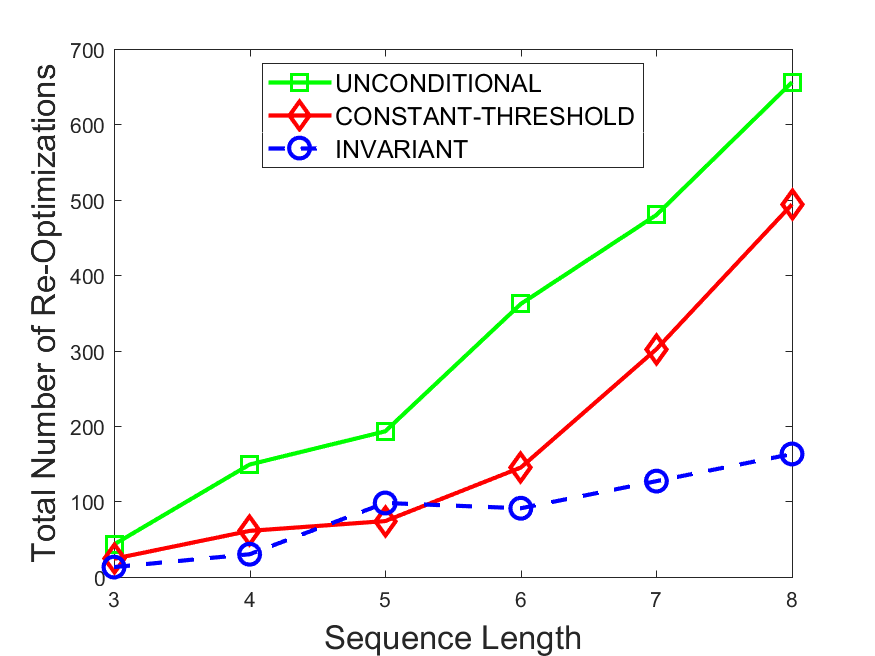}\label{fig:plan-switch}}
	\subfloat[]{\includegraphics[width=.25\linewidth]{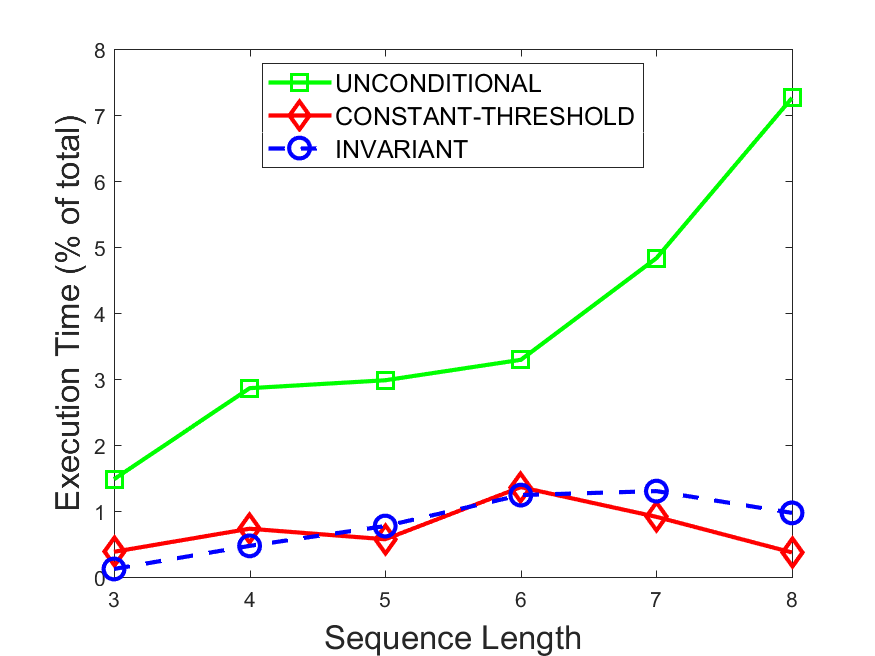}\label{fig:overhead}}
    \caption{Comparison of the adaptation methods applied on the traffic dataset in conjunction with ZStream algorithm (Kleene closure patterns): \protect\subref{fig:throughput} throughput (higher is better); \protect\subref{fig:relative-throughput} relative throughput gain over the non-adaptive method (higher is better); \protect\subref{fig:plan-switch} total number of plan reoptimizations; \protect\subref{fig:overhead} computational overhead (lower is better).}
	\label{fig:it-traffic-zstream}
\end{figure*}

\begin{figure*}
	\centering
	\subfloat[]{\includegraphics[width=.25\linewidth]{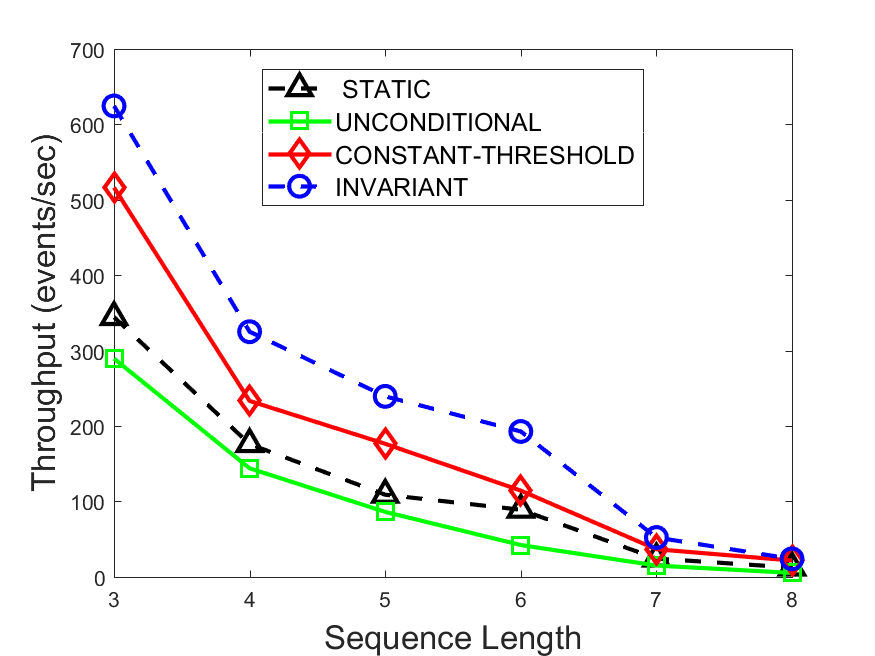}\label{fig:throughput}}
	\subfloat[]{\includegraphics[width=.25\linewidth]{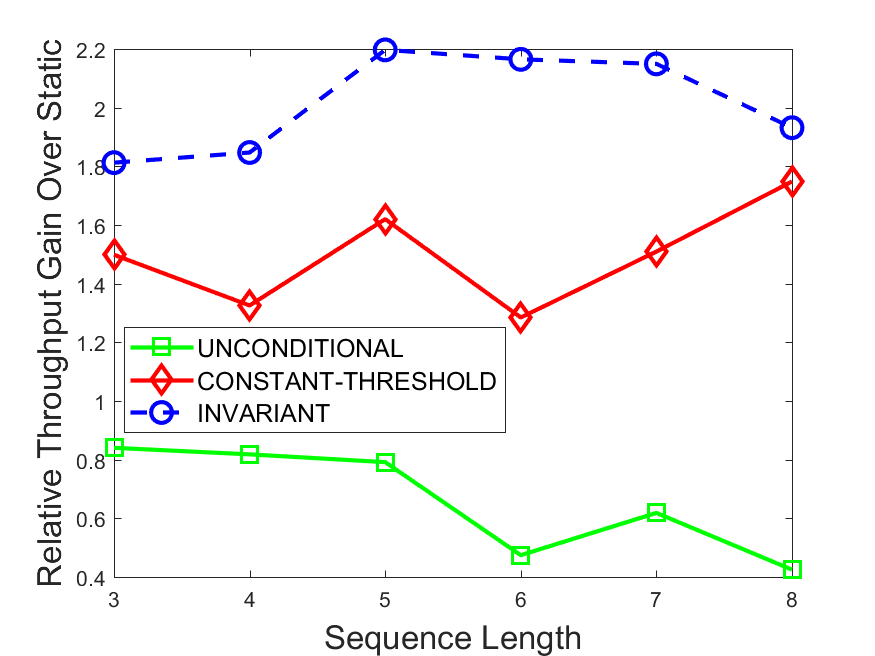}\label{fig:relative-throughput}}
	\subfloat[]{\includegraphics[width=.25\linewidth]{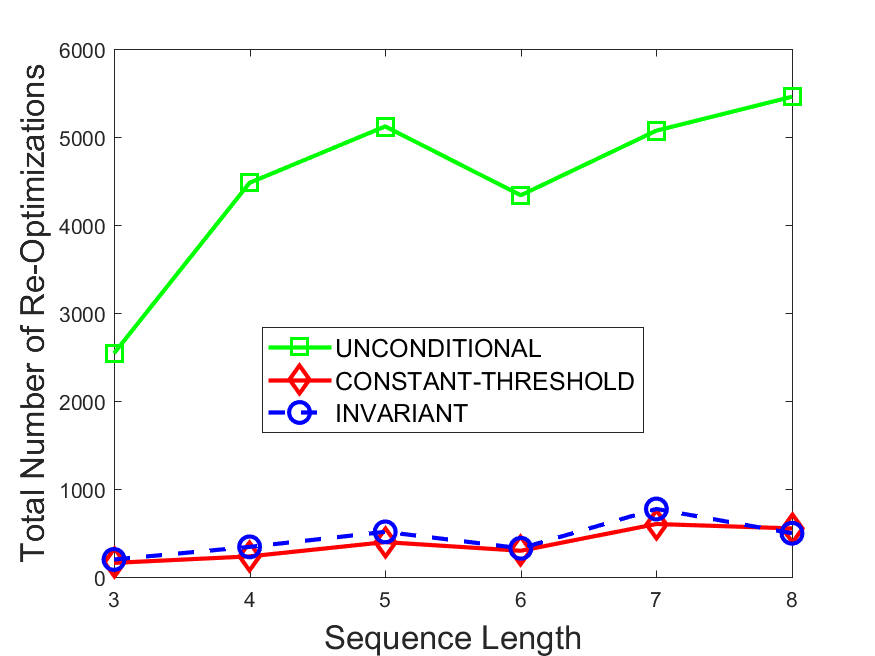}\label{fig:plan-switch}}
	\subfloat[]{\includegraphics[width=.25\linewidth]{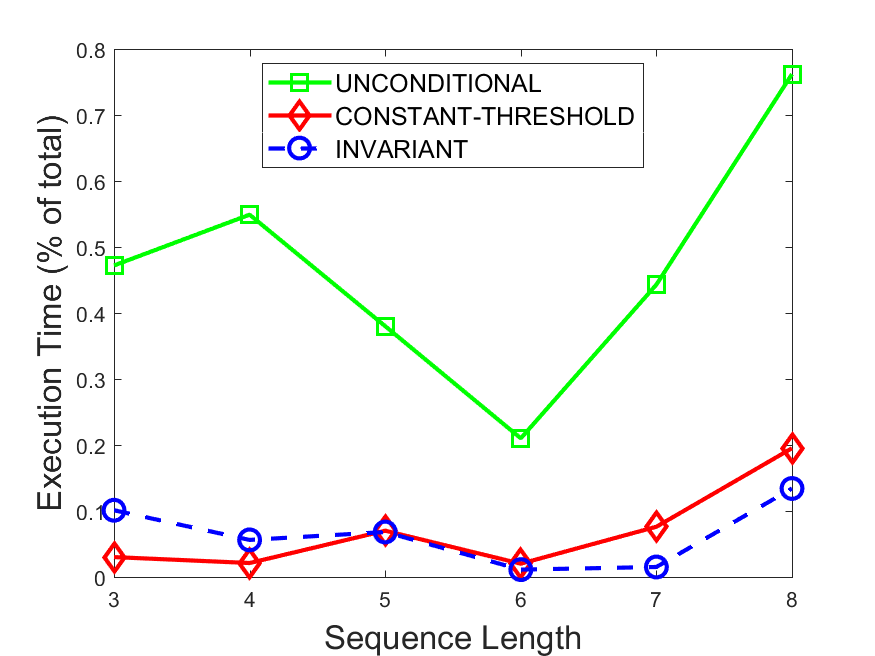}\label{fig:overhead}}
    \caption{Comparison of the adaptation methods applied on the stocks dataset in conjunction with the greedy algorithm (Kleene closure patterns): \protect\subref{fig:throughput} throughput (higher is better); \protect\subref{fig:relative-throughput} relative throughput gain over the non-adaptive method (higher is better); \protect\subref{fig:plan-switch} total number of plan reoptimizations; \protect\subref{fig:overhead} computational overhead (lower is better).}
	\label{fig:it-stocks-greedy}
\end{figure*}

\begin{figure*}
	\centering
	\subfloat[]{\includegraphics[width=.25\linewidth]{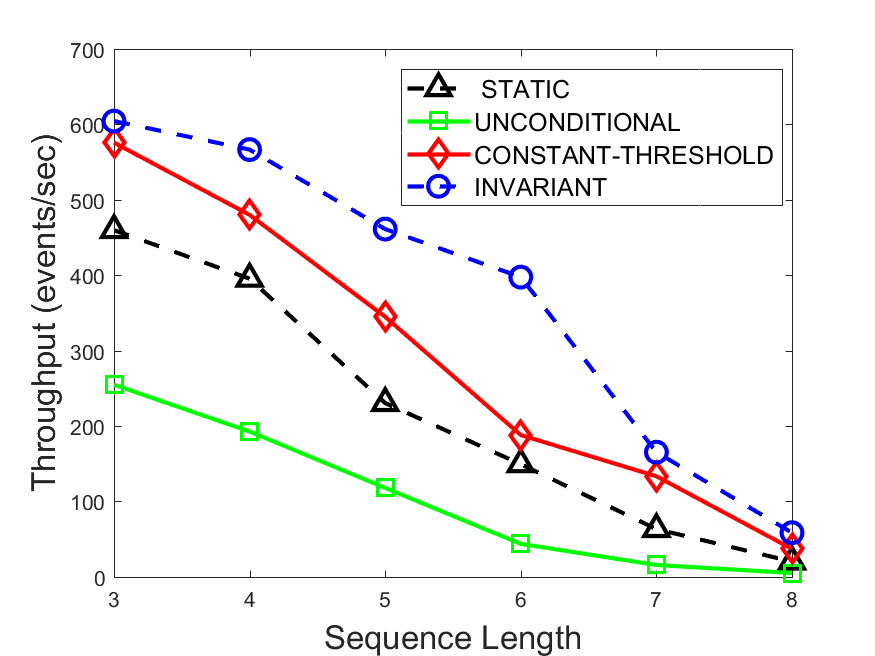}\label{fig:throughput}}
	\subfloat[]{\includegraphics[width=.25\linewidth]{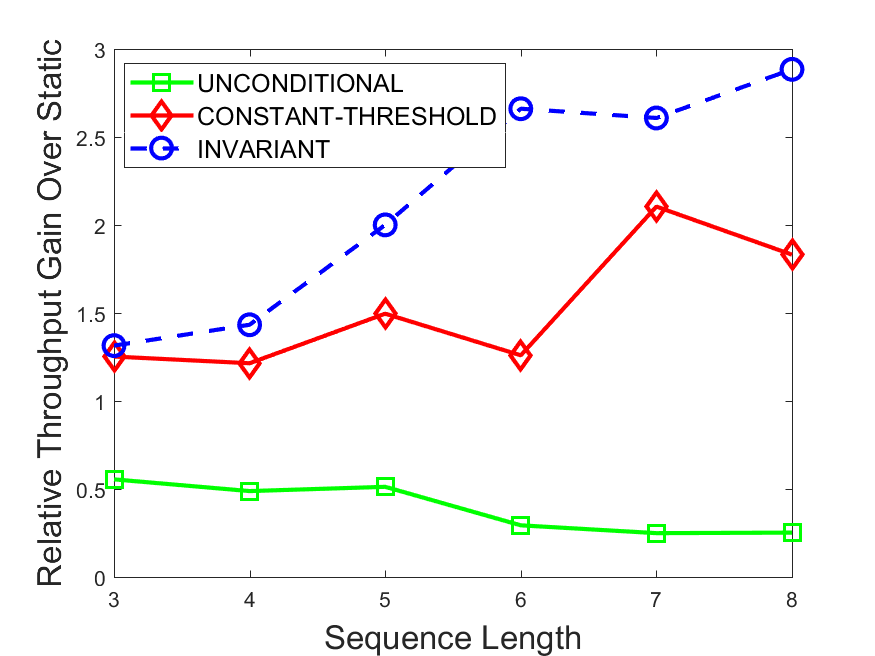}\label{fig:relative-throughput}}
	\subfloat[]{\includegraphics[width=.25\linewidth]{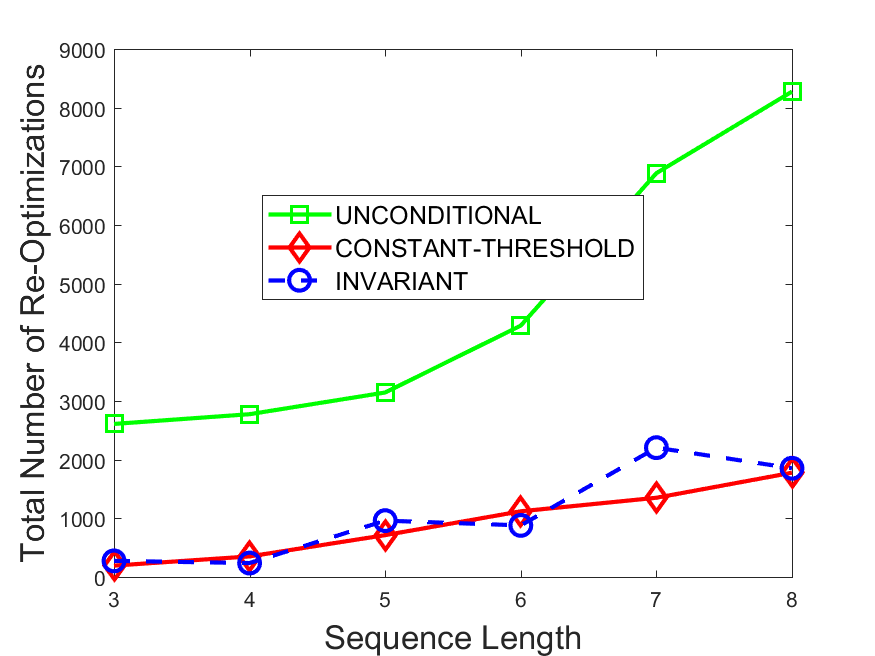}\label{fig:plan-switch}}
	\subfloat[]{\includegraphics[width=.25\linewidth]{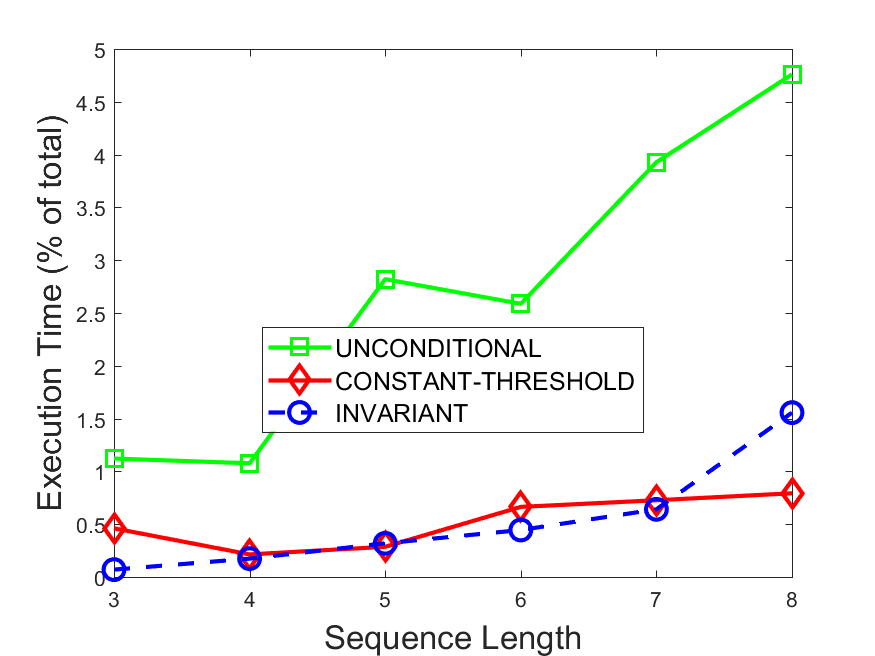}\label{fig:overhead}}
    \caption{Comparison of the adaptation methods applied on the stocks dataset in conjunction with ZStream algorithm (Kleene closure patterns): \protect\subref{fig:throughput} throughput (higher is better); \protect\subref{fig:relative-throughput} relative throughput gain over the non-adaptive method (higher is better); \protect\subref{fig:plan-switch} total number of plan reoptimizations; \protect\subref{fig:overhead} computational overhead (lower is better).}
	\label{fig:it-stocks-zstream}
\end{figure*}

\begin{figure*}
	\centering
	\subfloat[]{\includegraphics[width=.25\linewidth]{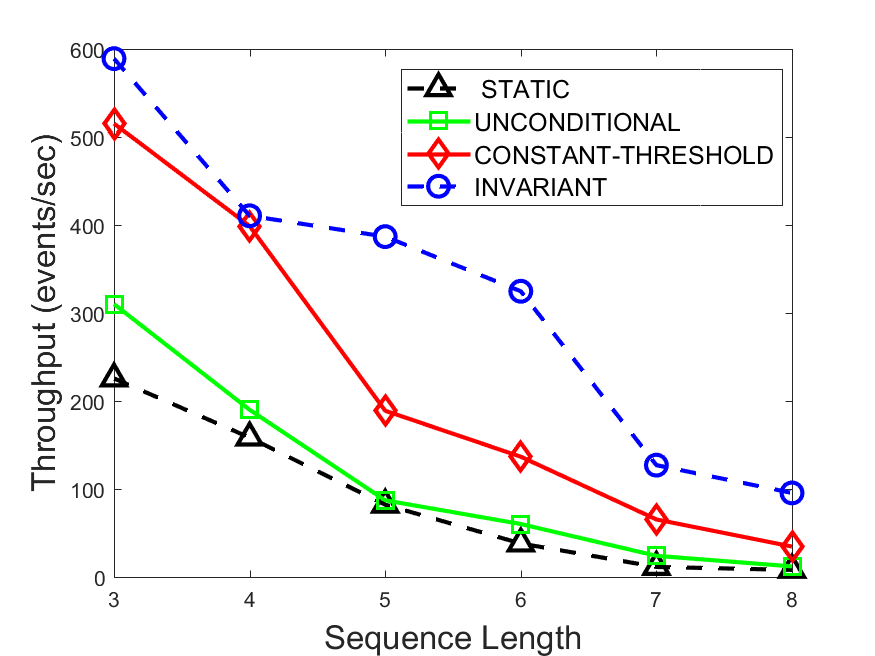}\label{fig:throughput}}
	\subfloat[]{\includegraphics[width=.25\linewidth]{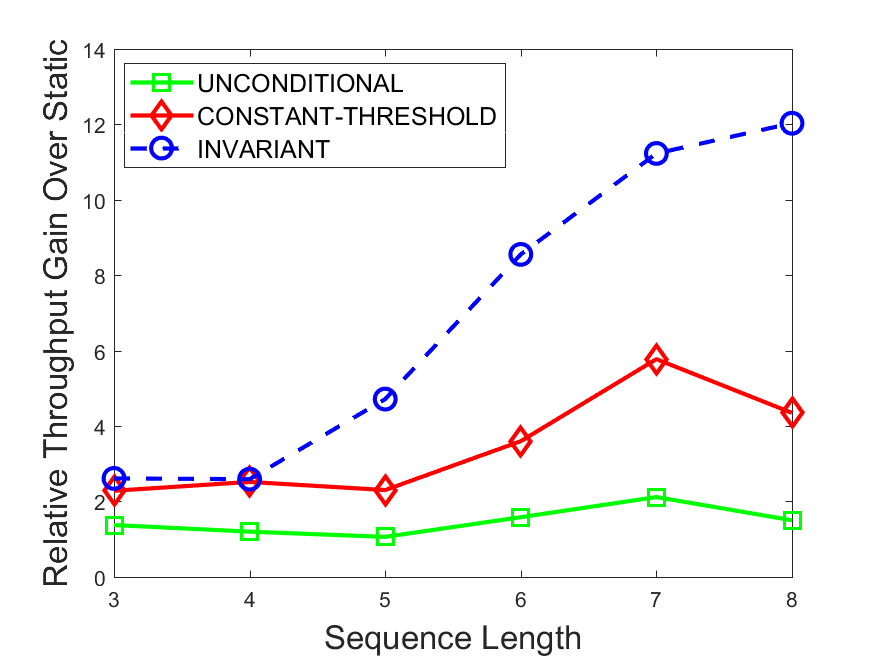}\label{fig:relative-throughput}}
	\subfloat[]{\includegraphics[width=.25\linewidth]{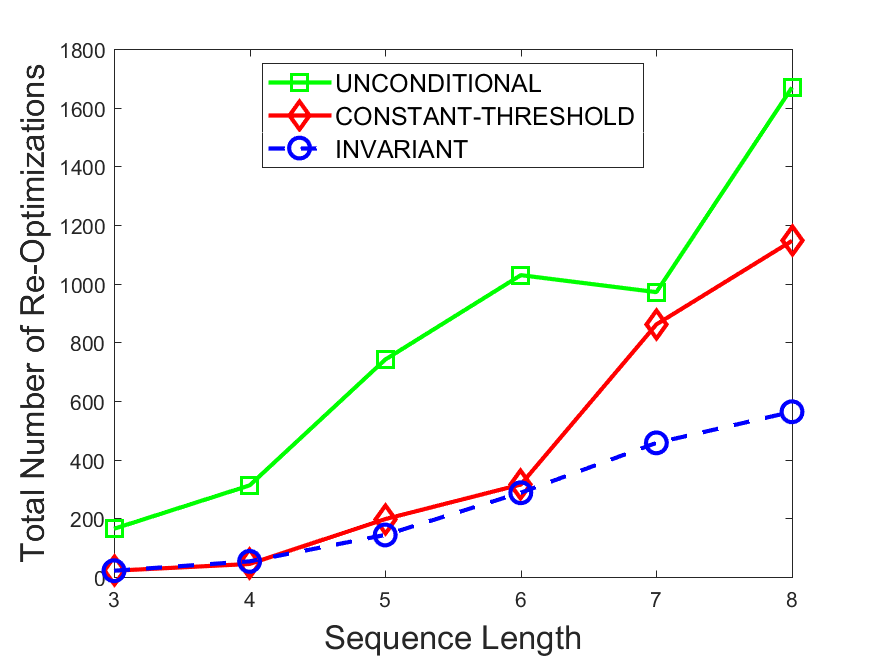}\label{fig:plan-switch}}
	\subfloat[]{\includegraphics[width=.25\linewidth]{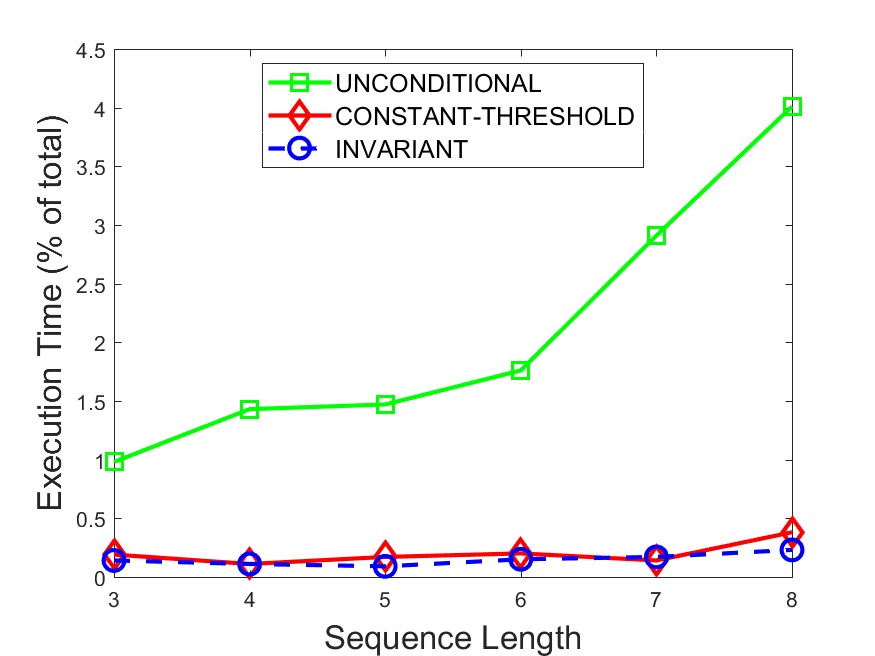}\label{fig:overhead}}
    \caption{Comparison of the adaptation methods applied on the traffic dataset in conjunction with the greedy algorithm (composite patterns): \protect\subref{fig:throughput} throughput (higher is better); \protect\subref{fig:relative-throughput} relative throughput gain over the non-adaptive method (higher is better); \protect\subref{fig:plan-switch} total number of plan reoptimizations; \protect\subref{fig:overhead} computational overhead (lower is better).}
	\label{fig:dis-traffic-greedy}
\end{figure*}

\begin{figure*}
	\centering
	\subfloat[]{\includegraphics[width=.25\linewidth]{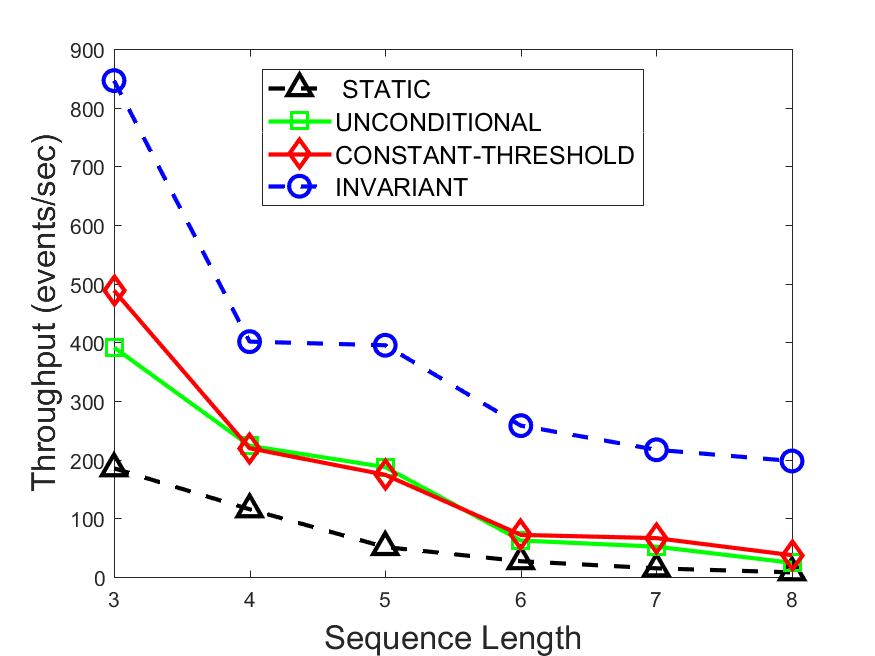}\label{fig:throughput}}
	\subfloat[]{\includegraphics[width=.25\linewidth]{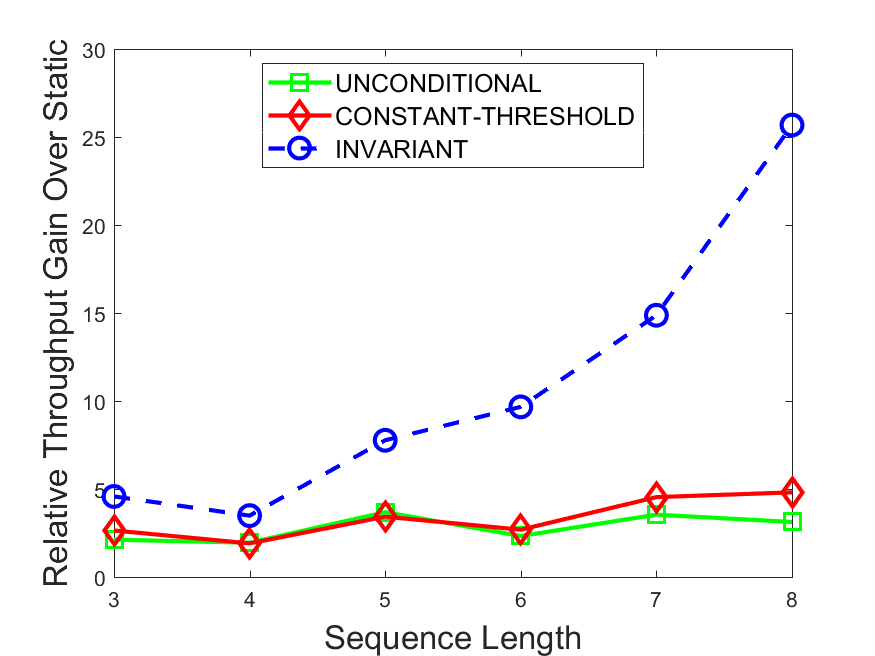}\label{fig:relative-throughput}}
	\subfloat[]{\includegraphics[width=.25\linewidth]{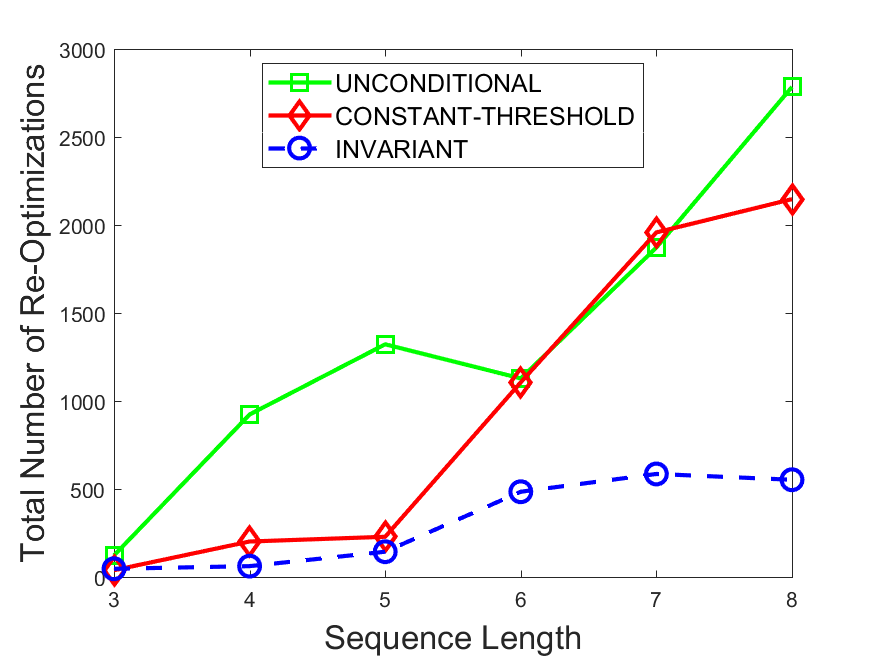}\label{fig:plan-switch}}
	\subfloat[]{\includegraphics[width=.25\linewidth]{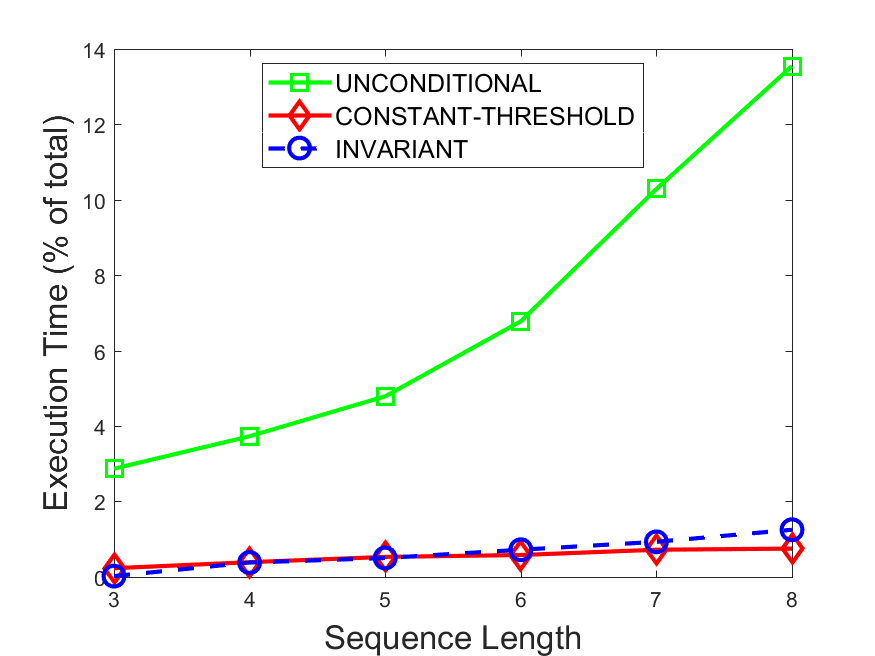}\label{fig:overhead}}
    \caption{Comparison of the adaptation methods applied on the traffic dataset in conjunction with ZStream algorithm (composite patterns): \protect\subref{fig:throughput} throughput (higher is better); \protect\subref{fig:relative-throughput} relative throughput gain over the non-adaptive method (higher is better); \protect\subref{fig:plan-switch} total number of plan reoptimizations; \protect\subref{fig:overhead} computational overhead (lower is better).}
	\label{fig:dis-traffic-zstream}
\end{figure*}

\begin{figure*}
	\centering
	\subfloat[]{\includegraphics[width=.25\linewidth]{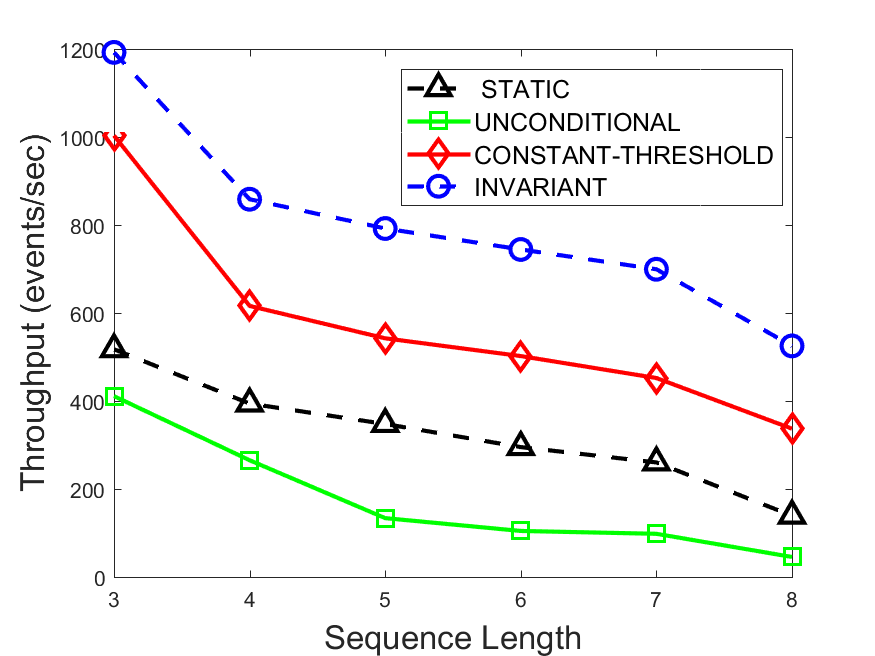}\label{fig:throughput}}
	\subfloat[]{\includegraphics[width=.25\linewidth]{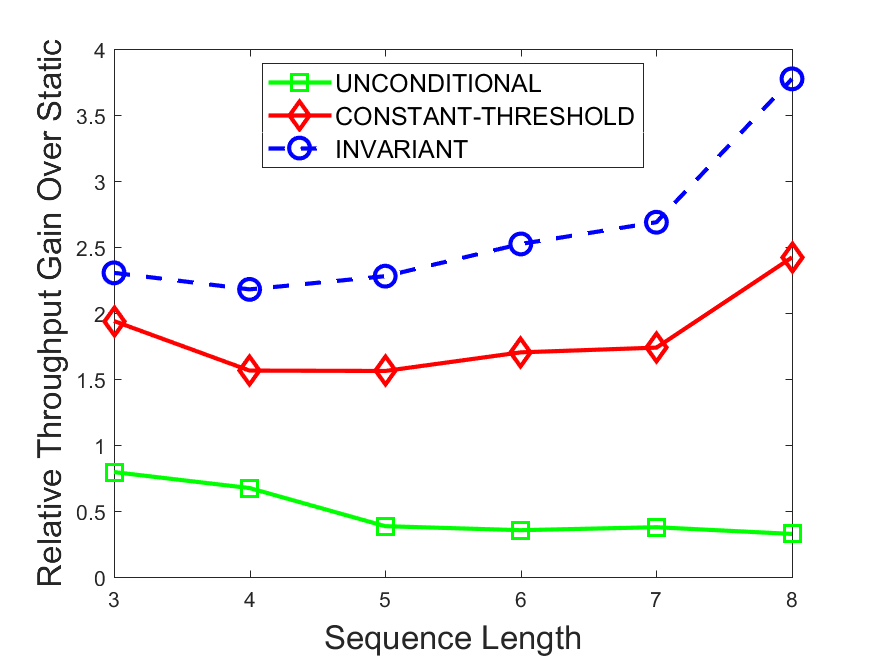}\label{fig:relative-throughput}}
	\subfloat[]{\includegraphics[width=.25\linewidth]{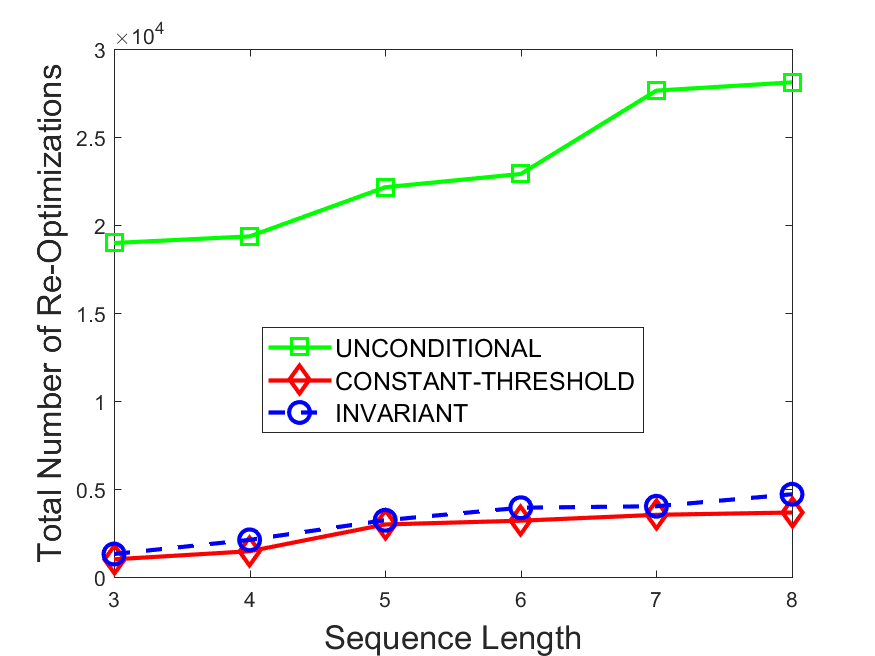}\label{fig:plan-switch}}
	\subfloat[]{\includegraphics[width=.25\linewidth]{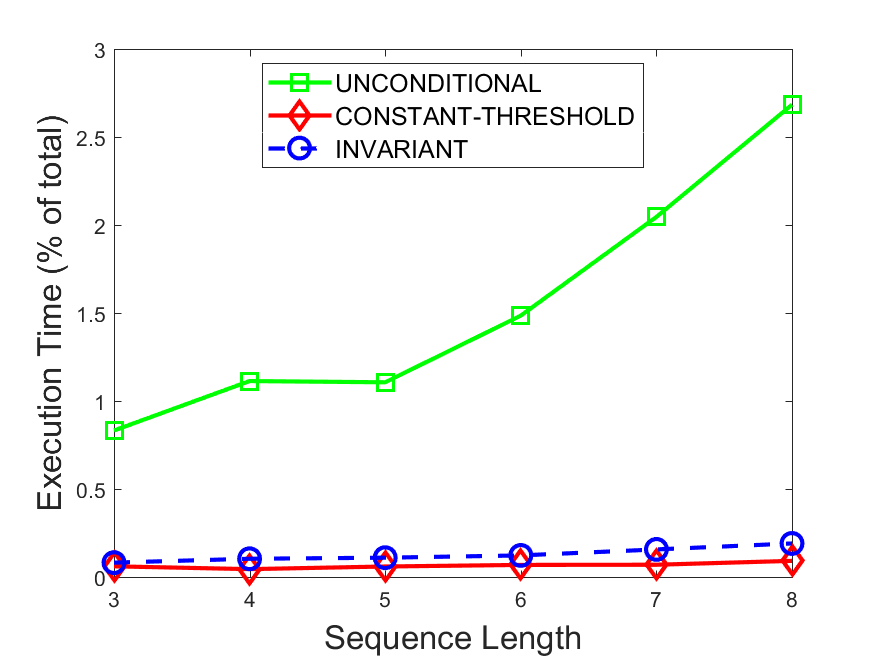}\label{fig:overhead}}
    \caption{Comparison of the adaptation methods applied on the stocks dataset in conjunction with the greedy algorithm (composite patterns): \protect\subref{fig:throughput} throughput (higher is better); \protect\subref{fig:relative-throughput} relative throughput gain over the non-adaptive method (higher is better); \protect\subref{fig:plan-switch} total number of plan reoptimizations; \protect\subref{fig:overhead} computational overhead (lower is better).}
	\label{fig:dis-stocks-greedy}
\end{figure*}

\begin{figure*}
	\centering
	\subfloat[]{\includegraphics[width=.25\linewidth]{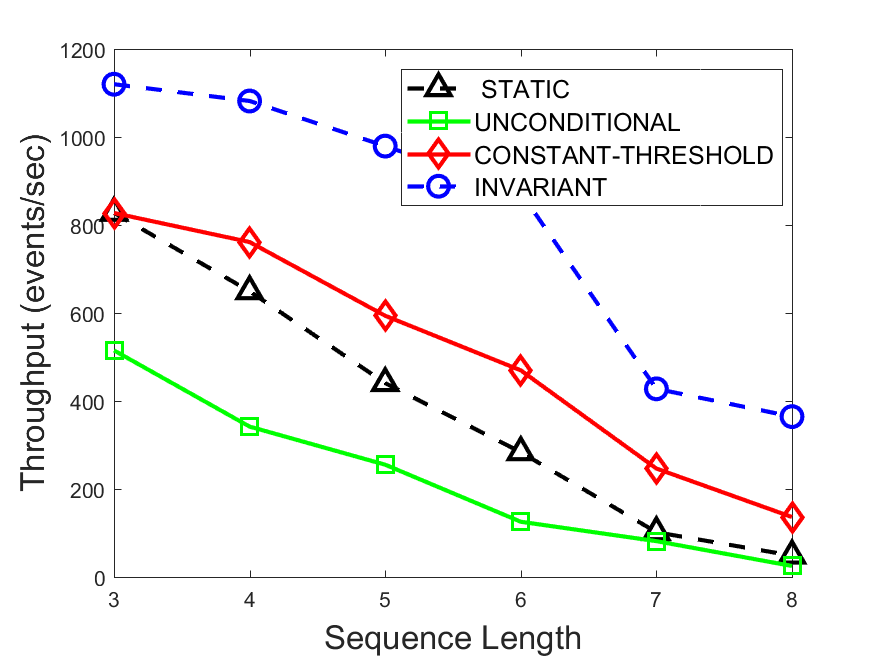}\label{fig:throughput}}
	\subfloat[]{\includegraphics[width=.25\linewidth]{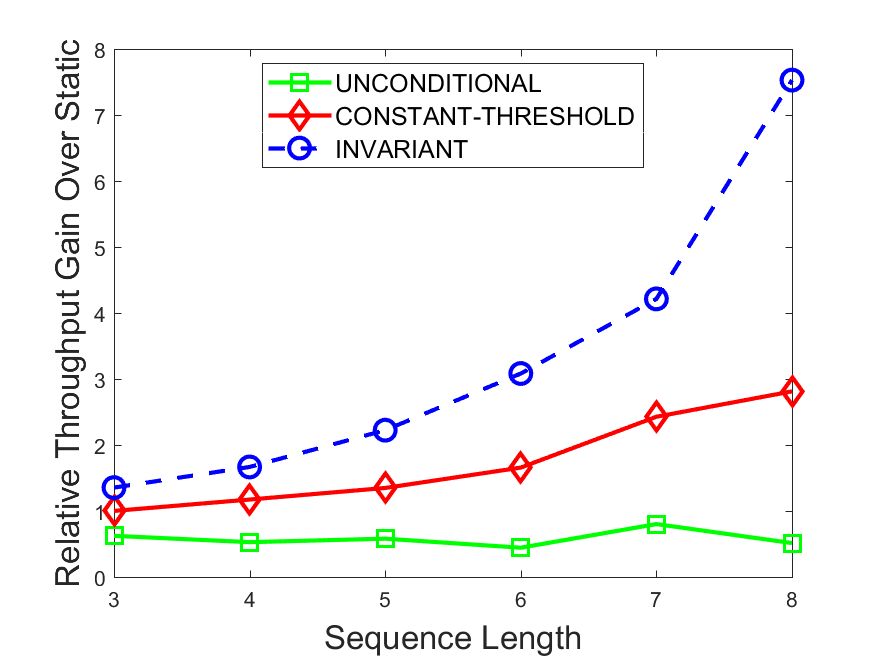}\label{fig:relative-throughput}}
	\subfloat[]{\includegraphics[width=.25\linewidth]{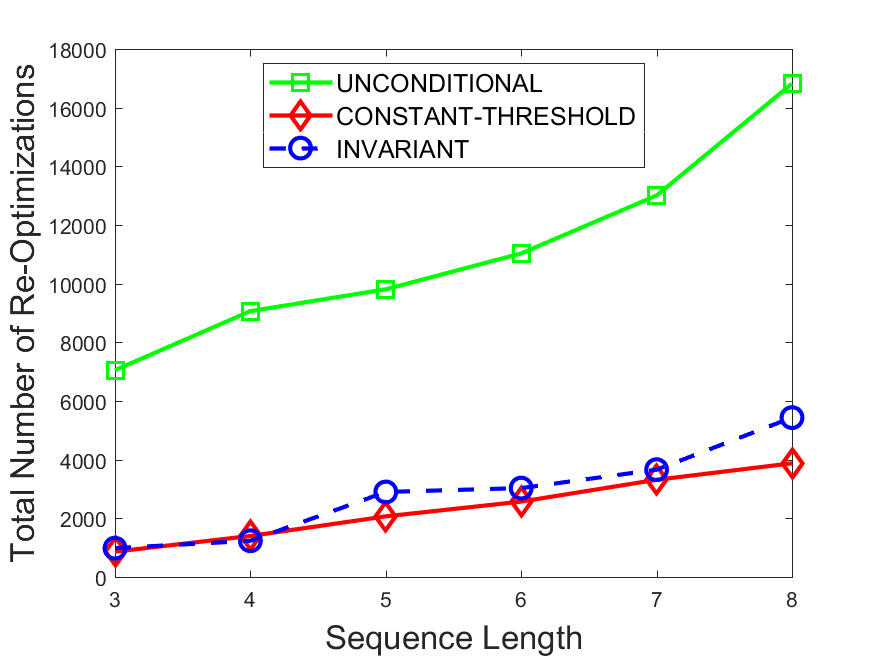}\label{fig:plan-switch}}
	\subfloat[]{\includegraphics[width=.25\linewidth]{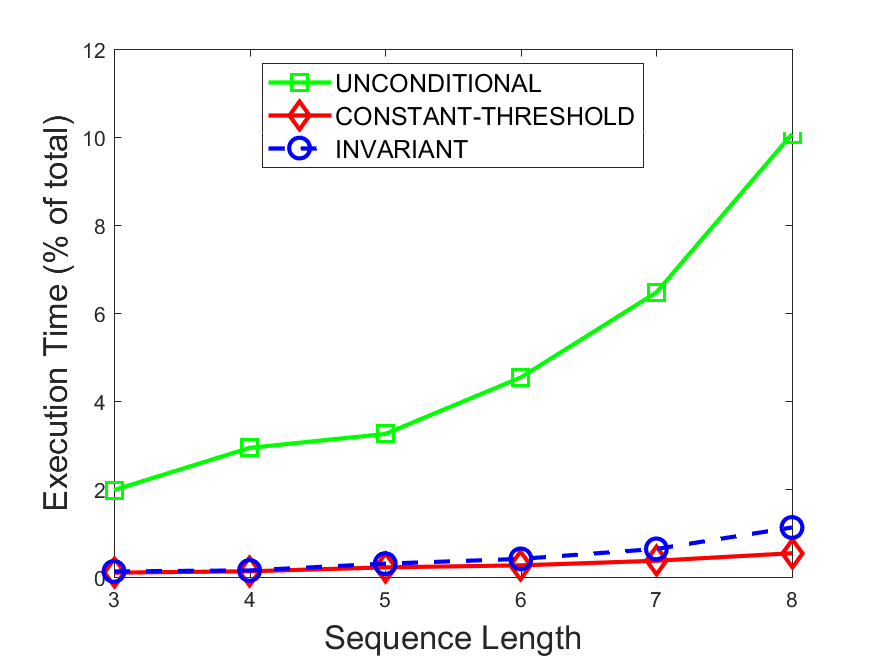}\label{fig:overhead}}
    \caption{Comparison of the adaptation methods applied on the stocks dataset in conjunction with ZStream algorithm (composite patterns): \protect\subref{fig:throughput} throughput (higher is better); \protect\subref{fig:relative-throughput} relative throughput gain over the non-adaptive method (higher is better); \protect\subref{fig:plan-switch} total number of plan reoptimizations; \protect\subref{fig:overhead} computational overhead (lower is better).}
	\label{fig:dis-stocks-zstream}
\end{figure*}

Five distinct set of patterns were used throughout the experiments
as specified below:
\begin{enumerate}
\item Sequence patterns - this set contains patterns with a single SEQ operator,
similar to the one demonstrated in Example 1. The results for sequence
pattern set follow the same trends as those discussed in Section \ref{sub:Experimental-Results}
and are displayed in Figures \ref{fig:seq-traffic-greedy}-\ref{fig:seq-stocks-zstream}.
\item Conjunction patterns - contains patterns with a single AND operator.
Each pattern in this set can be obtained by taking the pattern of
the same size from set 1 and removing the temporal constraints. For
this set, the relative gain of the considered adaptive methods was
higher than for other (non-composite) pattern sets. We can attribute
this result to the lower total selectivity of the inter-event conditions
(due to the lack of sequence constraints) and hence larger number
of intermediate partial matches, resulting in higher importance of
the correct adaptation decisions (Figures \ref{fig:con-traffic-greedy}-\ref{fig:con-stocks-zstream}).
\item Negation patterns - this set was produced from set 1 by adding a negated
event (an event under the negation operator) in a random position
in the pattern. Surprisingly, the introduction of this operator did
not significantly affect the experimental results, exhibiting nearly
identical relative throughput gains for all adaptive methods (Figures
\ref{fig:neg-traffic-greedy}-\ref{fig:neg-stocks-zstream}).
\item Kleene closure patterns - consists of sequence patterns containing
a single event under Kleene closure. This pattern demonstrated a significant
deviation from the rest in terms of the throughput measured for the
various adaptation methods. Due to the substantial complexity and
high cost of the Kleene closure operator regardless of its position
in the evaluation plan, the overall impact of the adaptation methods
was considerably lower as compared to other pattern sets. Still, the
invariant-based method was superior to the other algorithms in all
scenarios (Figures \ref{fig:it-traffic-greedy}-\ref{fig:it-stocks-zstream}).
\item Composite patterns - each pattern in this category is a disjunction
of three independent sequences. As each of the subsequences was evaluated
independently, the obtained results were very similar to those observed
for the sequence pattern set (Figures \ref{fig:dis-traffic-greedy}-\ref{fig:dis-stocks-zstream}).
\end{enumerate}
Each set contained 6 patterns varying in length from 3 to 8. For sets
1-4, the pattern size was defined as the number of events in a pattern.
Note that, while the events under the Kleene closure operator (set
4) are included in size calculation, while the negated events (set
) are excluded. For set 5, the definition was altered to reflect the
number of events in each subpattern.

\end{document}